\documentclass{iopjournal}

\usepackage{microtype}
\usepackage{ragged2e}

\AtBeginDocument{
  \justifying
  \emergencystretch=2em
}

\makeatletter
\long\def\@makecaption#1#2{
  \vskip\abovecaptionskip
  \small
  \noindent
  \begin{minipage}{\linewidth}
    \setlength{\parindent}{0pt}
    \setlength{\parskip}{0pt}
    \justifying
    \noindent #1. #2
  \end{minipage}
  \vskip\belowcaptionskip
}
\makeatother

\newacronym{SRMHD}{SRMHD}{special-relativistic magnetohydrodynamic}
\newacronym{EOS}{EOS}{equation of state}
\newacronym{GW}{GW}{gravitational wave}
\newacronym{BH}{BH}{black hole}
\newacronym{NSE}{NSE}{nuclear statistical equilibrium}
\newacronym[shortplural={CCSNe}, longplural={core collapse supernovae}]{CCSN}{CCSN}{core collapse supernova}
\newacronym{PNS}{PNS}{proto-neutron star}
\newacronym{SASI}{SASI}{standing accretion shock instability}
\newacronym{1D}{1D}{one-dimensional}
\newacronym{2D}{2D}{two-dimensional}
\newacronym{3D}{3D}{three-dimensional}
\newacronym{BNS}{BNS}{binary neutron star}
\newacronym{ET}{ET}{Einstein Telescope}
\newacronym{CE}{CE}{Cosmic Explorer}
\newacronym{EEMD}{EEMD}{ensemble empirical mode decomposition}
\newacronym{EMD}{EMD}{empirical mode decomposition}
\newacronym{IF}{IF}{instantaneous frequency}
\newacronym{IMF}{IMF}{intrinsic mode function}
\newacronym{sIMF}{sIMF}{significant IMF}
\newacronym{SNA}{SNA}{single nucleon approximation}
\newacronym{RMF}{RMF}{relativistic mean field}
\newacronym{NS}{NS}{neutron star}
\newacronym{ASD}{ASD}{amplitude spectral density}
\newacronym{PSD}{PSD}{power spectral density}
\newacronym{FWHM}{FWHM}{full width half maximum}
\newacronym{SNR}{SNR}{signal-to-noise-ratio}
\newacronym{LIGO}{LIGO}{Laser Interferometer Gravitational Wave Observatory}
\newacronym{KAGRA}{KAGRA}{Kamioka Gravitational Wave Detector}
\newacronym{SK}{SK}{Super-Kamiokande}
\newacronym{HK}{HK}{Hyper-Kamiokande}
\newacronym{JUNO}{JUNO}{Jiangmen Underground Neutrino Observatory}
\newacronym{DUNE}{DUNE}{Deep Underground Neutrino Experiment}
\newacronym{SNO+}{SNO+}{Sudbury neutrino observatory with liquid scintillator}
\newacronym{IBD}{IBD}{inverse beta decay}
\newacronym{MSW}{MSW}{Mikheyev-Smirnov-Wolfenstein}
\newacronym{LTWI}{LTWI}{low-$T/|W|$ instability}
\newacronym{TOV}{TOV}{Tolman-Oppenheimer-Volkoff}
\newacronym{MRI}{MRI}{magneto-rotational instability}
\glsdisablehyper
\newcommand{\de}{\mathrm{d}}
\newcommand{\Omegaaver}[1]{\langle{#1}\rangle_{\Omega}}

\begin{document}

\articletype{Paper}

\title{Impact of the equation of state on core collapse supernovae I: the low-$T/|W|$ instability}

\author{Marco Cusinato$^{1,2}$\orcid{0000-0003-4075-4539}, Martin Obergaulinger$^{1,3,*}$\orcid{0000-0001-5664-1382} and Miguel Ángel Aloy$^{1,3}$\orcid{0000-0002-5552-7681}}

\affil{$^1$Departament d'Astronomia i Astrof\'{\i}sica, Universitat de Val\`encia, Av. Vincent Andr\'es Estell\'es, 19, 46100 Burjassot, Spain}
\affil{$^2$Université Paris-Saclay, Université Paris Cité, CEA, CNRS, AIM, 91191 Gif-sur-Yvette, France}
\affil{$^3$Observatori Astronòmic, Universitat de València, 46980 Paterna, Spain}

\affil{$^*$corresponding author}

\email{marco.cusinato@uv.es, martin.obergaulinger@uv.es, miguel.a.aloy@uv.es}


\begin{abstract}

\begingroup\justifying\setlength{\parindent}{15pt}

\noindent
Rapidly rotating core-collapse supernovae are promising sources of multimessenger emission, as non-axisymmetric dynamics in the newly formed proto-neutron star can leave characteristic imprints on both gravitational waves and neutrinos. We present three-dimensional neutrino-magnetohydrodynamics simulations of the collapse of a rapidly rotating $\unit[35]{M_\odot}$ progenitor, performed with five different finite-temperature nuclear equations of state, to investigate how dense-matter physics affects the development of the \acrlong{LTWI} and its associated multimessenger signatures.

We find that the \acrlong{LTWI} develops in all \acrlong{EOS} models considered, indicating that its occurrence is robust for this rapidly rotating progenitor. However, its onset time, dominant azimuthal structure, lifetime, and characteristic multimessenger frequencies vary among models, reflecting differences in the evolving \acrlong{PNS} structure and rotation profile. The instability produces large-scale spiral modes that generate quasi-periodic \acrlong{GW} emission and modulate the neutrino luminosities, especially along directions close to the equatorial plane. 

The dominant \acrlong{GW} frequency associated with the instability correlates with the effective stiffness and compactness of the \acrlong{PNS}: models with more compact/stiffer configurations emit at higher frequencies. This suggests that, in rapidly rotating core-collapse supernovae, the frequency of the \acrlong{LTWI}-driven \acrlong{GW} signal may provide a diagnostic of the dense-matter \acrlong{EOS}, complementary to the information carried by the neutrino signal.
\par\endgroup

\end{abstract}

\section{\label{sec:introduction}Introduction}
\Glspl{CCSN} are among the most promising targets of multimessenger astrophysics. \Glspl{GW} and neutrinos, emitted from deep within the collapsing core, provide direct insight into the earliest phases of the dynamics and energetics of the system \cite{Burrows88,Kuroda16,Richers17,EggenbergerAndersen2021,Murphy2024}. In contrast, electromagnetic observations, originating at later times from the outer layers of the star, offer complementary constraints on the progenitor and on the properties of the ejecta \cite{Filippenko1997,Smartt2009, GalYam2017}. Likewise, the observation of supernova remnants offers the possibility of probing the circumstellar medium and nucleosynthesis products \cite{Vink_2012A&ARv..20...49}, and can probably be connected to the progenitor type and explosion mechanism \cite{Orlando2025}. 
The \glspl{GW} from Galactic events are expected to lie within the sensitivity of current ground-based detectors, such as Advanced \gls{LIGO} \cite{Aasi15}, Virgo \cite{Acernese15}, and \gls{KAGRA} \cite{Akutsu19}, while extragalactic explosions may become observable with future facilities, including \gls{CE} \cite{Srivastava22} and \gls{ET} \cite{Hild11}. Large underground neutrino experiments, such as \gls{SK} \cite{Abe2016}, \gls{HK} \cite{Dealtry2019}, \gls{JUNO} \cite{An2016}, \gls{DUNE} \cite{Abi2020}, IceCube \cite{Abbasi2011, Kopke2011}, and KM3NeT \cite{KM3NeT}, are expected not only to replicate the historic detection of the 25 neutrinos from SN~1987A \cite{Bionta87,Hirata87,Alexeyev88}, but to record thousands of events for a typical galactic \gls{CCSN}, with detectable signals also possible from supernovae in nearby satellite galaxies \cite{Suwa22}.

Massive stars ($M_\textnormal{ZAMS}\gtrsim \unit[8]{M_\odot}$) undergo nuclear fusion of progressively heavier elements, until they form an iron core  primarily supported by
the pressure of relativistic degenerate electrons. When the core surpasses the Chandrasekhar mass, it collapses to a \gls{PNS} \cite{Janka2016,Mueller_2020LRCA....6....3, Burrows2021, Yamada2024}. The sudden stiffening of the \gls{EOS} launches a shock-wave that initially stalls due to photodissociation of iron-group nuclei. The majority of successful explosions  are thought to be powered by the hydrodynamically aided neutrino-driven mechanism \cite{Wilson1982, Bethe1985, Janka2012}, in which a combination of  energy deposition by neutrinos behind the stalled shock and hydrodynamic instabilities such as convection and the \gls{SASI} revives the shock and powers the explosion. In more extreme cases, particularly in rapidly rotating and  magnetized progenitors, the explosion may instead be powered by magnetorotational processes \cite{Sawai_2016ApJ...817..153, Mosta2014, Obergaulinger2022, Bugli_2020MNRAS.492...58}. Conversely, if the shock fails to revive, the \gls{PNS} may collapse into a \gls{BH} \cite{Oconnor2011,Dessart2008, Obergaulinger2017,Burrows2021, Halevi2025}. In the last decade, these scenarios have been investigated by \gls{2D} and  \gls{3D} simulations \cite{Mosta2014, Matsumoto22, Wang24, Matsumoto24}.

Crucial to the evolution of the \gls{CCSN} is the \gls{EOS} of matter at supranuclear densities, which impacts the \gls{PNS} contraction, its thermal and compositional evolution, and the associated neutrino emission. The stiffness of the \gls{EOS} determines the radius, compactness, and maximum mass of the \gls{PNS}, thus influencing the likelihood of \gls{BH} formation and the timescale over which it may occur \cite{Pan2018, Powell2021, EggenbergerAndersen2025}. In recent years, several studies have investigated the impact of the \gls{EOS} on the explodability of the massive stars \cite{Janka2012, Suwa2013, Powell2025} and on the \gls{GW} emission using both \gls{2D} and \gls{3D} \cite{Marek2009, Richers17, EggenbergerAndersen2021, Jakobus2023, Murphy2024}. Such studies are motivated by and necessary due to the considerable theoretical and experimental uncertainties in the properties of matter at supranuclear densities. Experiments in terrestrial laboratories typically probe conditions that differ from those in \gls{CCSN} cores, and the complexity of the strong interaction makes theoretical calculations difficult. Hence, a large number of models for the nuclear \gls{EOS} have been proposed. Many of them are routinely used in numerical simulations of compact astrophysical objects and stellar explosions. The fact that the results of these models significantly depend on the chosen \gls{EOS} allows us to use them for putting additional constraints on the partly unknown parameters of nuclear-matter physics.

The \glspl{GW} emitted by \glspl{CCSN} are inherently stochastic, reflecting the wide range of physical mechanisms that contribute to their generation. In the early post-bounce phase, prompt convection excites oscillations with characteristic frequencies of a few hundred Hertz \cite{Marek2009, Murphy09, Yakunin10, Muller13, Yakunin15, Pan2018}, while convection within the \gls{PNS} excites $g$-modes that can reach frequencies of order $\gtrsim \unit[1]{kHz}$ \citep{TorresForne18, TorresForne19}. On longer timescales, post-shock convection \cite{Cusinato2025a} and \gls{SASI} \citep{Blondin2003, Blondin2007, CerdaDuran13, Kuroda16, Andresen2017, Mezzacappa2020, Mezzacappa2023} can develop, producing lower-frequency emission around $\sim\unit[100]{Hz}$. Rotation further modifies the \gls{GW} signal: moderate rotation can enhance the excitation of the \gls{PNS} $f$-mode \cite{Cusinato2025b}, while rapidly rotating cores may develop non-axisymmetric instabilities.

In particular, differentially rotating \glspl{PNS} can become unstable to the \gls{LTWI}, a shear-driven non-axisymmetric instability associated with the presence of strong differential rotation and an internal corotation point at which the frequency of one of the characteristic oscillation modes of the \gls{PNS} matches the local angular frequency of rotation \cite{Shibata2002,Shibata2003,Watts2005,Saijo2006,Shibagaki2020,Takiwaki2021,Bugli23}. 
Unlike the classical dynamical bar-mode instability, which requires a ratio between rotational and gravitational energy of $T/|W|\gtrsim 0.27$, the \gls{LTWI} can develop at substantially lower rotational energies, with typical thresholds of $T/|W|\sim0.04\text{--}0.06$, which can be reached in rapidly rotating \glspl{CCSN}. Strong differential rotation can lower this threshold even further 
\cite{Ott_2005ApJ...625L.119,Saijo2006,Scheidegger2008,Scheidegger2010b,Bugli23}. 
The instability leads to the growth of low-order spiral modes ($m=1,2$), generating long-lasting narrow-band \gls{GW} emission whose frequency is set by the rotational dynamics of the \gls{PNS}. Because the development of the \gls{LTWI} depends sensitively on the rotational profile, compactness, and internal structure of the \gls{PNS}, it is expected to be influenced by the nuclear \gls{EOS} through its influence on the contraction and differential rotation of the remnant.

In this study, we aim to characterise the impact of the nuclear \gls{EOS} on the multimessenger emission from \glspl{CCSN}, focusing on the \gls{GW} and neutrino signals produced during the shock-stall phase in rapidly rotating cores. We carry out a series of \gls{3D} neu\-tri\-no-mag\-ne\-to\-hy\-dro\-dy\-namics simulations of the core collapse of a single, rapidly rotating progenitor star with a zero-age main-sequence mass of $\unit[35]{M_{\odot}}$. Several previous studies of \gls{CCSN} have used the same stellar model as a progenitor \citep[e.g.,][]{Obergaulinger2017,Obergaulinger2022,Bugli23}, studying how variations of the magnetic field in the core at the moment of collapse affect the dynamics, the compact remnant, and the multimessenger signal. Among these previous works, we highlight the purely hydrodynamic model presented by \cite{Bugli23} in which the \gls{LTWI} led to the development of spiral modes and left an imprint in the \gls{GW} and neutrino emission. The present study is an extension of that one and analyses the results of five models of the same star with different nuclear \glspl{EOS}. We explore systematic differences in the dynamics, \gls{GW} signal, and neutrino emission between \glspl{EOS} differing in terms of the underlying input physics (Skyrme and \gls{RMF}-type approaches) and parameters such as their stiffness at supranuclear densities.  

The remainder of this paper is organized as follows. Section~\ref{sec:methods} outlines the numerical methods, progenitor model, and nuclear \glspl{EOS} employed in our \gls{CCSN} simulations. In Section~\ref{sec:results}, we present the main results of our study. Section~\ref{sec:discussion} discusses the implications of the \gls{EOS} for multimessenger detectability, and in Section~\ref{sec:conclusion} we summarise our conclusions.

\section{Methods}
\label{sec:methods}
\subsection{Numerical setup and progenitor model}
\label{sec:numerical_setup}
The \gls{3D} \gls{CCSN} simulations presented in this work were performed with the \texttt{Aenus-ALCAR} code \cite{Obergaulinger2008, Just2015, Just2018, Obergaulinger2018, Obergaulinger2022, Griffiths_2026a_arXiv260522927}. 
This code solves the equations of special-relativistic magnetohydrodynamics coupled to a spectral two-moment neutrino transport scheme. 
Specifically, it evolves the zeroth (energy density) and first (momentum density) moments of the Boltzmann equation, closed by the maximum-entropy Eddington factor \cite{Cernohorsky1994}, for electron neutrinos, $\nu_\textnormal{e}$, electron antineutrinos, $\overline{\nu}_\textnormal{e}$, and heavy-lepton neutrinos, $\nu_\textnormal{x} = (\nu_\mu,\,\overline{\nu}_\mu,\,\nu_\tau,\,\overline{\nu}_\tau)$.

The set of neutrino–matter interactions includes:
emission and absorption of electron neutrinos by protons and neutrons;
nucleonic absorption, emission, and scattering with weak-magnetism and recoil corrections;
emission and absorption of electron neutrinos by heavy nuclei;
coherent elastic scattering of all neutrino species off heavy nuclei;
inelastic scattering of neutrinos off electrons and positrons;
pair processes (electron–positron annihilation); 
and nucleonic bremsstrahlung. 
The contribution of gravity is included in the neutrino transport equations following the $\mathcal{O}(v/c)$-plus formulation of \cite{Endeve2012}. 
The nuclear composition is assumed to follow \gls{NSE} at high densities and temperatures, while at lower densities the transition to non-\gls{NSE} regimes is handled using the flashing scheme of \cite{Rampp2002}.

All simulations follow the evolution of a stellar model \texttt{35OC}, a Wolf–Rayet star with a zero-age main-sequence mass of $M_\textnormal{ZAMS} = \unit[35]{M_\odot}$ and subsolar metallicity ($Z = \unit[0.1]{Z_\odot}$).  This stellar model was originally evolved by \cite{Woosley2006} as a prototype  collapsar progenitor. However, subsequent studies (e.g.,  \cite{Dessart2008,Obergaulinger2017}) have shown that, in the context of \glspl{CCSN}, this model does not lead to prompt \gls{BH} formation, and if \gls{BH} forms at all, it does so only on timescales longer than $\mathcal{O}(\unit[1]{s})$. Moreover, a purely hydrodynamical simulation of this progenitor was shown to develop the \gls{LTWI} after core collapse \cite{Bugli23}.  Owing to its rapidly rotating core and its known tendency to sustain non-axisymmetric activity after bounce, \texttt{35OC} provides a suitable setup for studying the influence of the \gls{EOS} on the development of the \gls{LTWI}.

The simulations are evolved with the original rotation profile, but the progenitor magnetic field is replaced at the pre-supernova stage by a weak dipolar field. Modifying the magnetic field predicted by the stellar-evolution model is not fully self-consistent with the progenitor evolution. However, the magnetic fields in pre-supernova models are inferred from local hydrostatic properties rather than obtained from self-consistent MHD stellar-evolution calculations, and are typically available only in radiative stellar layers \cite{Aloy_2021MNRAS.500.4365, Griffiths_2026a_arXiv260522927, Griffiths_2026b_arXiv260522938}. The latter point is particularly important here, because it makes the magnetic-field structure provided by the pre-collapse stellar model spatially incomplete and lacking a physically well-defined global topology. 

For this reason, replacing the original magnetic-field profile by an analytically prescribed field is a physically motivated choice: it provides a regular magnetic configuration satisfying the solenoidal constraint, while also allowing us to reduce the initial field strength (see below).
The field geometry is defined by the vector potential $\mathbf{A}$ \cite{Suwa2007}:
\begin{equation}
    \label{eq:vector_potential}
    \left(A^r, A^\theta, A^\phi\right) = \frac{R_0^3 r}{2\left(r^3 + R_0^3\right)}\big(B_\textnormal{tor}\cos\theta, 0, B_\textnormal{pol}\sin\theta\big),
\end{equation}
where $B_\textnormal{tor} = B_\textnormal{pol} = \unit[10^6]{G}$ are the initial toroidal and poloidal magnetic field components, and $R_0 = \unit[10^8]{cm}$ is the characteristic length scale. The magnetic field strength of the pre-supernova progenitor is deliberately reduced because \cite{Obergaulinger2021} found no significant evidence for the development of the \gls{LTWI} in model \texttt{35OC} when using its original magnetic-field configuration, whereas \cite{Bugli23} did find the instability in the same progenitor assuming a vanishing magnetic field. While a strictly null magnetic field would suppress magnetic effects during the post-bounce evolution, adopting a weak seed field allows them to develop without making them dynamically important during the first $\unit[250]{ms}$ of evolution.

The simulations are performed in axisymmetry until $\unit[5]{ms}$ before core bounce, after which they are mapped onto \gls{3D} grids. The spherical grid consists of $n_\theta = 64$ zones in the $\theta$-direction and $n_\phi = 128$ zones in the $\phi$-direction, covering the full solid angle with an angular resolution of $2.8^\circ$ along each angular direction. In the radial direction the grid extends to $\unit[4\times10^{10}]{cm}$, and consists of two parts: a uniform grid from the centre to $\unit[16]{km}$ with a resolution of $\unit[5\times10^4]{cm}$, and a logarithmically stretched one from $\unit[16]{km}$ to the end of the domain. Except for model \texttt{LS}, which uses $n_r = 240$ zones, all models employ $n_r = 320$ radial zones. However, this choice of $n_r$ for model \texttt{LS} is not expected to have a significant impact on the results. The uniformly spaced inner part of the radial grid, where the \gls{PNS} forms and evolves, is essentially the same as in the other models. The differences are confined mainly to the logarithmically stretched outer region of the domain, at radii that are not affected by the dynamics over the simulated time span.

The energy domain is discretized into $n_\epsilon = 10$  bins extending from $\epsilon_\textnormal{min} = \unit[0]{MeV}$ to $\epsilon_\textnormal{max} = \unit[240]{MeV}$. The bin widths increase approximately logarithmically with energy, so that the low-energy bins are narrower and the high-energy bins progressively broader.

\subsection{Equation of state selection}
\label{sec:eos}
\begin{figure}
    \centering
    \includegraphics[width=0.7\textwidth]{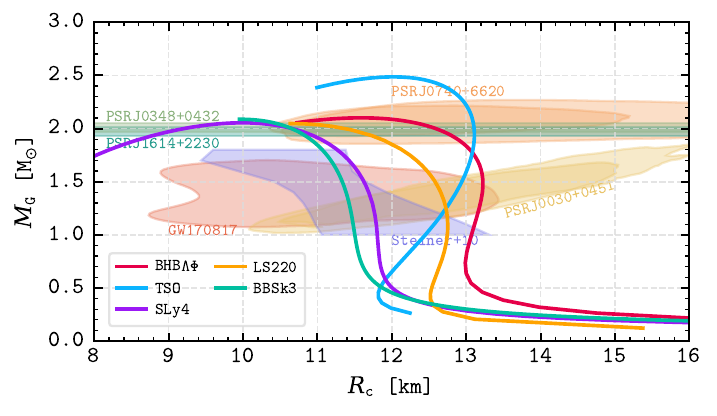}
    \caption{
    \label{fig:mr_eos}
    Gravitational mass-radius relations for cold, static, spherically symmetric, and non-rotating \glspl{NS} taken from \texttt{CompOSE} \cite{Typel2015,Oertel2017} using the \glspl{EOS} used in this work. The shaded regions represent astrophysical constraints at the $95\%$ confidence level. The yellow and orange regions display constraints from NICER observations \cite{Miller2019,Riley2019,Miller2021,Riley2021}; the red region corresponds to the detection of the GW170817 event \cite{LVK2018}; the purple region shows constraints from X-ray observations \cite{Steiner2010}; and the green regions indicate the maximum observed \gls{NS} masses \cite{Demorest2010,Antoniadis2013}.
    }
\end{figure}

The simulations employ two distinct \glspl{EOS}: one for high (nuclear) densities and one for low densities. The high-density \gls{EOS} is applied in regions with $\rho > \unit[10^6]{g/cm^{-3}}$ or where the temperature exceeds $T > \unit[0.5]{MeV}$.

The low-density \gls{EOS} (H.-Th. Janka, private communication) assumes nonrelativistic, nondegenerate nucleons and nuclei; arbitrarily relativistic and arbitrarily degenerate electrons and positrons; and a thermal photon bath, with Coulomb-lattice corrections.

To describe matter at high densities, we employ five finite-temperature, composition-dependent \glspl{EOS}: LS220 \cite{LS2201991}, SLy4 \cite{SLy2017}, BHB$\Lambda\phi$ \cite{BHBlp2014}, BBSk3 \cite{BBSk2025}, and TSO \cite{TSO2024}. The corresponding mass–radius relations for cold neutron stars in spherical symmetry are shown in Figure~\ref{fig:mr_eos}.

All these \glspl{EOS} are broadly consistent with current astrophysical constraints, including those derived from NICER observations \cite {Miller2019, Miller2021, Riley2019, Riley2021} and \gls{GW} detections \cite{Abbott2017, LVK19}. They can support \glspl{NS} more massive than the heaviest observed so far \cite{Demorest2010, Antoniadis2013}, and they agree with the constraints inferred from type-I X-ray bursters exhibiting photospheric radius expansion, as well as transient low-mass X-ray binaries \cite{Steiner2010}. Although LS220 does not satisfy all nuclear experimental constraints  \cite{Tews2017}, it is retained here to facilitate comparison with previous studies \cite{Takiwaki2021,Shibagaki2020,Bugli23}. 

Each \gls{EOS} accounts for protons, neutrons, electrons, positrons, photons, and $\alpha$ particles, as well as light nuclei (deuterium, tritium, and helium) and heavy nuclei in \gls{NSE}. The BHB$\Lambda\phi$ model further includes $\Lambda$ hyperons interacting through the $\phi$ meson.

The LS220 and SLy4 \glspl{EOS} are based on a nonrelativistic liquid-drop model with a Skyrme-type interaction. 
In particular, the LS220 model adopts a nuclear incompressibility modulus of $K = \unit[220]{MeV}$, which yields neutron star properties consistent with observational constraints. 

The BHB$\Lambda\phi$ \gls{EOS} adopts a \gls{RMF} framework, where the field equations are solved in the Hartree approximation. Additionally, at high densities it includes the full baryon octet—particularly $\Lambda$ hyperons—together with repulsive hyperon–hyperon interactions mediated by the $\phi$ meson.

The BBSk3 \gls{EOS} employs Brussels extended Skyrme interactions and was constructed within a Bayesian investigation of the dense-matter \gls{EOS} to ensure consistency with modern astrophysical constraints and with chiral effective field theory calculations of pure neutron matter.

Finally, the TSO \gls{EOS} follows the covariant density functional approach, including hypernuclear degrees of freedom, with baryonic coupling constants tuned such that the symmetry energy slope and skewness parameters are $L_\textnormal{sym} = \unit[30]{MeV}$ and $Q_\textnormal{sat} = \unit[400]{MeV}$, respectively.

\begin{table}
\centering
\resizebox{\textwidth}{!}{
\begin{tabular}{c|ccccccccc}
\hline
\multirow{ 2}{*}{\gls{EOS}} & $J$   & $L_{\rm sym}$ & $Q_{\rm sat}$ & $m^*/m$ & $B_1$ & $B_2$  & $M^\textnormal{grav}_\textnormal{max}$  & $M^\textnormal{bar}_\textnormal{max}$ & $\rho_0$\\
                            & [MeV] & [MeV]         & [MeV]         &         &  &[MeV$^{-1}$]   & [M$_\odot$]                             & [M$_\odot$]                           & [$10^{14}$ g/cm$^3$]\\
\hline\hline
SLy4 & 32.0 & 46.0 & $-365$ & 0.70 & 0.994 & 0.0311 & 2.05 & $\sim2.33$ & $2.66$ \\
BHB$\Lambda\phi$ & 31.9 & 55.0 & $+169$ & 0.56 & 1.031 & 0.0323 & 2.10 & $\sim2.40$ & $2.50$\\
BBSk3 & 29.6 & 39.6 & $-453$ & 0.461 & 1.621 & 0.0548 & 2.08 & $\sim2.38$ & $2.48$\\
TSO & 32.0 & 30.0 & $-400$ & 0.57 & 1.871 & 0.0585 & 2.47 & $\sim2.80$ & $2.50$\\
LS220 & 29.3 & 74.0 & $-411$ & 1.000 & 0.396 & 0.0135 & 2.05 & $\sim2.38$ & $2.60$\\
\hline
\end{tabular}
}

\caption{\label{tab:eoss_parameters}Cold \gls{EOS} indicators. Columns from left to right: \gls{EOS} name; symmetry energy at saturation density ($J$); slope of the symmetry energy at saturation density ($L_\textnormal{sym}$); skewness parameter ($Q_{\rm sat}$), taken or derived as follows: SLy4 from an explicit calculation using the Skyrme parameters of \cite{Chabanat_1998NuPhA.635..231} and formulae from \cite{Dutra_2012PhRvC..85c5201}; BBSk3 from \cite{Raduta2025}; LS220 from \cite{Oertel2017} (Table~IV; note that the quantity listed there corresponds to $Q_{\rm sat}$ in our convention); BHB$\Lambda\phi$ is mapped to the DD2 interaction in \cite{Oertel2017}; TSO from \cite{TSO2024}, adopting the sign convention consistent with the standard definition of $Q_{\rm sat}$; effective nucleon mass; heuristic \gls{EOS} parameters based on the symmetry-energy parameters, nucleon effective mass, and skewness parameter, $B_1$ Equation~\ref{eq:B1} and $B_2$  Equation~\ref{eq:B2}; maximum gravitational mass ($M^\textnormal{grav}_\textnormal{max}$), taken from the literature: SLy4 from \cite{Douchin_2001A&A...380..151,Oertel2017}, 
LS220 from \cite{Lattimer1991,Oertel2017}, 
BHB$\Lambda\phi$ from \cite{Banik2014}, 
BBSk3 from \cite{Raduta2025}, 
and TSO from \cite{TSO2024}; maximum baryonic mass ($M^\textnormal{bar}_\textnormal{max}$), estimated from typical \gls{TOV} solutions assuming $M^\textnormal{bar} - M^\textnormal{grav} \simeq 0.15$--$\unit[0.30]{M_\odot}$, depending on the \gls{EOS} stiffness; nuclear saturation density ($\rho_0$).}
\end{table}

 In Table~\ref{tab:eoss_parameters}, we summarise a set of nuclear-matter and \gls{NS} properties that provide insight into the behaviour of the different \glspl{EOS} employed in this work.
The parameters $J$ and $L_{\rm sym}$ characterise the magnitude and density dependence of the nuclear symmetry energy at saturation, while $Q_{\rm sat}$ encodes the skewness of symmetric nuclear matter and thus constrains the high-density behaviour of the pressure. The effective nucleon mass $m^*/m$ influences both the thermal properties of the \gls{EOS} and the neutrino-matter interaction rates. In addition, we report the maximum gravitational and baryonic masses of cold neutron stars, which provide integral constraints on the overall stiffness of the \gls{EOS}.

For later use, we introduce the combinations
\begin{align}
B_1 &= \frac{1}{m^*/m}\frac{J}{L_{\rm sym}}\label{eq:B1},   \\
B_2 &= \frac{1}{m^*/m}\frac{1}{L_{\rm sym}} \label{eq:B2},
\end{align}
These quantities are heuristic proxies rather than fundamental \gls{EOS} parameters. They are designed to combine, in a compact way, the overall scale of the symmetry energy ($J$), its density dependence ($L_{\rm sym}$), and the effective mass, which influences the thermal response and neutrino-matter interactions. Their role in this chapter is therefore purely comparative: they provide a convenient way to rank the adopted \glspl{EOS} and to test whether broad trends in the simulations correlate with simple combinations of cold nuclear-matter properties. Any such correlation should, however, be interpreted as suggestive only, since the effective stiffness of the \gls{PNS} during the evolution depends on the full hot and composition-dependent \gls{EOS}.

During the \gls{CCSN} evolution, the \gls{PNS} continuously accretes mass and contracts, probing a wide range of densities and thermodynamic conditions. As a consequence, its effective stiffness is not a fixed property but evolves in time and cannot be uniquely characterised by nuclear-matter parameters defined at saturation density alone.
To capture this behaviour, we adopt the tidal Love number $\kappa_2$ as an effective diagnostic of the \gls{PNS} stiffness. Here, $\kappa_2$ is used only as a comparative indicator of the instantaneous structure of the hot, evolving \gls{PNS}, and not as a direct characterisation of the corresponding cold-\gls{NS} \gls{EOS}. We compute it following \citet{Hinderer2010}. However, we stress that the procedure to compute $\kappa_2$ formally applies to cold, static, and spherically symmetric neutron stars in hydrostatic equilibrium. In the \gls{CCSN} context, the \gls{PNS} is instead hot, accreting, and rotating. Therefore, the resulting $\kappa_2$ should be regarded as an effective indicator of the \gls{PNS} stiffness. In practice, we evaluate it using spherically averaged profiles, which allows us to track relative differences among \glspl{EOS} during the evolution. Despite these limitations, $\kappa_2$ provides a useful proxy for comparing the compressibility of the \gls{PNS} across different models. Therefore, when we compare two  \glspl{EOS} we refer to one as \emph{stiffer} (\emph{softer}) than the other if the value of $\kappa_2$ is smaller (larger). 
Here we use $\kappa_2$ only as an effective structural proxy for comparing our hot, evolving \gls{PNS} models. This usage should not be confused with the standard cold-\gls{NS} literature, where \gls{EOS} stiffness is more commonly discussed in terms of radius or tidal deformability $\Lambda= \frac{2}{3}  \kappa_2 C^{-5}$ at fixed mass (where $C=GM^{\rm grav}/(Rc^2)$ denotes the compactness of the star, with $G$ and $c$ the gravitational constant and the speed of light in vacuum, respectively), rather than through a monotonic ranking of $\kappa_2$ alone.

\subsection{Gravitational wave extraction}
\label{sec:GWextraction}
The \glspl{GW} signal is extracted  with the standard quadrupole formula \cite{MTW1973,Thorne1980},
\begin{equation}
    \label{eq:quadrupole_formula}
    h_{ij}(t,\mathcal{D})=\frac {2G}{c^{4}\mathcal{D}}{\ddot {Q}}_{ij}(t-\mathcal{D}/c),
\end{equation}
where $\mathcal{D}$ is the distance to the source, $t$ is the time and $Q_{ij}$ are the mass quadrupole moments. In spherical coordinates, the mass quadrupole moments are given by
\begin{equation}
    \label{eq:simfra_quad_sph}
    {Q}_{lm}=\int \rho (r, \theta, \phi, t )r^2Y^*_{lm}(\theta, \phi)\textnormal{d}V,
\end{equation}
where $Y^*_{lm}$ are the spherical harmonics and $\rho$ the density. By applying the continuity equation and performing a partial integration, the first time derivative of the quadrupole moment can be rewritten in terms of spatial derivatives.
Rather than computing the contribution from the entire volume occupied by the emitting source, we first consider the contribution from a spherical shell bounded by an inner radius $r_1$ and an outer radius $r_2$. The contribution from the whole source is then recovered straightforwardly by taking the limits $r_1 \to r_{\rm min}$ and $r_2 \to r_{\rm max}$, where $r_{\rm min}$ and $r_{\rm max}$ denote the innermost and outermost radial boundaries of the source, respectively.

Applying the continuity equation and performing a partial integration, the first time derivative of the quadrupole moment associated with this shell can be written as
\begin{equation}
    \label{eq:simfra_dtdQ_shell}
    \left.\dot{Q}_{lm}\right|_{r_1}^{r_2}
    =
    \int_{r_1}^{r_2}\de r\, r^2
    \int \de\Omega\,
    \rho\,\mathbf{v}\cdot
    \nabla\!\left(r^2Y^*_{lm}(\theta,\phi)\right)
    -
    \left[
    \int \de\Omega\,
    \rho\,v_r\,r^4Y^*_{lm}(\theta,\phi)
    \right]_{r_1}^{r_2},
\end{equation}
where $\mathbf{v}$ is the fluid velocity, $v_r$ its radial component, and
\[
    \left[F(r)\right]_{r_1}^{r_2} \equiv F(r_2)-F(r_1).
\]
The first term in Equation~\eqref{eq:simfra_dtdQ_shell} is the volume contribution, while the second term is the surface contribution arising from the radial boundaries of the shell.

This surface term is often irrelevant when the quadrupole moment is evaluated over the full computational domain and the boundary contribution vanishes. However, \citet{Zha2024} showed that--in \gls{2D} simulations--when decomposing the \gls{GW} signal into contributions from subdomains of the simulation volume, neglecting this term leads to an incorrect spatial attribution of the emission. It must therefore be retained in the shell decomposition used here.

Finally, following \citet{Raynaud2022} the \gls{GW} strain generated by a subdomain of the star can be expressed as
\begin{equation}
    h(t, \mathcal{D})\big|_{r_1}^{r_2} = h_+-ih_\times = \frac{1}{\mathcal{D}}\frac{G}{c^2}\frac{8\pi}{5}\sqrt{\frac{2}{3}}\sum_{m=-2}^{+2} \frac{\partial}{\partial t}(\dot{Q}_{2m})\big|_{r_1}^{r_2}{}_{-2}Y_{2m}(\Theta,\Phi),
\end{equation}
where $\Theta$ and $\Phi$ are the angles associated with the location of the source in the sky, and ${}_{-2}Y_{2m}$ are the spin-weighted spherical harmonics.

\subsection{Ensemble Empirical Mode Decomposition}
\label{sec:eemd}
To decompose our set of \gls{GW} signals into a finite number of simple oscillatory components, known as \glspl{IMF}, we employ the \gls{EEMD} method \cite{Huang98,Huang99,Wu09}, following the approach described in \citet{Cusinato2025a}. Previous studies have employed the \gls{EEMD} to identify \gls{SASI}-induced modulations and to reconstruct \glspl{GW} emitted from \gls{CCSN} embedded in detector noise \cite{Takeda2021, Hu2022, Yuan2024}. In  this work we apply the \gls{EEMD} to extract and characterise the \gls{GW} signal associated to the \gls{LTWI}. 
This technique mitigates mode mixing by adding random Gaussian noise to the signal before each decomposition; the resulting \glspl{IMF} are then averaged to obtain a noise-cancelled ensemble.

The decomposition of each \gls{GW} signal in the post-bounce time interval $t \in [\unit[-0.005]{\rm s},\,\unit[0.25]{\rm s}]$ is performed as follows:
\begin{enumerate}[i.]
    \item Filter the signal to remove frequencies above \unit[3000]{Hz};
    \item Apply standard \gls{EMD} to extract the \glspl{IMF} and the residual;
    \item Subtract the residual from the filtered signal;
    \item Perform \gls{EEMD}, limiting the number of \glspl{IMF} to ten, with parameters $\sigma_0 = 1$ and $n = 2\times10^6$, where $\sigma_0^2$ is the variance of the random Gaussian noise, and $n$ the number of realizations of the signal plus Gaussian noise used to produce a noise-cancelled ensemble.
\end{enumerate}

The \gls{EMD} and \gls{EEMD} calculations are carried out using the \texttt{PyEMD} package \cite{pyemd}, employing the \texttt{akima} interpolation method \cite{Akima1970} to determine the local extrema.

\subsection{Criteria to identify the LTWI in our models}
\label{sec:LTWI_criteria}

Since there is no analytic theory defining the exact conditions under which the \gls{LTWI} develops, in this section, we show that our models, in addition to the low magnitude of $T/|W|$, satisfy several other diagnostics supporting the interpretation of the observed spiral patterns as manifestations of the \gls{LTWI}. 
\paragraph{Pattern frequency} The instability exhibits an approximately constant pattern frequency during its active phase. This behaviour is characteristic of shear-type non-axisymmetric instabilities associated with a well-defined corotation point, in contrast to transient convective or stochastic spiral structures \cite{Watts2005,Passamonti2015}. The nearly time-independent frequency observed in our models is consistent with this interpretation (see Section~\ref{sec:corotation}). 
For a non-axisymmetric perturbation with azimuthal dependence $\propto \cos(m\phi-\omega_{\rm mode} t)$, the quantity $\omega_{\rm mode}$ defines the mode angular frequency, while
\begin{equation}
    \label{eq:lowTW_pattern}
    \Omega_{\rm p} = \frac{\omega_{\rm mode}}{m}
\end{equation}
is the corresponding pattern angular frequency, i.e., the angular rate at which the spiral structure rotates as a whole. 

\paragraph{Corotation region} A corotation radius, $r_\textnormal{cor}$, exists inside the \gls{PNS}, defined by the condition
\begin{equation}
    \label{eq:lowTW_cor_radius_condition}
    \Omega_{\rm rot}(r_{\rm cor}) = \Omega_\textnormal{p}.    
\end{equation}
The presence of such a corotation point is widely regarded as a necessary condition for the development of the \gls{LTWI}, as shown in both idealized equilibrium models and \gls{CCSN} simulations \cite{Shibata2002,Shibata2003,Watts2005,Saijo2006}. As we shall see, in our models, the measured mode frequency lies within the range of rotational frequencies inside the \gls{PNS}, ensuring the existence of an internal corotation region.

\paragraph{Differential rotation} The instability develops only while significant differential rotation is present. The growth of the spiral coincides with phases in which the fraction of differential rotational energy remains substantial, and weakens as differential rotation is reduced. This dependence on shear is a defining property of the \gls{LTWI} \cite{Saijo2006,Passamonti2015,Bugli23}.

\paragraph{Convective region} The instability preferentially grows when the corotation region overlaps with a convectively unstable layer. Previous studies have shown that buoyancy and shear can cooperate to facilitate mode growth in differentially rotating stars \cite{Saijo2006,Takiwaki2021,Bugli23}. In our simulations, the spiral amplitude peaks when the corotation region lies within a zone where the square of the Brunt--Väisälä frequency 
\begin{equation}
    N^2 =
\frac{1}{\rho}\frac{\partial\Phi}{\partial r}
\left(
\frac{1}{c_s^2}\frac{\partial p}{\partial r}
-
\frac{\partial\rho}{\partial r}
\right),
\end{equation}
satisfies $N^2 < 0$ (see Section~\ref{sec:corotation}), further supporting this interpretation.

\paragraph{Dominant azimuthal modes} The dominant non-axisymmetric modes correspond to low-order azimuthal numbers ($m=1,\,2$), which are characteristic signatures of the \gls{LTWI} reported in both equilibrium studies and \gls{CCSN} simulations \cite{Shibata2002,Shibagaki2021,Bugli23}. The observed morphology and modal hierarchy are therefore consistent with the expected behaviour of the \gls{LTWI}.

\section{Results}
\label{sec:results}

\begin{table}[!ht]
    \centering
    \begin{tabular}{c|ccccc}
    \hline
         Model name                              & \texttt{BH}      & \texttt{TS}      & \texttt{SL}      & \texttt{LS}      & \texttt{BB} \\\hline\hline
         \gls{EOS}                               & BHB$\Lambda\phi$ & TSO              & SLy4             & LS220            & BBSk3 \\\hline
         $T/|W|_\textnormal{ic,b}$  [$10^{-2}$]               & 5.2              & 6.0              & 5.8              & 5.8              & 6.0 \\
         $M_\textnormal{ic,b}$ [$M_\odot$]         & 0.56             & 0.48             & 0.57             & 0.46             & 0.50  \\
         $R_\textnormal{ic,b}$ [km]                & 12               & 17               & 12               & 19               & 19 \\
         $J_\textnormal{ic,b}$ [$10^{48}$ erg$\cdot$s] & 1.66         & 1.26             & 1.75             & 1.20             & 1.39 \\
         $E_\textnormal{diff,ic,b}/T_\textnormal{ic,b}$ & 0.65         & 0.68             & 0.73             & 0.69             & 0.69 \\\hline
         $T/|W|_\mathrm{b}$ [$10^{-2}$]          & 5.23             & 6.01             & 5.76             & 5.78             & 5.97\\
         $\kappa_{2,[1,15]}$ $[10^{-3}]$          & $24.8^{\pm2.9}$  & $27.6^{\pm2.6}$  & $26.0^{\pm2.4}$  & $23.9^{\pm2.3}$  & $26.6^{\pm2.3}$\\
         $\Delta h_\mathrm{+, eq}$ [cm]          & 441              & 439              & 363              & 384              & 491\\
         $L_{\nu_e,\textnormal{peak}}$ [$10^{53}$ erg/s] & $5.02$   & $5.88$           & $6.97$           & $4.28$           & $6.02$ \\*[1pt]\hline
         $\kappa_2$ $[10^{-3}]$                  & $18.2^{\pm1.1}$  & $20.6^{\pm1.3}$  & $18.9^{\pm0.7}$  & $16.8^{\pm1.5}$  & $20.3^{\pm1.1}$\\
         $f_{11}$ [Hz]                           & $159^{\pm30}$    & $181^{\pm40}$    & $185^{\pm22}$    & $181^{\pm30}$    & $174^{\pm20}$\\
         $f_{22}$ [Hz]                           & $312^{\pm18}$    & $291^{\pm30}$    & $322^{\pm35}$    & $367^{\pm40}$    & $278^{\pm36}$\\
         $f_{\nu, 1}$ [Hz]                       & $152^{\pm30}$    & $176^{\pm35}$    & $192^{\pm31}$    & $184^{\pm35}$    & $168^{\pm44}$\\
         $f_{\nu, 2}$ [Hz]                       & $320^{\pm47}$    & $296^{\pm35}$    & $336^{\pm55}$    & $376^{\pm49}$    & $288^{\pm31}$\\    
         $f_\textnormal{GW}$ [Hz]                & $312^{\pm18}$    & $296^{\pm14}$    & $328^{\pm12}$    & $360^{\pm15}$    & $288^{\pm17}$\\
         $f_{h_\textnormal{char}/\sqrt{f}}$ [Hz] & $314$            & $294$            & $333$            & $373$            & $290$ \\\hline
         $\kappa_{2,250}$ $[10^{-3}]$            & $16.9$           & $18.6$           & $16.1$           & $13.3$           & $19.1$\\
         $\Omegaaver{R_\textnormal{PNS}}$ [km]   & $62^{+10}_{-14}$ & $65^{+9}_{-13}$  & $62^{+10}_{-12}$ & $59^{+12}_{-13}$ & $66^{+9}_{-11}$ \\
         $M_\mathrm{PNS}$ [M$_\odot$]            & $2.02$           & $1.98$           & $1.98$           & $1.85$           & $1.98$\\
         $\Omegaaver{R_\mathrm{shock}}$ [km]     & $184$            & $204$            & $209$            & $258$            & $196$\\
         $E_\mathrm{GW}$ [$10^{47}$ erg]         & $2.13$           & $1.90$           & $3.33$           & $3.30$           & $2.03$\\
         $E_\nu$ [$10^{52}$  erg]                & $6.62$           & $6.28$           & $6.22$           & $5.93$           & $6.23$ \\
         $P_\textnormal{rot}$ [ms]               & $4.14^{\pm1.35}$ & $4.91^{\pm1.92}$ & $4.44^{\pm1.65}$ & $4.02^{\pm1.43}$ & $5.74^{\pm1.97}$ \\\hline
    \end{tabular}
    \caption{Models summary and quantitative results. Rows from top to bottom provide: 
    \gls{EOS} employed in the simulation;
    ratio between innercore rotational and gravitational energy at core bounce;
    innercore mass at core bounce;
    innercore radius at core bounce;
    angular momentum at core bounce;
    ratio between differential rotation and rotational energy for the innercore at core bounce;
    ratio between \gls{PNS} kinetic rotational energy and gravitational energy at bounce ($T/|W|$); 
    \gls{PNS} average and standard deviation (superscript) of the tidal Love number ($\kappa_{2,[1,15]}$) in the interval $\unit[1-15]{ms}$;
    \gls{GW} strain range width at bounce time of the $+$ polarization seen by an equatorial observer ($\Delta h_{+,\textnormal{eq}}$);
    peak electron neutrino luminosity at $\unit[250]{ms}$; 
    average and standard deviation (superscript) of the tidal Love number ($\kappa_2$) in the interval $\unit[50-250]{ms}$;
    peak frequency of the density spherical harmonics decomposition for $\ell=1$, $m=1$ component ($\tilde{\rho}_{11}$) computed as the  Fourier transform and its \gls{FWHM} in the $\unit[50-250]{ms}$ interval ($f_{11}$); 
    peak frequency of the $\tilde{\rho}_{22}$ Fourier transform and its \gls{FWHM} in the $\unit[50-250]{ms}$ interval ($f_{22}$);
    peak frequency below $\unit[250]{Hz}$ of the Fourier transform and its \gls{FWHM} for equatorial equivalent neutrino luminosity in the $\unit[50-250]{ms}$ interval (the three flavours have the same peak frequency), $f_{\nu, 1}$;
    peak frequency above $\unit[250]{Hz}$ of the Fourier transform and its \gls{FWHM} for equatorial equivalent neutrino luminosity in the $\unit[50-250]{ms}$ interval (the three flavours have the same peak frequency), $f_{\nu, 2}$;
    peak frequency of the \gls{GW} Fourier transform and its \gls{FWHM} in the $\unit[50-250]{ms}$ interval  ($f_{\rm GW}$);
    peak frequency of the characteristic strain, $h_\textnormal{char}/\sqrt{f}$ (the two polarisations have the same peak frequency);
    \gls{PNS} tidal Love number ($\kappa_{2,\rm 250}$) at $\unit[250]{ms}$;
    average radius of the \gls{PNS} (computed as an angular average, $\Omegaaver{ R_\textnormal{PNS}} = \frac{1}{4\pi}\int \de \Omega R_\textnormal{PNS}(\Omega)$) with subscripts and superscripts representing the maximum and minimum deviation from the average at $\unit[250]{ms}$, corresponding to the equatorial and polar radius, respectively; 
    mass of the \gls{PNS};
    average shock radius;
    energy carried away in the form of \glspl{GW}; 
    cumulative energy carried away as neutrinos; 
    average rotation period of the \gls{PNS}, computed as $P_\textnormal{rot}=\int_\textnormal{PNS}2\pi\Omega_{\rm rot}^{-1}\de V/\int_\textnormal{PNS}\de V$, and its standard deviation (superscript).
    }
    \label{tab:quantitative_res}
\end{table}

The five \glspl{CCSN} models were evolved in \gls{3D} for $\unit[250]{ms}$ after core bounce to investigate the impact of the nuclear \gls{EOS} on the early post-bounce evolution and on observable signatures such as \gls{GW} and neutrino emission.  The key quantitative properties of the \gls{PNS}, \gls{GW} and neutrino signals, and shock development during this phase   are summarised in Table~\ref{tab:quantitative_res}.

At $\unit[250]{ms}$ post-bounce, the \gls{PNS} in all models is still accreting matter and the shock remains stalled at radii of order $\sim\unit[200]{km}$. 
Due to the rapid rotation of the \gls{PNS}, all models reach values of $T/|W|$ within the regime where the \glspl{LTWI} develop ($T/|W|_\textnormal{b}\approx0.05-0.06$). Owing to the large amount of rotational energy, all simulations develop strong non-axisymmetric modes characterised by the emergence of large-scale spiral patterns expanding from within the \gls{PNS} through the post-shock region.

\subsection{Pre-core-bounce evolution}
\label{sec:prebounce_evo}
\begin{figure}[t]
    \centering
    \includegraphics[width=\linewidth]{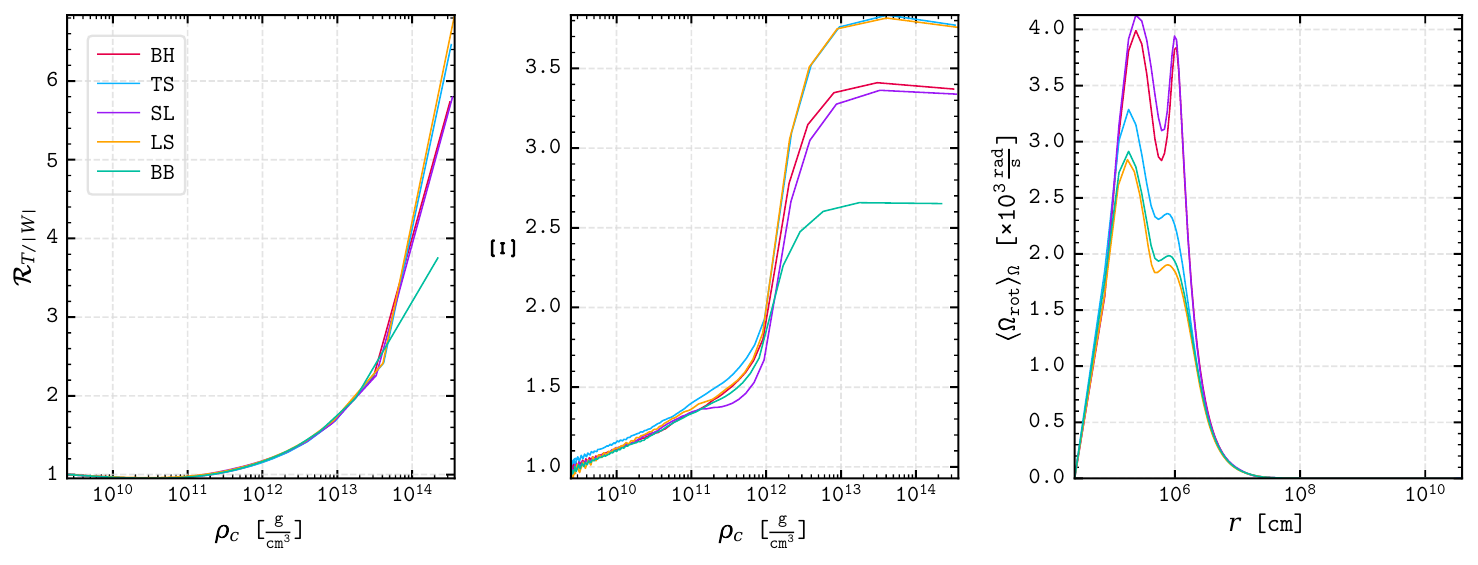}
    \caption{Normalised rotational parameter, $\mathcal{R}_{T/|W|}$, (left panel), homology deviation parameter, $\Xi$ (middle) as a function of the central density, $\rho_c$. Right panel: radial profile of the angle-averaged rotational frequency at core bounce.}
    \label{fig:collapse}
\end{figure}

The choice of different \glspl{EOS} may influence how rotation is amplified and redistributed during the evolution prior to core bounce. Specifically, identical initial rotation profiles do not guarantee identical rotational states at bounce, as the mapping between radius and enclosed mass evolves differently depending on the \gls{EOS}. For this reason, we analyse the evolution of the collapsing core up to bounce\footnote{Since the simulations are evolved in axisymmetry up to $\unit[5]{ms}$ before core bounce, this analysis is based on \gls{2D} simulations, unless stated otherwise.
}.

At bounce, we define the inner core as the region enclosed between the centre of the star and the outer sonic point ($|v_r|<c_s$, with $c_s$ the sound speed).  Although the inner core is defined at bounce, we reconstruct its previous evolution during collapse by identifying its enclosed mass $M_{\mathrm{ic,b}}$ and tracing this mass shell backward in time. In this way, all inner core quantities are evaluated on a fixed Lagrangian mass coordinate corresponding to the material that forms the inner core at bounce.

We compare the value of $T/|W|_{\mathrm{ic,b}}$ at bounce, computed as
\begin{equation}
    \label{eq:inn_TW}
    \left(\frac{T}{|W|}\right)_{\rm ic, b} = \frac{\frac{1}{2}\int_{\rm ic, b} \de V \rho v_\phi^2}{|\int_{\rm ic, b} \de V \rho\Phi|},
\end{equation}
where $v_\phi$ is the azimuthal velocity and $\Phi$ the gravitational potential.
This quantity exhibits only moderate relative variations ($\lesssim 13\%$) across the models. The ratio between rotational kinetic and gravitational energy can be approximated, to leading order, as 
\begin{equation}
    \left(\frac{T}{|W|}\right)_{\rm ic, b} \sim \frac{J_{\rm ic,b}^2}{G\,M_{\rm ic,b}^3\,R_{\rm ic,b}},
\end{equation}
where $R_{\mathrm{ic,b}}$, and $J_{\mathrm{ic,b}}$ are the inner core radius and angular momentum at bounce, respectively.
Despite the similar values of $T/|W|_{\mathrm{ic,b}}$, the inner core properties differ significantly across models: $M_{\mathrm{ic,b}}$ ranges from $0.46$ to $\unit[0.56]{M_\odot}$, $R_{\mathrm{ic,b}}$ from $12$ to $\unit[19]{km}$, and $J_{\mathrm{ic,b}}$ from $\unit[1.20\times10^{48}]{erg\cdot s}$ to $\unit[1.75\times10^{48}]{erg\cdot s}$.
We also compute the fraction of differential rotation energy relative to the total rotational energy, obtained by subtracting the rotational energy of an equivalent rigidly rotating configuration as
\begin{equation}
    \label{eq:T_diff}
    E_\textnormal{diff} = T - \frac{1}{2}\frac{J_z^2}{I},
\end{equation}
where $I$ is the moment of inertia and $J_z$ the $z$-component of the angular momentum. We find that restricting to the inner core and evaluated at bounce, the ratio  $E_\textnormal{diff,ic,b}/T_\textnormal{ic,b}$ varies between $65\%$ and $73\%$.

To remove timing differences during collapse, we compare the evolution of the normalised rotational parameter of the entire system,
\begin{equation}
    \mathcal{R}_{T/|W|}(\rho_c)=\frac{T/|W|(\rho_c)}{T/|W|(t=0)},
\end{equation}
as a function of the central density $\rho_c$.

Assuming angular momentum conservation and homologous contraction, one obtains
\begin{equation}
    T \propto \frac{J^2}{I} \propto \frac{1}{R^2}, \qquad |W| \propto \frac{GM^2}{R} \propto \frac{1}{R}, \qquad \frac{T}{|W|} \propto \frac{1}{R}.
\end{equation}
The inferred dependence of $T/|W|$ under homologous contraction motivates the definition of a homology deviation parameter,
\begin{equation}
    \Xi(\rho_c) =\frac{T/|W|_\textnormal{ic}(\rho_c)}{T/|W|_\textnormal{ic}(t=0)}\frac{R_\textnormal{ic}(\rho_c)}{R_\textnormal{ic}(t=0)}.
\end{equation}
Here, $R_{\mathrm{ic}}$ and $T/|W|_{\mathrm{ic}}$ during collapse are evaluated on a Lagrangian grid defined by the inner core mass at bounce, and assuming that the collapse is spherically symmetric.

The left panel of Figure~\ref{fig:collapse} shows the evolution of $  \mathcal{R}_{T/|W|}$. Model \texttt{BB} departs from the common trend at $\rho_c \sim \unit[2\times10^{13}]{g/cm^3}$ and reaches bounce with a value approximately a factor of two smaller than the other models. The remaining models separate into two groups, \texttt{TS}–\texttt{LS} and \texttt{BH}–\texttt{SL}, exhibiting approximately sevenfold and sixfold increases, respectively.
The same grouping is reflected in the evolution of $\Xi$ (middle panel), where  deviations from a common trend emerge at $\rho_c \sim \unit[10^{12}]{g/cm^3}$, i.e., at densities above which neutrinos become trapped.
The right panel illustrates that these differences translate into qualitatively distinct angular rotational frequency profiles at bounce. Models \texttt{BH} and \texttt{SL} develop two comparable rotational maxima, and correspondingly stronger differential rotation, while \texttt{BB}, \texttt{LS}, and \texttt{TS} exhibit a single dominant peak. Moreover, the appearance of the outer peak in 
$\Omegaaver{\Omega_{\rm rot}}$
\footnote{Hereafter, we denote with $\Omegaaver{A}$ the solid angle average of a quantity $A$, i.e., $\Omegaaver{A}=(4\pi)^{-1}\int_{4\pi} \de\Omega A$.} 
creates a region outside the inner core where 
$\de\Omegaaver{\Omega_{\rm rot}}/\de r>0$. 
Since the standard local criterion for the axisymmetric \gls{MRI} requires 
$\de\Omega_{\rm rot}/\de r<0$, 
this positive gradient hinders (on average) the growth of the instability in that layer, i.e., in the region where the \gls{MRI} would otherwise be expected to develop most easily in a \gls{PNS} (see, e.g., \citet{Rembiasz2016}).

Overall, this analysis demonstrates that the choice of \gls{EOS} significantly affects the stratification and rotational structure of the core prior to bounce, even when starting from identical progenitor and initial rotation profiles.

\subsection{Overview}
\label{sec:overview}
We describe the evolution of all models, focusing primarily on the equatorial plane, where the non-axisymmetric dynamics (spiral structures associated with the \gls{LTWI}) are most prominent. Additionally, we assess the degree to which the flow preserves reflection symmetry about the equatorial plane.

\subsubsection{Equatorial plane.}
\label{sec:eq_plane}
\begin{figure}[t]
    \centering
    \includegraphics[width=\linewidth]{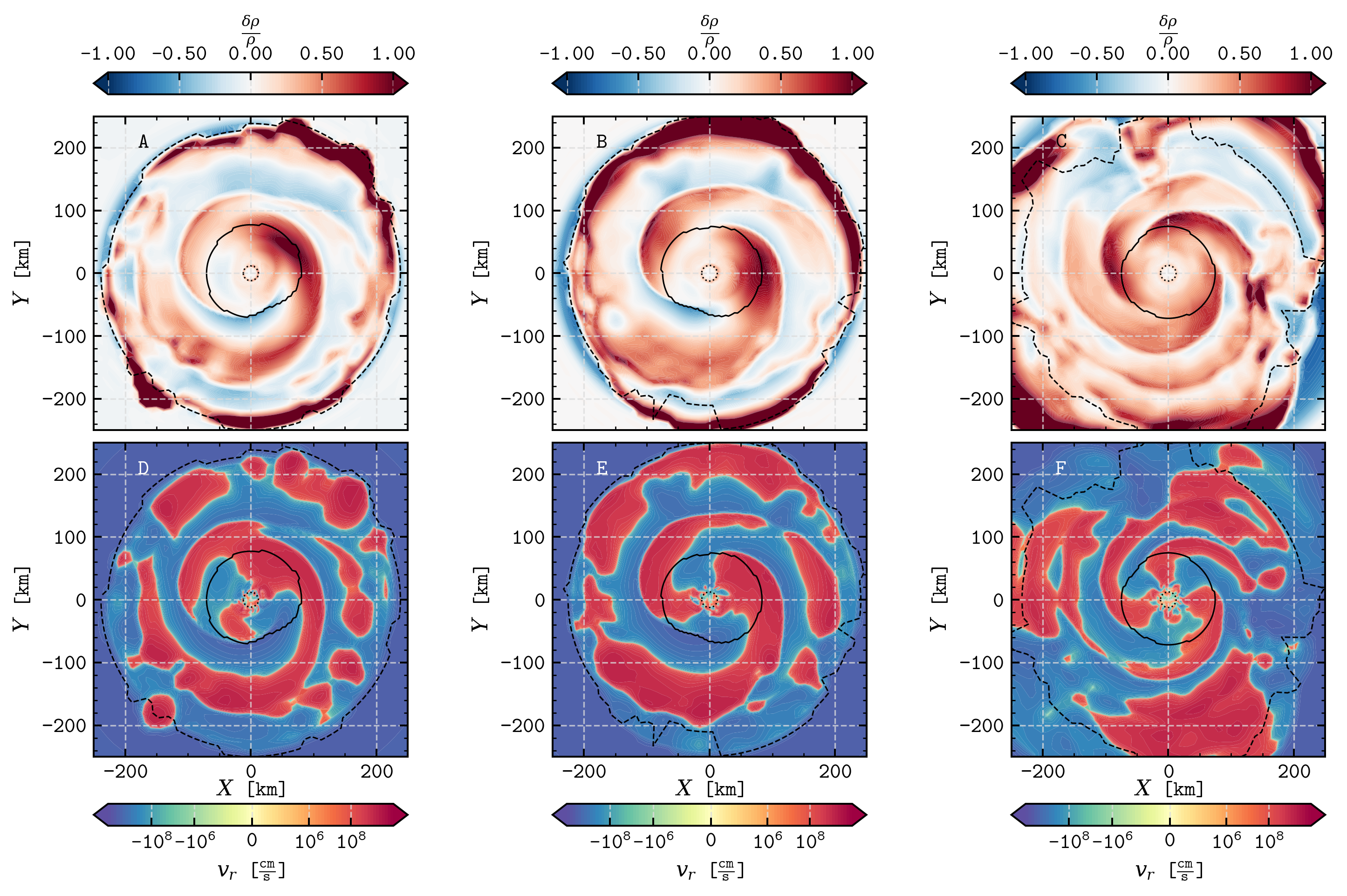}
    \caption{Equatorial slices of the relative deviation from spherical density distribution, $\delta\rho/\rho$ (top row) and radial velocity, $v_r$ (bottom row). Left column shows a one-arm spiral at $\unit[0.101]{s}$ in model \texttt{LS}. Middle and right columns display model \texttt{BH} two- and three-arm spirals at $\unit[0.100]{s}$ and $\unit[0.160]{s}$, respectively. Dashed, solid, and dotted lines represent the isosurfaces at $\unit[10^9]{g/cm^3}$, $\unit[10^{11}]{g/cm^3}$ (as a proxy for the location of the \gls{PNS}),  and $\unit[10^{14}]{g/cm^3}$, respectively.
    }
    \label{fig:spiral}
\end{figure}
All models develop large-scale spiral patterns once the \gls{LTWI} sets in, although the detailed morphology and lifetime of the spirals vary significantly across the \gls{EOS} sample. In all cases, the non-axisymmetric structure originates within the highly deformed \gls{PNS} and extends into the post-shock region, typically reaching the $\unit[10^9]{g/cm^{3}}$ isosurface. As shown in Figure~\ref{fig:spiral}, the non-axisymmetric structure is clearly visible in the equatorial plane, both in the radial velocity and in the density deviation from spherical symmetry, defined as
\begin{equation}
    \label{eq:density_deviation}
    \frac{\delta\rho}{\rho} = \frac{\rho - \Omegaaver{\rho}}{\Omegaaver{\rho}}.
\end{equation}

The simplest evolution is found in model \texttt{SL}, in which the instability develops into a long-lived two-armed spiral that persists until the end of the simulated interval. By contrast, models \texttt{TS} and \texttt{BB} show a more sequential evolution, in which the spiral first appears as a one-armed pattern and subsequently develops two-armed and three-armed phases before being disrupted. Model \texttt{BH} also exhibits a relatively ordered evolution, forming first a two-armed structure (Figures~\ref{fig:spiral}.B and \ref{fig:spiral}.E), and transitioning from a two-armed to a three-armed (Figures~\ref{fig:spiral}.C and \ref{fig:spiral}.F) configuration before disappearing. The most complex behaviour is found in model \texttt{LS}, where the morphology changes repeatedly between one-armed (Figures~\ref{fig:spiral}.A and \ref{fig:spiral}.D), two-armed, and later multi-armed configurations, with transient additional arms appearing during the nonlinear phase.

The onset time of the spiral activity also differs among models. The earliest spiral develops in model \texttt{BH} at about $\unit[50]{ms}$ after bounce, whereas the latest onset is found in model \texttt{BB}, where the pattern becomes visible only at about $\unit[100]{ms}$. The duration of the strongly non-axisymmetric phase varies accordingly, ranging from relatively short-lived episodes in \texttt{BH}, \texttt{TS}, and \texttt{BB} to a more persistent spiral activity in \texttt{SL} and a more irregular but long-lasting evolution in \texttt{LS}.

Overall, despite these morphological differences, all models show coherent low-order spiral structure whose growth and evolution are consistent with the development of the \gls{LTWI}. The differences among models are therefore best understood not as a distinction between the presence and absence of the instability, but rather as variations in its dominant azimuthal structure, onset time, and nonlinear evolution.

\subsubsection{Equatorial symmetry.}
\label{sec:eq_symm}

\begin{figure}
    \centering
    \includegraphics[width=\linewidth]{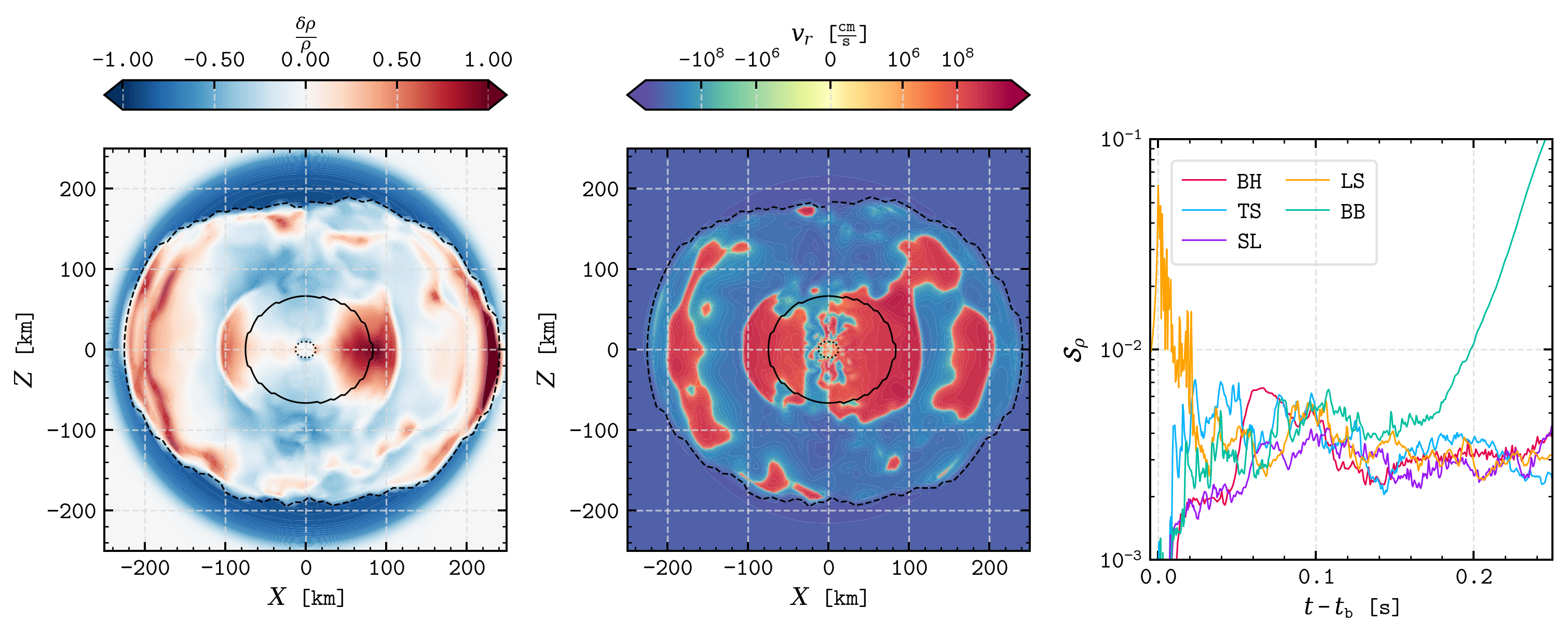}
    \caption{Left and middle panels: polar slices of the relative deviation from the axisymmetric density distribution, and radial velocity, respectively for model \texttt{BH} at $\unit[0.101]{s}$. Solid, dashed, and dotted lines have the same meaning as in Figure~\ref{fig:spiral}.  Right panel, shows the evolution of the density asymmetry score, $\mathcal{S}_\rho$ (Equation~\eqref{eq:symmetry_score}), for all models.}
    \label{fig:symmetry}
\end{figure}

While the spiral modes cause a strong non-axisymmetric deformation, the system retains an approximate reflection symmetry about the equatorial plane. The spiral arms develop primarily within the equatorial plane, where they extend up to two hundred kilometres from the \gls{PNS} centre, but no significant asymmetry is observed between the northern and southern hemispheres. The left and middle panels of Figure~\ref{fig:symmetry} show the distribution of the density deviation from spherical symmetry and the radial velocity along a vertical slice at a time in which the instability is active. The spiral extends for approximately $\unit[100]{km}$ in the vertical direction, i.e., its vertical size ir roughly comparable to the \gls{PNS} polar diameter, it is approximately symmetric in density and radial velocity, while it is antisymmetric in azimuthal velocity (not shown).

To quantify the degree of equatorial symmetry, we define the density asymmetry score as:
\begin{equation}
    \label{eq:symmetry_score}
    \mathcal{S}_\rho = \sqrt{\frac{\int\textnormal{d}V (\rho(r, \theta, \phi)-\rho(r, \pi-\theta, \phi))^2}{\int\textnormal{d}V\rho^2}},
\end{equation}
where we limit the integral to the volume within the $\unit[10^9]{g/cm^3}$ isosurface to capture only the deformation associated with the spiral formation.
The right panel of Figure~\ref{fig:symmetry} shows the evolution of this quantity for each model. The asymmetry value remains below $1\%$ throughout the evolution in all cases except \texttt{LS} and \texttt{BB}. For model \texttt{LS} the core bounce and subsequent shock formation are somewhat asymmetric. This leads to an initially large asymmetry score, that quickly drops below 1\% after $\sim \unit[30]{ms}$. In model \texttt{BB}, a slightly denser region develops just below the shock front in the northern hemisphere at $\sim\unit[45]{^\circ}$ from the equatorial plane, making it unrelated to the development of the instability. 
However, even here the asymmetry ratio reaches at most $\sim10\%$. Thus, the \gls{LTWI}-driven flow remains approximately equatorially symmetric in most cases, although departures from perfect symmetry do occur. This fact  has important implications for the emission of \glspl{GW}, and, in particular, on the $\times$-polarised equatorial component (see Section~\ref{sec:GWs}).

\subsection{PNS Structure and Energetics}
\label{sec:PNS_hydro}
\begin{figure}[t]
    \centering
    \includegraphics[width=\linewidth]{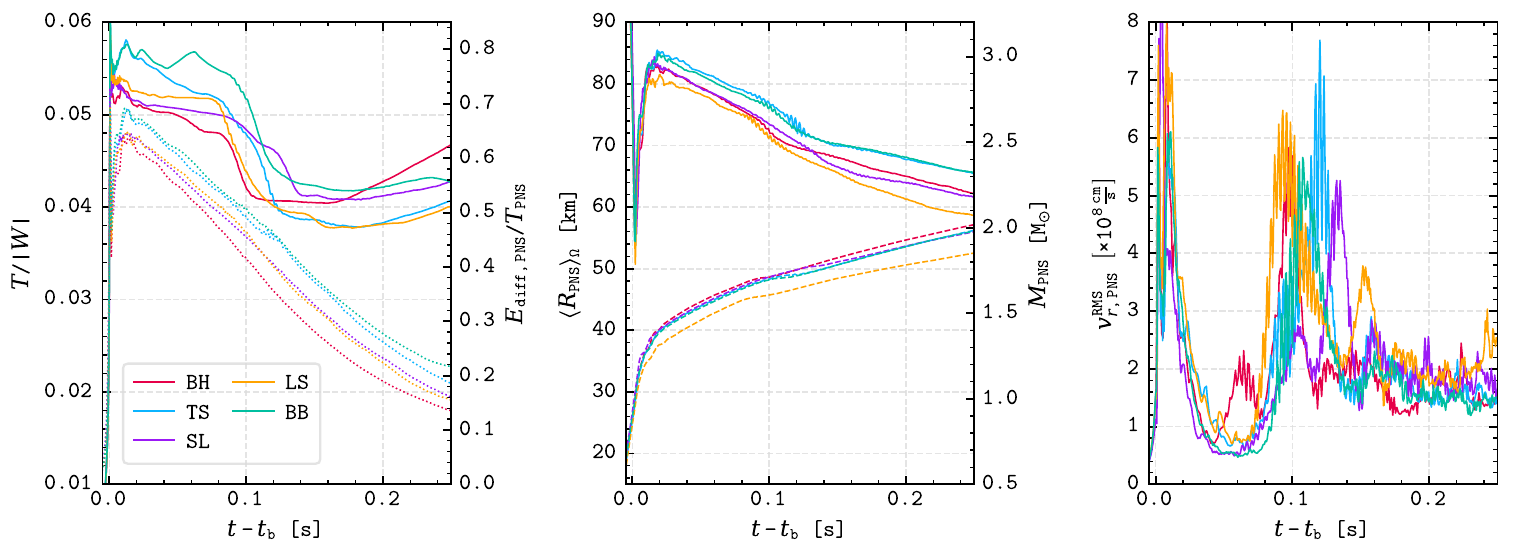}
    \caption{Left panel: evolution of the ratio between rotational and gravitational energy, $T/|W|$, of the \gls{PNS} (solid) and degree of differential rotation $E_\textnormal{diff}/E_\textnormal{rot}$ (dotted). Middle panel: \gls{PNS} average radius (solid) and mass (dashed). Right panel: average residual of the radial velocity inside of the \gls{PNS}.
    }
    \label{fig:PNS_evolution}
\end{figure}

Following the overview of the instability, we now analyse the evolution of the global \gls{PNS} properties that regulate the development of the \gls{LTWI}. The development of spirals in the equatorial plane is associated with the \gls{LTWI} and induces dynamical changes in the \gls{PNS} rotational profile and in the shear energy. For this reason, regarding the energetics of the \gls{PNS}, we analyse the evolution of the ratio between rotational and gravitational energy of the \gls{PNS} computed as in Equation~\eqref{eq:inn_TW}, but extended to the whole \gls{PNS}  defined as the isodensity surface at $\unit[10^{11}]{g/cm^3}$, and the amount of differential rotational energy (Equation~\eqref{eq:T_diff}). Additionally, we compare the evolution of the \gls{PNS} masses and radii for our models, and to measure the asymmetric radial motion within the \gls{PNS}, we define the \emph{\gls{PNS} residual of the radial velocity}  as
\begin{equation}
    \label{eq:vr_PNS}
    v^\textnormal{RMS}_{r,\textnormal{PNS}} = \sqrt{\langle \left(v_r - \langle v_r\rangle_\Omega \right)^2\rangle_\textnormal{PNS}},
\end{equation}
where $v_r$ is the radial velocity and $\langle\cdot\rangle_{\rm PNS}$ denotes averaging  over a spherical shell (subscript $\Omega$) or throughout the \gls{PNS} volume.
Finally, to quantify the development of the spiral we employ the time evolution of the normalised azimuthal Fourier amplitudes of the density,
\begin{equation}
\label{eq:fourier_coeff}
A_m(t)=\frac{\left|\int \de V\rho(r,\phi,z,t)\,e^{im\phi}\right|}{\int \de V\rho(r,\phi,z,t)},
\end{equation}
evaluated within the isosurface at $\unit[10^9]{g/cm^3}$. 

In the left panel of Figure~\ref{fig:PNS_evolution}, we show the evolution of the \gls{PNS} $T/|W|$, as solid lines, which closely connects with the development of the spiral instability.

As reported by \cite{Shibagaki2021,Bugli23}, a decrease in $T/|W|$ is associated with the onset of spiral patterns in the context of \gls{LTWI}. 
In our simulations, $T/|W|$ drops from $\sim5\%$ to $\sim4\%$ on timescales of approximately $\sim 50\,$ms. This decrease is slightly lower than the $5\%$ found by \cite{Shibagaki2021}, but higher than the $2.3\%$ observed in model \texttt{H} studied by \cite{Bugli23}. Additionally, in \cite{Bugli23} 
the $T/|W|$ reduction in the model that develops \gls{LTWI} compared to the \gls{2D} counterpart that does not develop it is less sharp than in the other cases.
Across all models, variations in $T/|W|$ are correlated with the development and decay of the spiral instability. In particular, local decreases of $T/|W|$ coincide with the emergence of spiral arms. The instability shows greater activity (as quantified by $A_m$; see below) when $T/|W|$ reaches its minimum, after which the ratio remains approximately constant while the spiral pattern is sustained. At later times, the increase of $T/|W|$ coincides with the weakening and eventual disruption of the spiral structure.

An illustrative example is model \texttt{BH}: the accelerated decrease observed at $\approx\unit[50]{ms}$ marks the initial emergence of a two-armed spiral. The second more pronounced decay happening at $\approx\unit[100]{ms}$, corresponds to its full development. Finally, by $\unit[160]{ms}$ the spiral is totally disrupted as $T/|W|$ rises  back to approximately its original value.

However, a low value of $T/|W|$ is not a sufficient condition for the development of the instability. Another crucial requirement is that the  \gls{PNS} be differentially rotating \cite{Watts2005, Passamonti2015}. Using Equation~\eqref{eq:T_diff}, we compute the fraction of differential-rotation energy relative to the total rotational energy (left panel of Figure~\ref{fig:PNS_evolution}, dotted line). As in \cite{Bugli23}, the energy at bounce is approximately equally distributed between rigid and differential rotation. After bounce, the fraction of differential-rotation energy increases to $\sim70\%$ at $\sim\unit[20]{ms}$, before decreasing nearly linearly, reaching $15-20\%$ of the total by $\unit[250]{ms}$.  Notably, the instability develops when the differential-rotation energy remains relatively high, at least $45\%$ of the total.

The formation of the spiral and the onset of the \gls{LTWI} are tied to the rotation rate of the models and also have consequences on the \gls{PNS} evolution. In all models, the \gls{PNS} continues to accrete mass throughout the simulation. The stiffer 
\gls{EOS} of model~\texttt{LS} produces a \gls{PNS} with a mass consistently lower by approximately $\unit[0.15]{M_\odot}$ compared to the other models (dashed line in the mid panel of Figure~\ref{fig:PNS_evolution}). After about $\unit[40]{ms}$, however, the accretion rates among all considered \glspl{EOS} converge to similar values. Over the same period, the \gls{PNS} average radii show only minor differences, differing by less than $\unit[5]{km}$.

Between $\unit[75]{ms}$ and $\unit[125]{ms}$, all models undergo a brief phase lasting approximately $\unit[10]{ms}$ during which the \gls{PNS} accretion rate temporarily drops to nearly zero before recovering to its previous level. In this interval, the \gls{PNS} contraction also accelerates, leading to a more rapid decrease in radius. This transient behaviour coincides with the onset of the \gls{LTWI}. The spiral mode development transports angular momentum outward from the \gls{PNS} into the post-shock region. The associated increase in centrifugal support drives an outward mass flux that temporarily counteracts accretion onto the \gls{PNS}, producing the observed drop in the accretion rate.

The right panel of Figure~\ref{fig:PNS_evolution} shows relatively broad peaks in $v^\textnormal{RMS}_{r,{\rm PNS}}$. Looking, e.g., at equatorial maps of $\delta\rho/\rho$, these peaks coincide with the formation of one or more spiral arms, with the largest values occurring when the 
instability is fully developed. In some models (e.g., \texttt{BH}), these peaks coincide with the small drops in $T/|W|$ shown in Figure~\ref{fig:PNS_evolution}, supporting a connection between at least some of the local decreases in $T/|W|$ and the emergence of spiral arms. More generally, we find a broad correlation whereby larger peaks in $v^\textnormal{RMS}_{r,{\rm PNS}}$ are associated with larger drops in $T/|W|$.

\begin{figure}
    \centering
    \includegraphics[width=\linewidth]{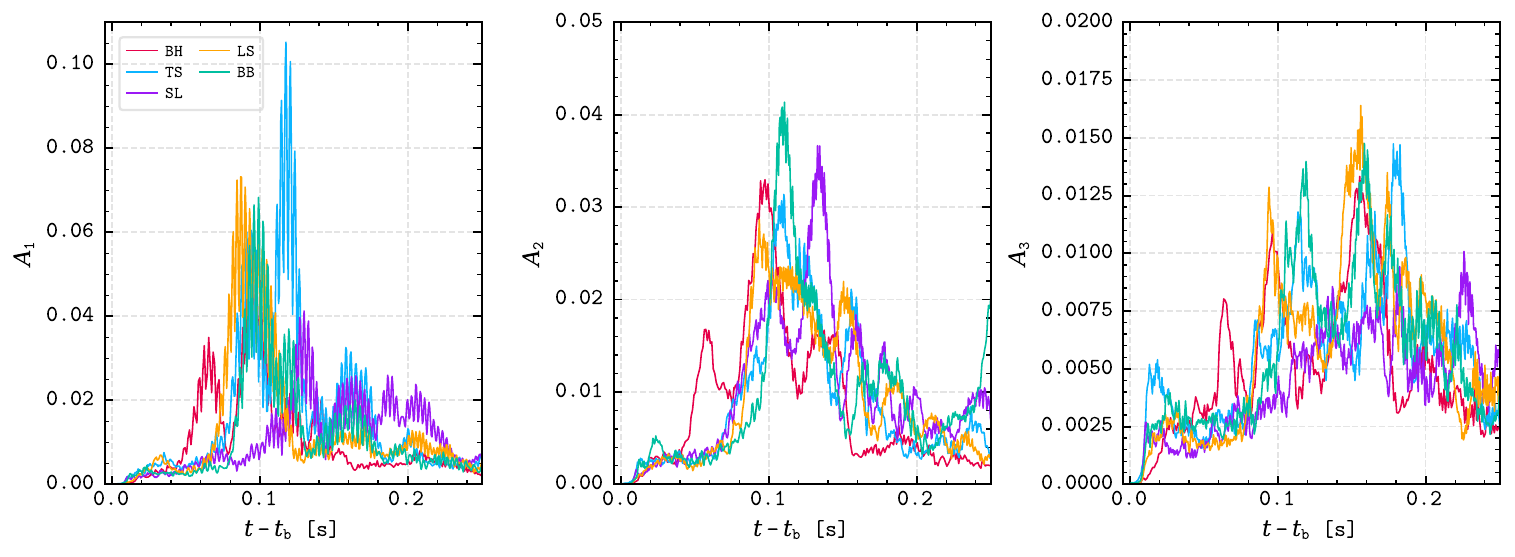}
    \caption{Evolution of the normalised azimuthal Fourier amplitudes of the density for $m=1$ (left panel), $m=2$ (middle panel) and $m=3$ (right panel). Note the scale difference on the $y$-axis. }
    \label{fig:fourier_deco}
\end{figure}

Figure~\ref{fig:fourier_deco} shows the evolution of the normalised Fourier amplitudes for $m=1$, $2$, and $3$ in our model sample. The growth of these amplitudes indicates the development of non-axisymmetric spiral structure, with the dominant mode identified by the largest $A_m$.  
In all models except \texttt{SL}, the $m=1$ (left panel) component dominates over $m=2$ (middle), and $m=3$ (right) during the first $\unit[100]{ms}$ post-bounce. At later times, either the $m=1$ or $m=2$ mode becomes dominant  depending on the model, while $m=3$ remains subdominant. 
In model \texttt{SL}, the $m=2$ component dominates throughout the \gls{LTWI} evolution, consistent with the presence of a sustained two-armed spiral, as discussed in Section~\ref{sec:overview}. 
We take the maximum amplitude of the dominant Fourier component as a proxy for the saturation of the \gls{LTWI} and the peak development of the spiral. Moreover, we find that the largest peaks in $A_m$ coincide in time with the largest peaks in $v^\textnormal{RMS}_{r,{\rm PNS}}$ (Figure~\ref{fig:PNS_evolution}), which supports the use of $v^\textnormal{RMS}_{r,{\rm PNS}}$ as a proxy for the onset and relative strength of the instability.
In the following, we employ a spherical harmonics decomposition rather than a Fourier decomposition, as it provides a more complete characterization of the angular structure and enables a direct morphological comparison with the \gls{GW} signal.
\begin{figure}[t]
    \centering
    \includegraphics[width=\linewidth]{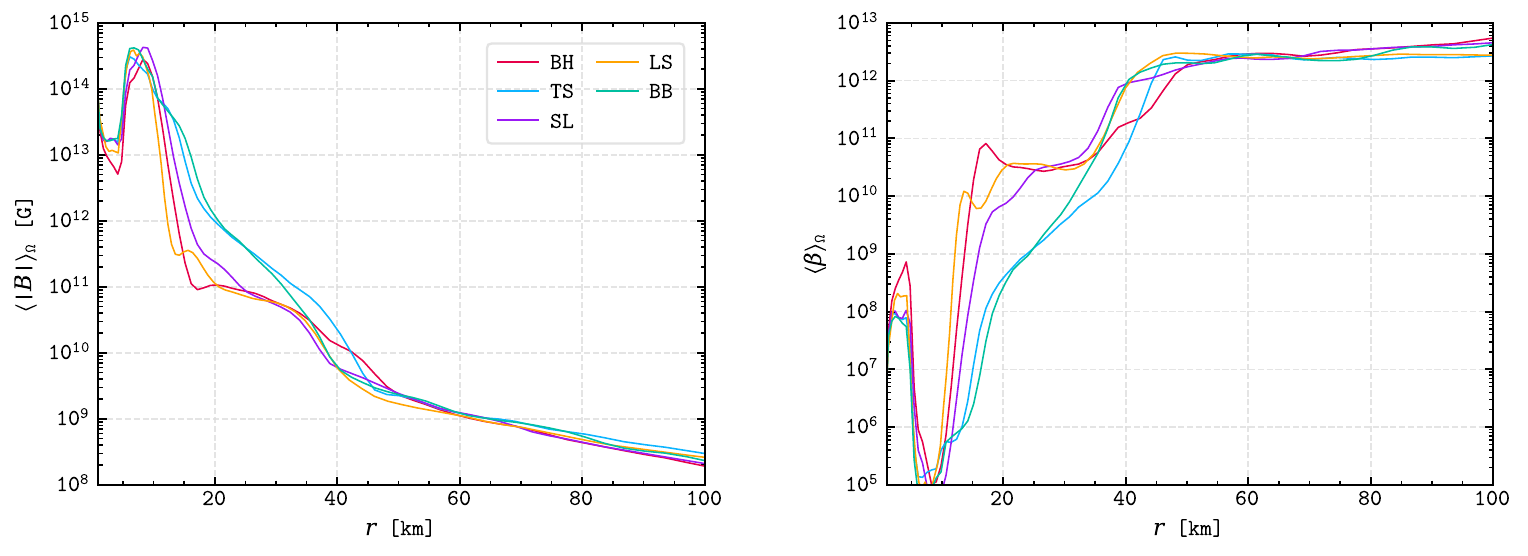}
    \caption{Radial profiles of the magnetic-field strength, $|\mathbf{B}|$ (left), and plasma-$\beta$ (right) at $\unit[75]{ms}$ after core bounce for all models.}
    \label{fig:bfield}
\end{figure}

Figure~\ref{fig:bfield} shows the radial profiles of the magnetic-field strength and plasma-$\beta$ within the inner $\unit[100]{km}$ of the \gls{CCSN} at $\unit[75]{ms}$ after core bounce, shortly before the onset of the \gls{LTWI}, for all models. The left panel displays the magnetic-field strength,
\begin{equation}
    |\mathbf{B}|=\sqrt{B_r^2+B_\theta^2+B_\phi^2},
\end{equation}
while the right panel shows the plasma-$\beta$,
\begin{equation}
    \beta=\frac{P_{\rm gas}}{|\mathbf{B}|^2/2},
\end{equation}
where $P_{\rm gas}$ is the gas pressure. Despite minor differences among the models, they all exhibit the same qualitative behaviour. The magnetic field reaches strengths of approximately $\sim\unit[10^{9}]{G}$ near the \gls{PNS} surface ($r\sim\unit[75]{km}$), while the plasma remains strongly pressure dominated, with $\beta\sim10^{12}$. Thus, magnetic stresses are dynamically negligible compared with the thermal pressure during the development of the instability. These results suggest that the magnetic field is unlikely to play a significant role in the onset of the \gls{LTWI} in the present models.

\subsection{Spherical harmonics decomposition}
\label{sec:sph_harm_deco}
To quantitatively characterise the development of the \gls{LTWI}, we analyse the growth of its non-axisymmetric modes using a spherical-harmonic decomposition of the density field. The spherical harmonics coefficients of the density on a sphere of radius $r$ are
\begin{equation}
    \label{eq:rho_sp}
    \tilde{\rho}_{\ell m}(r) =  \int\textnormal{d}\Omega \rho(r,\theta, \phi)Y_{\ell m}(\theta, \phi),
\end{equation}
where $Y_{\ell m}(\theta,\phi)$  are the (real) spherical harmonics, defined as 
\begin{equation}
    \label{eq:real_sph_harm}
    Y_{\ell m}(\theta,\phi) = 
        \begin{cases}
        \sqrt{2} (-)^m \Im[Y_{\ell m}^\textnormal{cpx}(\theta, \phi)],\textnormal{ if }m < 0\\
        \Re[Y_{\ell m}^\textnormal{cpx}(\theta, \phi)],\textnormal{ if } m=0\\
        \sqrt{2} (-)^m \Re[Y_{\ell m}^\textnormal{cpx}(\theta, \phi)],\textnormal{ if }m > 0\\
        \end{cases}.
\end{equation}
where $Y_{\ell m}^{\mathrm{cpx}}$ are the complex spherical harmonics,
\begin{equation}
    \label{eq:sph_harm}
    Y_{\ell m}^{\rm cpx}(\theta,\phi)=\sqrt{\frac{2\ell + 1}{4\pi} \frac{(\ell - m)!}{(\ell + m)!}}P_{\ell m}(\cos\theta)e^{im\phi},
\end{equation}
and $P_{\ell m}$ are the associated Legendre polynomials, computed as
\begin{equation}
    \label{eq:legendre}
    P_{\ell m}(\cos\theta) = (-)^m 2^\ell(1-\cos^2\theta)^{m/2}\sum_{k=m}^{\ell}\frac{k!}{(k-m)!}\cos^{k-m}\theta\binom{\ell}{k}\binom{(\ell+k-1)/2}{\ell}.
\end{equation}

The spiral deformation is associated with the $\ell=m$ components. Specifically, the number of azimuthal arms in the equatorial deviation from the axisymmetric density and radial velocity corresponds to the order $m$ of the dominant spherical harmonic.

\begin{figure}[t]
    \centering
    \includegraphics[width=\linewidth]{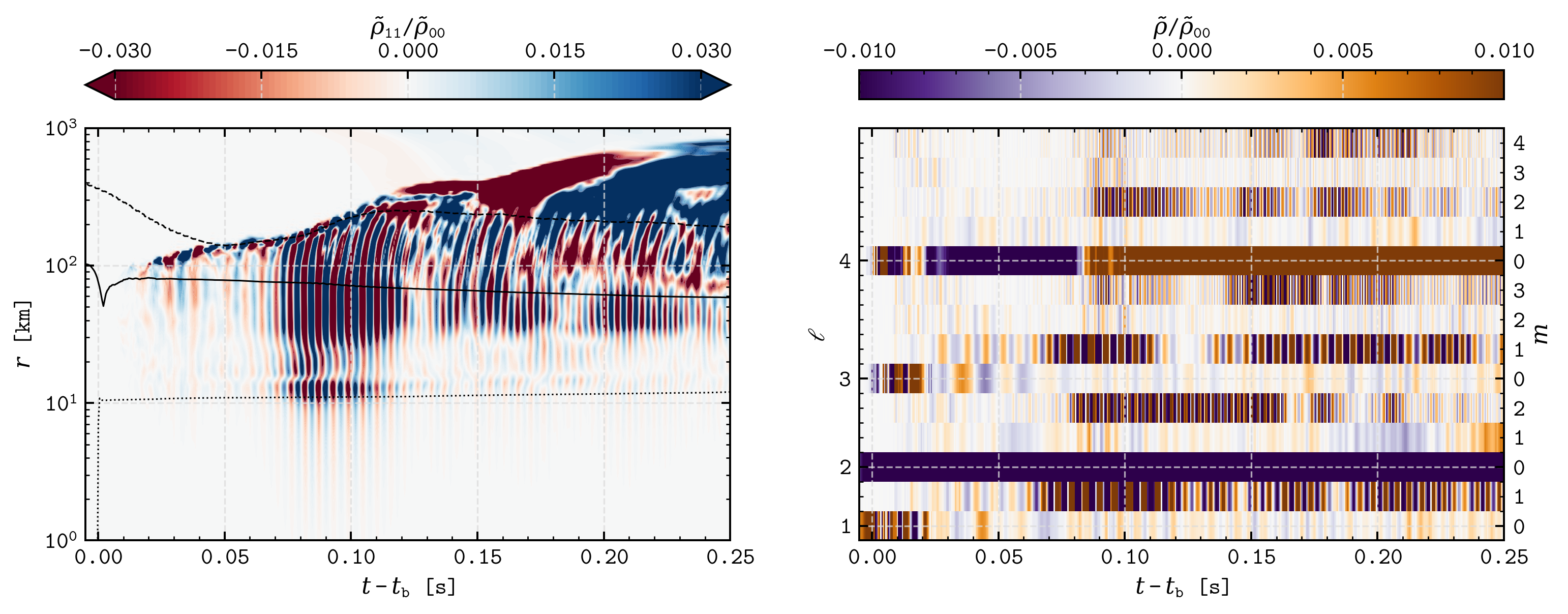}
    \caption{Left panel: space-time evolution of the spherical harmonics decomposition of the density for $\ell=1$, $m=1$, normalised by the $\ell=0$, $m=0$ component. Dotted, solid, and dashed lines represent the isosurfaces at $10^{14}$, $10^{11}$, and $\unit[10^9]{g/cm^3}$, respectively. 
    Right panel: evolution of the normalised spherical harmonics decomposition of the density at $r=\unit[30]{km}$, for $0<\ell\leq4$ and $m\geq 0$. Both panels refer to model \texttt{LS}.}
    \label{fig:barcode}
\end{figure}

Figure~\ref{fig:barcode} shows this decomposition for model \texttt{LS}, which we select as it exhibits the richest mode evolution, thus providing a clear link between the large-scale structure described in Section~\ref{sec:overview} and its harmonic content. 
The left panel of Figure~\ref{fig:barcode} shows the space–time evolution of the normalised $\tilde{\rho}_{11}/\tilde{\rho}_{00}$ amplitude. 
The $(\ell,m)=(1,1)$ component emerges in two distinct radial regions that become active during the initial development of the \gls{LTWI} at approximately $\unit[75]{ms}$. The first is an inner region, where the $(\ell,m)=(1,1)$ component remains confined within $r=\unit[20]{km}$ and disappears around $\unit[110]{ms}$, i.e., while the spiral mode is active. For this reason and because Figure~\ref{fig:spiral} indicates that the spiral structure originates at $r>\unit[20]{km}$, we suggest that the inner region does not contribute to the development of the large-scale spiral.
The second is an outer region above $\sim\unit[25]{km}$ extending to the \gls{PNS} surface. This outer component, however, is not restricted to the \gls{PNS} interior, but propagates outward at a reduced  speed, eventually reaching the shock. These two zones are separated by a transition layer in which their contribution overlaps.

To determine the hierarchy of unstable modes, in the right panel of Figure~\ref{fig:barcode}, we display the time evolution of all spherical-harmonic amplitudes with $\ell\leq 4$ and $m\geq0$ evaluated at a radius of $\unit[30]{km}$, which lies within the instability region. Throughout the evolution we observe contributions from  all $\ell=m$ modes, together with $(\ell,m)=(3,1)$, $(4,2)$, and $(4,0)$. 

Two purely axisymmetric components arise from the changing deformation due to the rapid rotation rather than any instability.
The axisymmetric quadrupole, $(2, 0)$, is negative throughout the evolution, consistent  with the high degree of equatorial flattening produced by the rapid rotation.
The behaviour of the axisymmetric hexadecapole, $(4,0)$, is more complex. Its sign change indicates an evolution of the higher-order structure of the deformation, reflecting variations not only between the polar and equatorial regions but also at intermediate latitudes.

The $(\ell, \,m)=(1,\,1)$ mode appears at $\sim\unit[70]{ms}$ and persists through the simulation, indicating the formation of a one-armed spiral.
Around $\sim\unit[80]{ms}$ the $(\ell,m)=(2,2)$ component emerges, marking the development of a second arm.
Between $\sim\unit[150]{ms}$ and $\sim\unit[220]{ms}$ the appearance of the $\ell=m=3$ and $\ell =m=4$ components signals the growth of additional arms and the progressive fragmentation of the pattern.
The presence of $(3,1)$ and $(4,2)$ modes shows that the spiral has a three-dimensional character and couples to the background rotational deformation; these mixed-order modes indicate vertical structure and mode–mode coupling. However, their amplitudes remain subdominant, and the overall flow preserves north–south symmetry to a high degree.

Overall, the modal evolution recovered from the spherical–harmonic analysis matches the qualitative behaviour described in Section~\ref{sec:overview}.
The appearance, ordering, and spatial origin of the modes depend on the specific flow geometry of each model and reflect the non-linear development of the \gls{LTWI}.

\begin{figure}[t]
    \centering
    \includegraphics[width=\linewidth]{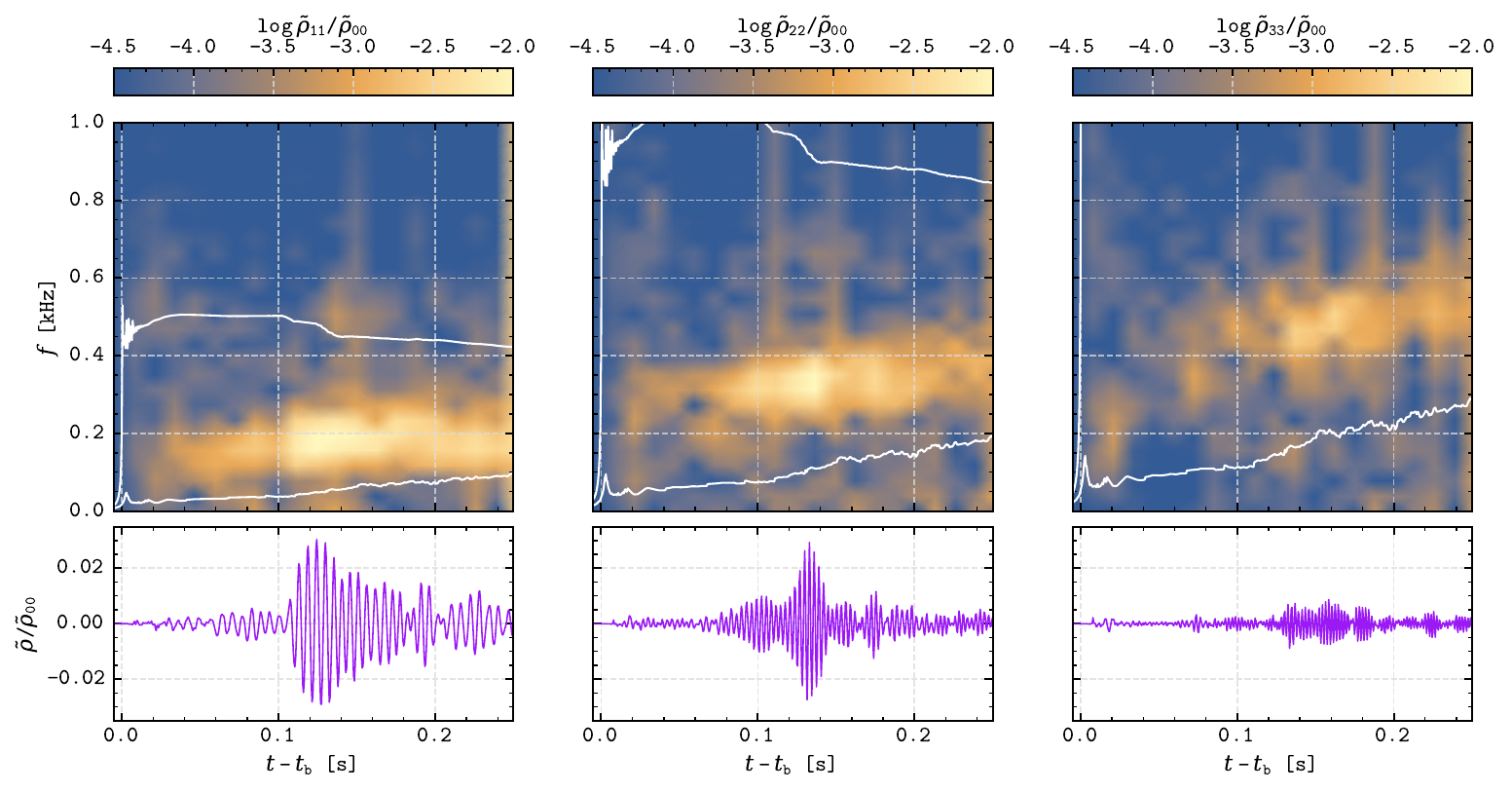}
    \caption{Spectrograms of the normalised spherical harmonics decomposition of the density performed with a time window of $\unit[20]{ms}$  (top row) and corresponding time series (bottom row) for $(\ell, m)$ equal to $(1,1)$, $(2,2)$, and $(3,3)$ for model \texttt{SL} at $r=\unit[30]{km}$. White lines represent the maximum rotation frequency within the \gls{PNS} and the surface-averaged rotation frequency, both multiplied by $m=1,2,$ and 3 from left to right.}
    \label{fig:sph_spectro}
\end{figure}

In order to characterise the frequency of the instability, we compute the spectrograms of $\tilde{\rho}_{\ell m}/\tilde{\rho}_{00}$ at a radius of $r=\unit[30]{km}$, within the region where the instability produces the strongest pattern. In Figure~\ref{fig:sph_spectro} we show the spectrograms of the $(\ell,m)=(1,1)$, $(2,2)$, and $(3,3)$ components in the top row, together with the corresponding time series in the bottom row, for model \texttt{SL}.

A rotating pattern with azimuthal number $m$ produces $m$ oscillation cycles per rotation period, and the characteristic signal frequency is
\begin{align}
f \simeq m\,f_{\rm rot},
\end{align}
where $f_{\rm rot}=\Omega_{\rm rot}/2\pi$ is the linear rotational frequency. Therefore, plotting $m f_{\rm rot}$ in spectrograms provides a useful reference for identifying the frequencies associated with rotating non-axisymmetric modes.
The $\ell=m=1$ mode, present from (at least) $\unit[50]{ms}$, becomes stronger at approximately $\unit[100]{ms}$ 
with a frequency of about $\unit[170]{Hz}$ (left panel of Figure~\ref{fig:sph_spectro}). In contrast to previous works~\cite{Shibagaki2021,Takiwaki2021,Bugli23}, in our models, irrespective of the \gls{EOS}, the mode frequency remains almost constant throughout the entire instability phase. 
The physical origin of this difference with respect to previous simulations is not yet clear. 
In the present models, the nearly constant mode frequency suggests that the pattern speed is set by a corotation region whose location and local rotation rate evolve only slowly during the active phase of the instability. 
We do not attribute this behaviour solely to the approximate treatment of gravity, since simulations performed with the same code framework and progenitor model have reported a different temporal evolution of the \gls{LTWI} frequency \cite{Bugli23}. 
A systematic study varying the treatment of gravity, progenitor structure, rotation profile, magnetic-field strength, and post-bounce evolution time would be required to isolate the origin of this difference.

According to \cite{Watts2005,Passamonti2015,Passamonti2020}, the \gls{LTWI} is triggered when the frequency of one of the \gls{PNS} modes matches the local rotation frequency, defining a corotation point. The maximum rotation frequency of the differentially rotating \gls{PNS} is located near the centre, while the minimum is found near the surface. As indicated by the white lines, which represent these extrema, the $\ell=m=1$ mode possesses at least one corotation point inside the \gls{PNS}, since its frequency lies between the two curves.

In the middle panel of Figure~\ref{fig:sph_spectro}, we present a similar analysis for the $\ell=m=2$ mode, which is directly associated with the two-armed spiral structure. This mode appears at approximately $\unit[80]{ms}$ with a frequency roughly twice that of the $\ell=m=1$ mode, and it remains essentially constant during the time in which the amplitude of the mode is enhanced ($\unit[0.1-0.2]{s}$).  Again, a corotation point lies within the \gls{PNS}, as the mode frequency is enclosed by the maximum and minimum rotational frequencies multiplied by $m=2$.

In the right panel of Figure~\ref{fig:sph_spectro} we show the $\ell=m=3$ mode, which becomes visible at later times, around $\unit[120]{ms}$. Its frequency is roughly three times that of the $\ell=m=1$ mode, and it is also enclosed between three times the maximum and minimum \gls{PNS}  frequencies, indicating once again a corotation point.

The bottom panels of Figure~\ref{fig:sph_spectro} compare the time-series amplitudes of these modes. This comparison clearly shows that the  $\ell=m=2$ mode dominates the early dynamics of the \gls{LTWI} phase in model \texttt{SL}, consistently with the observed two-armed spiral. The $\ell=m=1$ mode  becomes relevant only at later times, and  $\ell=m=3$ mode remains subdominant.

Despite the differences introduced in the rotation profile and in the stratification of the inner core during collapse evolution (Section~\ref{sec:prebounce_evo}), and consequently in the \gls{PNS} structure, the \gls{LTWI} develops in all models considered. In particular, the instability sets in within a comparable post-bounce time interval of $\unit[50-90]{ms}$ in all cases, which indicates that its onset is a feature largely independent from the  choice of the \gls{EOS}. 
At the same time, quantitative properties of the instability exhibit a dependence on the \gls{EOS}. In particular, the pattern frequency varies between models, reflecting differences in the structure of the \gls{PNS}. Additionally, variations are observed in the dominant azimuthal structure of the spiral modes, as well as in the associated \gls{GW} emission (see Section~\ref{sec:GWs}). 
These results suggest that the development of the \gls{LTWI} is a robust outcome in our model set, whose characteristics are sensitive to the underlying microphysics encoded in the \gls{EOS}.

\subsection{Characterization of the corotation regions}
\label{sec:corotation}
\begin{figure}[t]
    \centering
    \includegraphics[width=\textwidth]{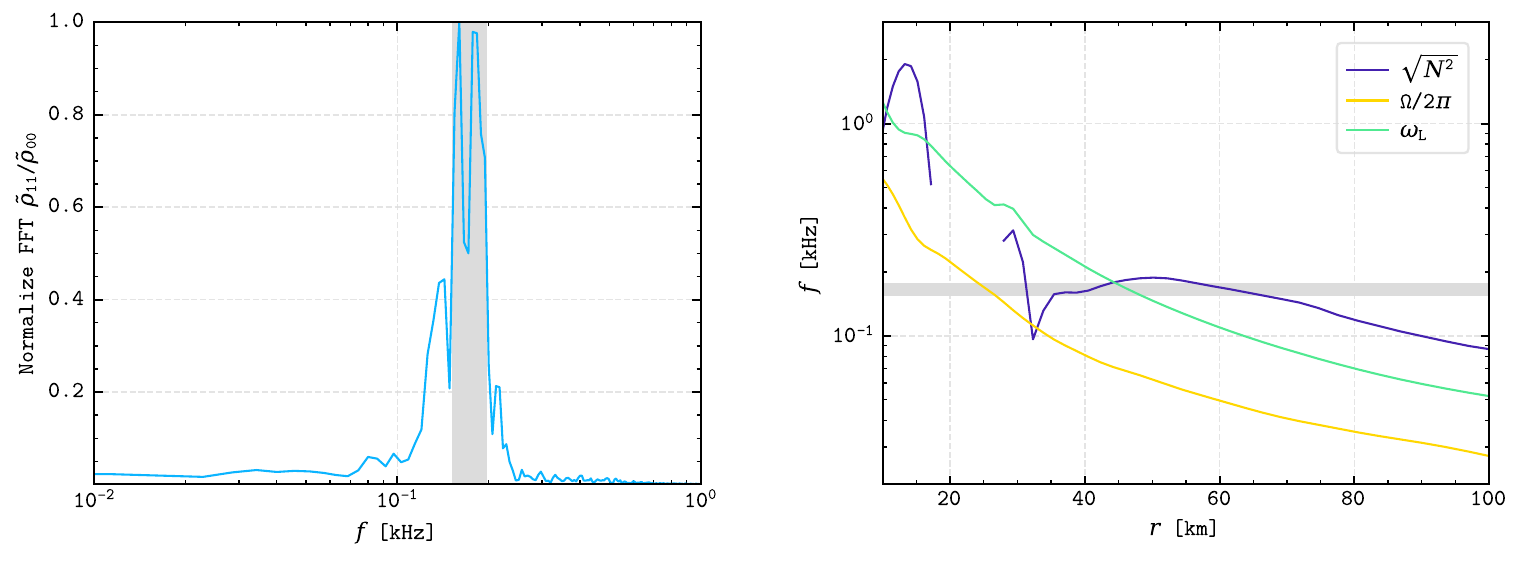}
    \caption{
    \label{fig:corotation}
     Left panel: normalised Fourier transform of the $\tilde{\rho}_{11}/\tilde{\rho}_{00}$ averaged between $r=\unit[25]{km}$ and $r=\unit[45]{km}$ 
     for the time interval $\unit[75-170]{ms}$ for model \texttt{TS}. The shaded rectangle represents the \gls{FWHM}, computed from the locations at which the normalised Fourier transform is equal to half of its maximum value.  
     Right panel: propagation diagram of the equatorial profiles of rotation (yellow), Brunt--Väisälä (blue),  
     and Lamb (green) frequencies at $\unit[110]{ms}$. The band represents the \gls{FWHM} of the dominant frequency of the mode $(1,1)$.}
\end{figure}

Building upon the results presented in Section~\ref{sec:sph_harm_deco}, we determine the temporal evolution of the corotation radius, $R_\textnormal{cor}$. Since the \gls{LTWI} exhibits an approximately constant pattern frequency during its saturated phase, we estimate this frequency from the Fourier spectrum of the $(\ell,m)=(1,1)$ mode over the time interval in which the instability is active. We compute the Fourier transform at radii between $\unit[25]{km}$ and $\unit[45]{km}$, in $\unit[1]{km}$ increments, and average the resulting spectra over this radial interval.  The pattern frequency is then identified with the maximum of this radially averaged spectrum, and the spectral width is estimated from its \gls{FWHM}. This procedure, illustrated for model \texttt{TS} in the left panel of Figure~\ref{fig:corotation}  provides a robust estimate of the global pattern speed while reducing the sensitivity to local fluctuations at individual radii. 

Since we have extracted a band of frequencies, we define a corotation region, $R_\textnormal{cor}$, which spans several kilometers, as the radial interval for which the equatorial rotation frequency satisfies
\begin{equation}
    \label{eq:cor_region}
    f_\textnormal{rot} \in \left[f_\mathrm{mode}-\frac{\Delta f}{2},\,f_\mathrm{mode}+\frac{\Delta f}{2}\right],
\end{equation}
with $\Delta f$ being the \gls{FWHM}.

To assess where the corotation region lies within the \gls{PNS}, in the right panel of Figure~\ref{fig:corotation} we show the  radial profiles of the rotation,
Brunt--Väisälä angular frequency,
\begin{equation}
    \label{eq:BV_freq}
    \omega_\textnormal{BV} = \frac{|N|}{2\pi},
\end{equation}
and Lamb frequency
\begin{equation}
    \label{eq:lamb}
    \omega_\textnormal{L} = \frac{1}{2\pi}\frac{\sqrt{\ell(\ell+1)}c_s}{r},
\end{equation}
evaluated for $\ell=1$, and where $c_s$ is the sound speed.

We use the Brunt--Väisäla frequency as a proxy to locate zones of convective instability, as $N^2<0$. Consistent with previous work \cite{Saijo2006, Takiwaki2021, Bugli23}, the growth of the \gls{LTWI} is strongest when the corotation region lies within such a convectively unstable zone. Furthermore, previous studies \cite{Passamonti2015, Passamonti2020} indicate that the unstable mode associated with the \gls{LTWI} is not a $p$-mode, which requires $f_\textnormal{mode} \gtrsim \omega_\textnormal{L}$ for acoustic propagation, but is instead consistent with $f$-, $g$-, or $i$-modes.

\begin{figure}
    \centering
    \includegraphics[width=1\linewidth]{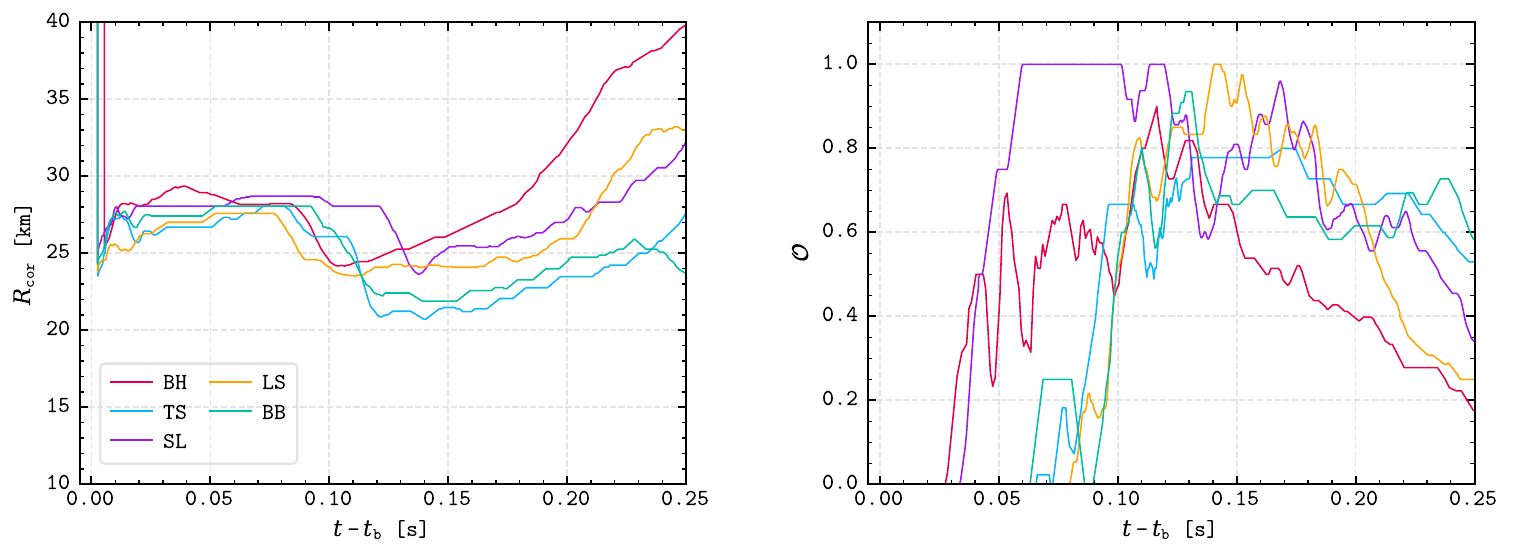}
    \caption{The left and right panels show the evolution of the average corotation radius, and overlap between the corotation and convective regions for all the models, respectively.}
    \label{fig:cor_region_overlap}
\end{figure}

We define the radial extent of the convectively unstable region (where $N^2(r,t)<0$) at each time $t$ as the interval
\begin{equation}
    [r_-(t),\,r_+(t)].
\end{equation}
In a similar fashion, the spectral width of the corotating mode taken as the instantaneous \gls{FWHM} defines a corotation region
\begin{equation}
    [r_{\rm cor}^-(t),\,r_{\rm cor}^+(t)].
\end{equation}
Since, in the models considered here, the convectively unstable layer and the corotation region each form a single connected radial interval, we measure the fractional radial overlap between the corotation region and the convectively unstable layer with the \emph{overlap function}
\begin{equation}
    \label{eq:overlap}
    \mathcal{O}(t)=
    \frac{
    \max\!\left[
        0,\,
        \min\!\big(r_+(t),r_{\rm cor}^+(t)\big)
        -
        \max\!\big(r_-(t),r_{\rm cor}^-(t)\big)
    \right]
    }{
    r_{\rm cor}^+(t)-r_{\rm cor}^-(t)
    },
\end{equation}
which, by construction, is $0$ if there is no overlap between the  two regions, while $\mathcal{O}=1$ if the corotation region is fully embedded in the convective layer.

Figure~\ref{fig:cor_region_overlap} shows the evolution of the corotation radius and of the overlap function for all models, in the left and right panels, respectively. In each case, $R_\mathrm{cor}$ initially moves inward  entering the convectively unstable zone of the \gls{PNS}. 
As the corotation region continues to shrink, its overlap with the convectively unstable zone increases. In our limited sample of models, the onset of the \gls{LTWI} happens when the overlap reaches $\gtrsim50\%$, while the instability strength is maximum (as indicated by the peak of the Fourier amplitudes in Figure~\ref{fig:fourier_deco}) when the corotation radius reaches its minimum value and the overlap between the corotation and convective regions reaches $\gtrsim75\%$. 
After the instability peak, with the loss of differential rotation, the corotation region gradually moves outward, eventually exiting the convective zone. 
Model \texttt{SL} constitutes an exception: the two regions show complete overlap $\sim\unit[50]{ms}$ before the instability begins to develop. This discrepancy may stem from the limitations of the  definition of the convective region, which is based on the equatorial average of the Brunt--Väisälä frequency, and may not fully capture the multidimensional structure of convection.

Based on the results presented so far, we conclude that the observed dynamics is due to the development of a \gls{LTWI}. The presence of substantial differential rotation and the evolution of the $T/|W|$ ratio place the system in a regime susceptible to shear instabilities \cite{Saijo2006,Passamonti2015,Bugli23}. The development of large-scale non-axisymmetric spiral patterns and the growth of $m=1$ and $m=2$ components in the Fourier amplitudes and the $\ell=m$ coefficients in the spherical harmonics decompositions, indicates the emergence of a coherent global mode \cite{Shibata2002,Shibagaki2021,Bugli23}.
The approximately constant pattern frequency, together with the fact that the dominant emission frequencies lie within the range of the angular velocity inside the \gls{PNS}, are consistent with the presence of a corotation region. The corotation region is also located within a convectively unstable layer, a condition that has been identified in previous studies as necessary for the development of the \gls{LTWI} \cite{Shibata2002,Shibata2003,Watts2005,Saijo2006}. 

We now turn to a description of the impact of the \gls{LTWI} on the multimessenger emission, more specifically on the \gls{GW} and neutrino signals.

\subsection{Gravitational waves}
\label{sec:GWs}
\subsubsection{Overall evolution}
\label{sec:GWs_over_evolution}
\begin{figure}[t]
    \centering
    \includegraphics[width=\textwidth]{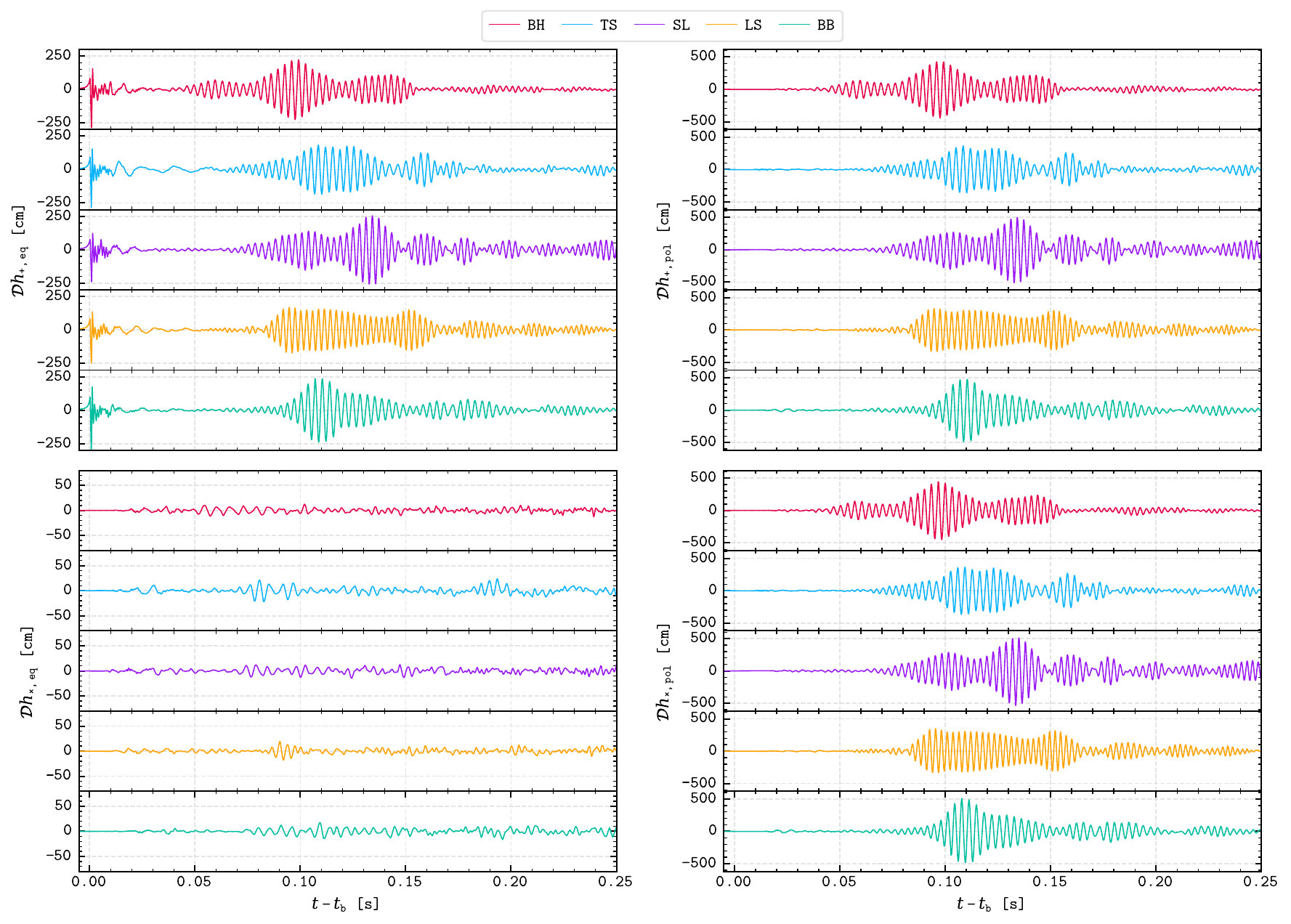}
    \caption{
    \label{fig:GWs_strains}
     \gls{GW} strains emitted by a source at a distance $\mathcal{D}$. The top and bottom left panels show the strains measured at the equator for the $+$ and $\times$ polarisations, respectively. The top and bottom right panels show the corresponding polarisations measured along the polar axis. Each panel contains results from models \texttt{BH}, \texttt{TS}, \texttt{SL}, \texttt{LS}, and \texttt{BB}.
    }
\end{figure}

In this section we focus on the impact of the \gls{LTWI} on the \gls{GW} emission. Figure~\ref{fig:GWs_strains} shows the two independent polarisations, $+$ and $\times$, emitted by a source at a distance $\mathcal{D}$ measured along the polar axis (\texttt{pol}) and the equator (\texttt{eq}).

Before discussing the behaviour of individual models, we summarise the features common to all of them.
All models emit only a weak signal with $\times$ polarization in equatorial directions, with an amplitude $|\mathcal{D}h_{\times,\textnormal{eq}}|\lesssim\unit[20]{cm}$. This is a direct consequence of the system being approximately symmetric about the equator, as described in Section~\ref{sec:eq_symm}. In contrast, the $+$ polarization for equatorial observers exhibits the characteristic bounce signal of rapidly rotating progenitors: a slow rise followed by a sharp drop, with a strain width (i.e., the difference between the highest and lowest points in the bounce signal) of $\unit[360-490]{cm}$. Once the \gls{LTWI} develops, the signal amplitude increases to $|\mathcal{D} h_{+, \mathrm{eq}}| \sim\unit[250]{cm}$, with a morphology that depends on the specific model and is described below.

Along the polar direction, the two polarisations show morphologies very similar to each other and to $h_{+,\textnormal{eq}}$, aside from the absence of the bounce feature. The \gls{GW} signal grows to about $|\mathcal{D} h_{+,\times, \mathrm{pol}}| \sim\unit[500]{cm}$, i.e., twice as high as the emission in an equatorial direction, $|\mathcal{D} h_{+, \mathrm{eq}}|$. The two polar polarization components are phase-shifted by $\pi/2$; otherwise they are morphologically equivalent.

Furthermore, the  morphology of the polarisations, once the \gls{LTWI} develops, is similar to the time series of the $(\ell,m)=(2,2)$ mode of the density decomposition  (middle panel of Figure~\ref{fig:sph_spectro}).
To quantify this correspondence we use the matching score, $\mathcal{M}$, \citep{Suvorova19}:
\begin{equation}
    \label{eq:matching}
    \mathcal{M} = \frac{\langle S_1(t) | S_2(t) \rangle}{\sqrt{\langle S_1(t) | S_1(t) \rangle \langle S_2(t) | S_2(t) \rangle}},
\end{equation}
where $S_i(t)$ denotes the two signals, and $\langle\cdot|\cdot\rangle$ represents the inner product. Specifically, we compare $h_{\times,\textnormal{pol}}$ with the $(\ell,m)=(2,2)$ spherical-harmonic component of the density at $\unit[40]{km}$, during the period of activity of the instability ($\unit[50-250]{ms}$). This analysis reveals that each model has a $\mathcal{M}\geq0.7$, indicating that the evolution of $h_{\times,\textnormal{pol}}$ closely tracks that of the $(2,2)$ mode.

We now turn to a model-by-model description of the detailed \gls{GW} morphology during the growth and saturation of the \gls{LTWI}. Since the morphology of the polarisations is similar to one another, we focus on $h_{+,\textnormal{pol}}$, referring to the upper right panel of Figure~\ref{fig:GWs_strains}.

The temporal evolution of the \gls{GW} enhancement differs across the models. 
Model \texttt{BH} shows an enhanced phase lasting $\sim\unit[100]{ms}$ ($\unit[50-150]{ms}$): the signal grows over several tens of milliseconds, reaches a peak, and then gradually weakens until disappearing at $\unit[150]{ms}$. 
In model \texttt{TS}, the amplitude increases progressively, reaching a maximum at $\sim\unit[100]{ms}$ and remaining enhanced for $\sim\unit[40]{ms}$ before weakening and fragmenting into a final clump of emission. 
The signal in model \texttt{SL} rises more slowly, peaking at $\unit[135]{ms}$ and then dropping sharply, after which only weak, isolated pulses remain. 
Model \texttt{LS} instead shows a rapid increase in amplitude, with the peak level sustained for more than $\unit[100]{ms}$ before declining; the subsequent emission appears in short bursts of moderately enhanced strength. 
Finally, model \texttt{BB} displays a rapid increase followed by a gradual decay of the signal until the end of the simulation.

To summarise, the five models are similar in terms of the maximum amplitude and the duration of the time during which the emission is strongest. They differ when it comes to the time at which the enhanced emission starts and how fast it rises as well as the activity in the form of late pulses. 

\begin{figure}[t]
    \centering
    \includegraphics[width=\linewidth]{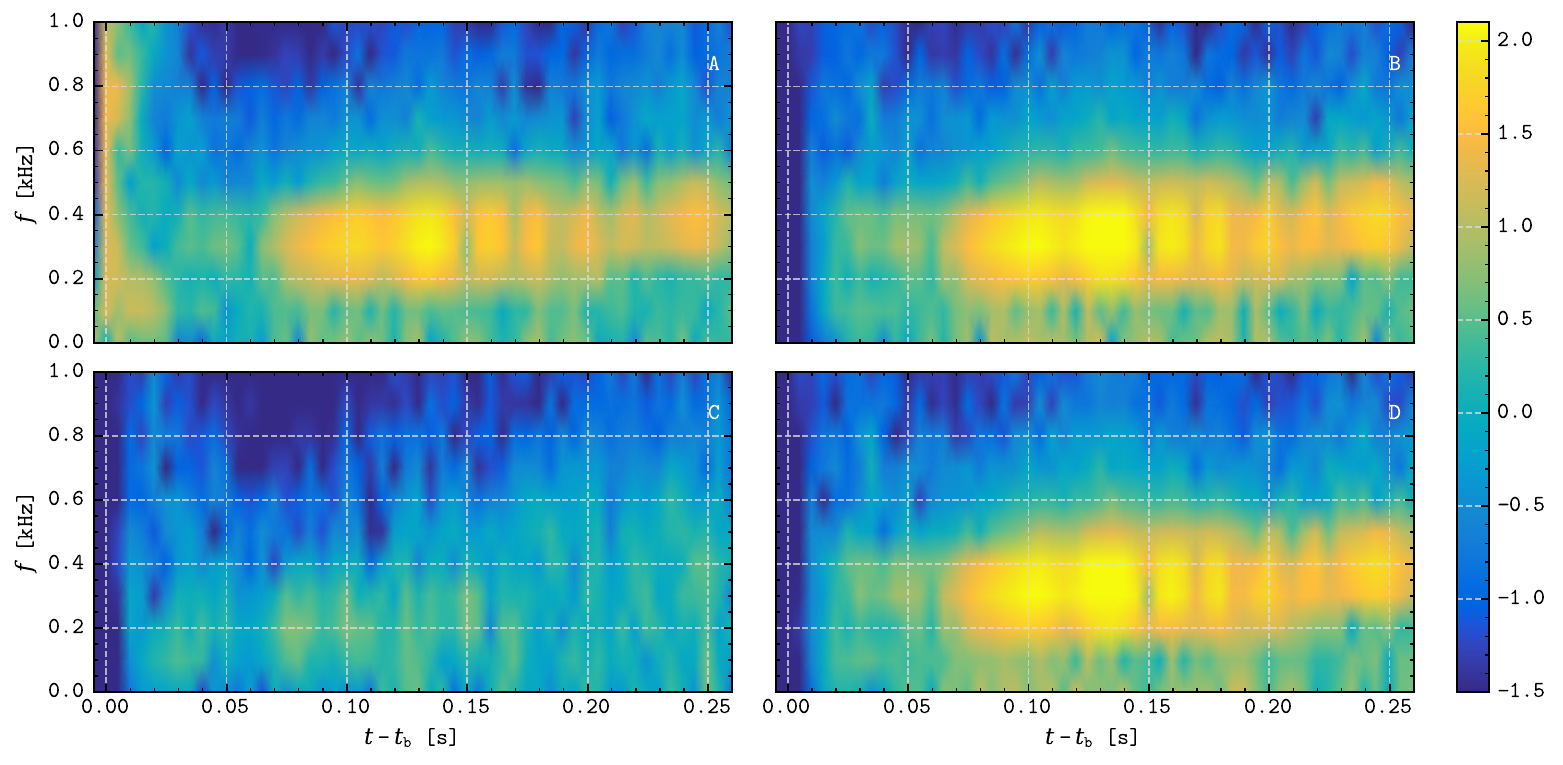}
    \caption{Spectrograms of the \gls{GW} strains measured at the equator (panels \textbf{A}: $+$ polarization, and \textbf{C}: $\times$ polarization) and along the polar axis (panels \textbf{B}: $+$ polarization, and \textbf{D}: $\times$ polarization) for model \texttt{SL}.}
    \label{fig:GWs_spectrograms}
\end{figure}

A complementary view of the aforementioned features in the time-frequency domain is provided by the spectrogram of the \gls{GW} signal. In Figure~\ref{fig:GWs_spectrograms} we display time-frequency maps of the two polarisations for polar and equatorial observers of model \texttt{SL} performed with a time window of $\unit[15]{ms}$. 

The equatorial symmetry is reflected in the suppression of  the $\times$ polarization for equatorial observers (panel C). The emitted power remains weak throughout the simulation and no specific emission frequency is observed. On the other hand,  the $+$ polarization on the equator (panel A) shows the broad-band, loud emission associated with core bounce \citep{Richers17}. This signal is followed by a low-frequency ($f\leq\unit[250]{Hz}$) weaker component lasting for $\sim\unit[30]{ms}$, which corresponds to prompt convection (described in \cite{Cusinato2025a} for axisymmetric models).

Once the \gls{LTWI} develops, a strong feature emerges. A strong \gls{GW} signal begins at $\sim\unit[75]{ms}$ with a frequency of $\sim\unit[320]{Hz}$ in all polarisations except $h_{\times,\textnormal{eq}}$. This feature persists until the end of the simulation, maintaining an approximately constant frequency.
Panels A, B, and D show that this component is present in  $h_{+,\textnormal{eq}}$, $h_{\times,\textnormal{pol}}$, and $h_{+,\textnormal{pol}}$, and it is consistent with the $(\ell, m)=(2,2)$ character of emission. Comparing these spectrograms with the spherical-harmonic decomposition of the density (see middle panel of Figure~\ref{fig:sph_spectro}) we find that the $(2,2)$  mode exhibits the same frequency and temporal evolution as the \gls{GW} signal. Moreover, we remark that not only the \gls{GW} and $(2,2)$  mode frequencies closely align, but also the   overall morphology of the waveform resembles that of the time series of the density mode.

Since the \gls{GW} emission associated with the development of the \gls{LTWI} shows an approximately constant frequency throughout the active phase, we can extract its dominant frequency using the same procedure employed in Section~\ref{sec:corotation} for the $(\ell,m)=(1,1)$ mode. Specifically, we compute the Fourier transform of $h_{\times,\textnormal{pol}}$ over the interval in which the instability is active ($\unit[50-250]{ms}$). We then identify the peak frequency, with an uncertainty given by the \gls{FWHM}. These peak frequencies coincide with those of the $(2,2)$ density mode (see Table~\ref{tab:quantitative_res}).

To assess how the choice of the \gls{EOS} influences this emission, we use the tidal Love number of the \gls{PNS}.
For each model, we construct the time series of $\kappa_2$ and compute its mean and standard deviation over the same $\unit[50{-}250]{ms}$ window.

\begin{figure}
    \centering
    \includegraphics[width=\linewidth]{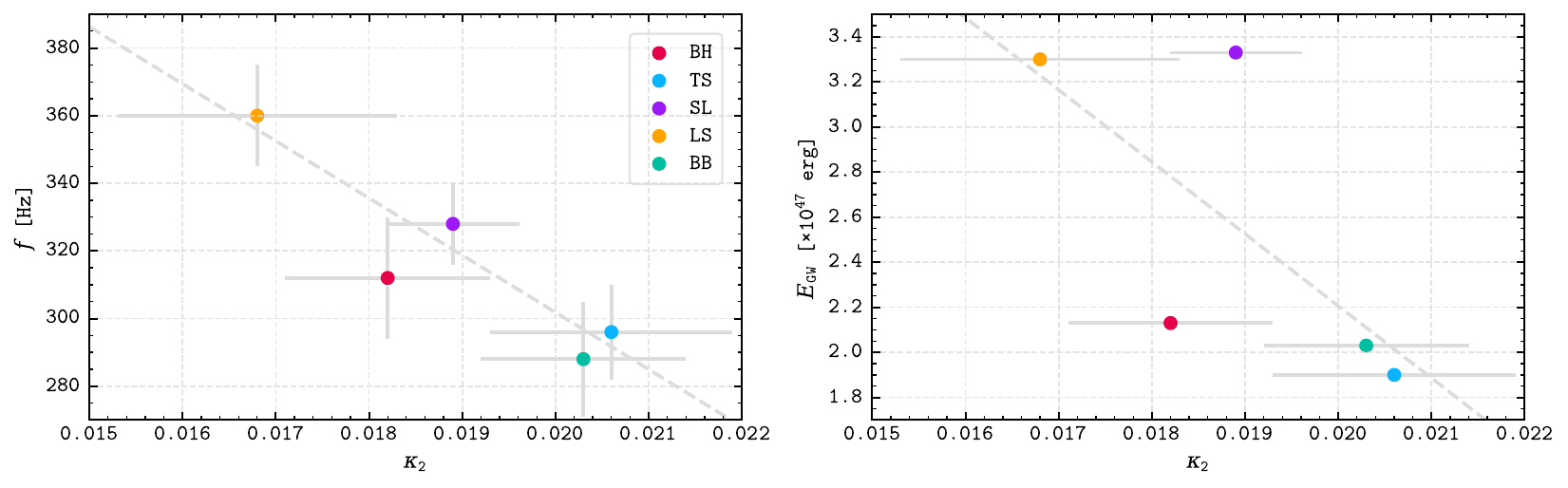}
    \caption{Left panel: dominant frequency of the \gls{GW} signal associated with the \gls{LTWI} against the average tidal Love number $\kappa_2$. Right panel: energy emitted in \glspl{GW} until \unit[250]{ms} against $\kappa_2$. Errors on the x-axis represent the standard deviation from the average value of $\kappa_2$. Errors on the y-axis represent half of the \gls{FWHM}. The dashed line represents the linear fit.}
    \label{fig:GWs_peak_freq}
\end{figure}

The left panel of Figure~\ref{fig:GWs_peak_freq} shows the dominant \gls{GW} frequency associated with the \gls{LTWI} plotted against the average tidal Love number. We observe that a clear trend emerges: models with a smaller Love number, which in our convention correspond to more compact/stiffer \gls{PNS} configurations, emit at higher frequencies. On the contrary, 
larger values of $\kappa_2$, correlated in our models with softer \glspl{EOS}, produce lower frequencies. 
This behaviour may be explained by the fact that smaller values of $\kappa_2$ correspond to more compact \glspl{PNS} with higher mean densities, implying shorter dynamical timescales and therefore higher characteristic oscillation frequencies. Similar correlations between tidal deformability and \gls{GW} spectral features have also been reported in binary neutron-star merger studies, where the post-merger peak frequency is closely related to the \gls{EOS}-dependent stellar compactness and tidal deformability \cite{Bernuzzi2015,Rivieccio2024}.
A linear fit to the data yields
\begin{equation}
\label{eq:fit}
    f=(-17.4\pm4.0)\times10^3  \kappa_2 + (6.4\pm8)\times 10^2,
\end{equation}
with a coefficient of determination  $R_\textnormal{fit}^2=0.83$. However, we stress that this relation should not be interpreted as a universal relation. Here, $\kappa_2$ is employed only as a structural proxy for the hot, evolving \gls{PNS}, since its formal definition applies to cold, non-rotating \glspl{NS}. Additionally, our analysis is based on a single progenitor and one initial rotation rate. The observed correlation should therefore be regarded as suggestive, motivating future investigations over a broader range of progenitors and rotational configurations.

The right panel of Figure~\ref{fig:GWs_peak_freq} shows the total energy emitted in the form of \glspl{GW}, $E_\textnormal{GW}$, at $\unit[250]{ms}$ as a function of the tidal Love number. Contrary to the peak frequency, $E_\textnormal{GW}$ does not exhibit a clear monotonic dependence on the \gls{EOS} stiffness. Instead, the models cluster into two distinct subsets. Models \texttt{LS} and \texttt{SL} emit $\sim\unit[3\times10^{47}]{erg}$, while models \texttt{BH}, \texttt{BB}, and \texttt{TS} emit roughly $\unit[2\times10^{47}]{erg}$. This separation is consistent with the morphology of the signals: the higher-energy models are precisely those in which the \gls{GW} emission shows a series of pulses of moderate amplitude after the main period of enhanced emission has passed. On the contrary, the lower-energy models show a shorter phase of strong non-axisymmetric activity, which peaks rapidly before disappearing, leading to a correspondingly smaller integrated emission (see previous discussion on the morphology of the signals).
Moreover, this clustering might also reflect differences in the underlying microphysics of the \glspl{EOS}. In particular, the \glspl{EOS} of models \texttt{SL} and \texttt{LS}, adopt a \gls{SNA}  for heavy nuclei below nuclear saturation density, while those of models \texttt{BH}, \texttt{BB}, and \texttt{TS} describe  matter in \gls{NSE} with an ensemble of nuclei.
The fluid in the region where the \gls{LTWI} develops is treated with the high-density \gls{EOS}, which may affect its thermodynamic structure. Therefore, the different treatment of the matter, may have an impact on  the convection and the differential rotation profile, potentially leading to the observed  separation between the energy carried away as \glspl{GW}.

\subsubsection{Space-time evolution of the GW emitting regions}
\label{sec:ts_evo_GW}
\begin{figure}[t]
    \centering
    \includegraphics[width=1\linewidth]{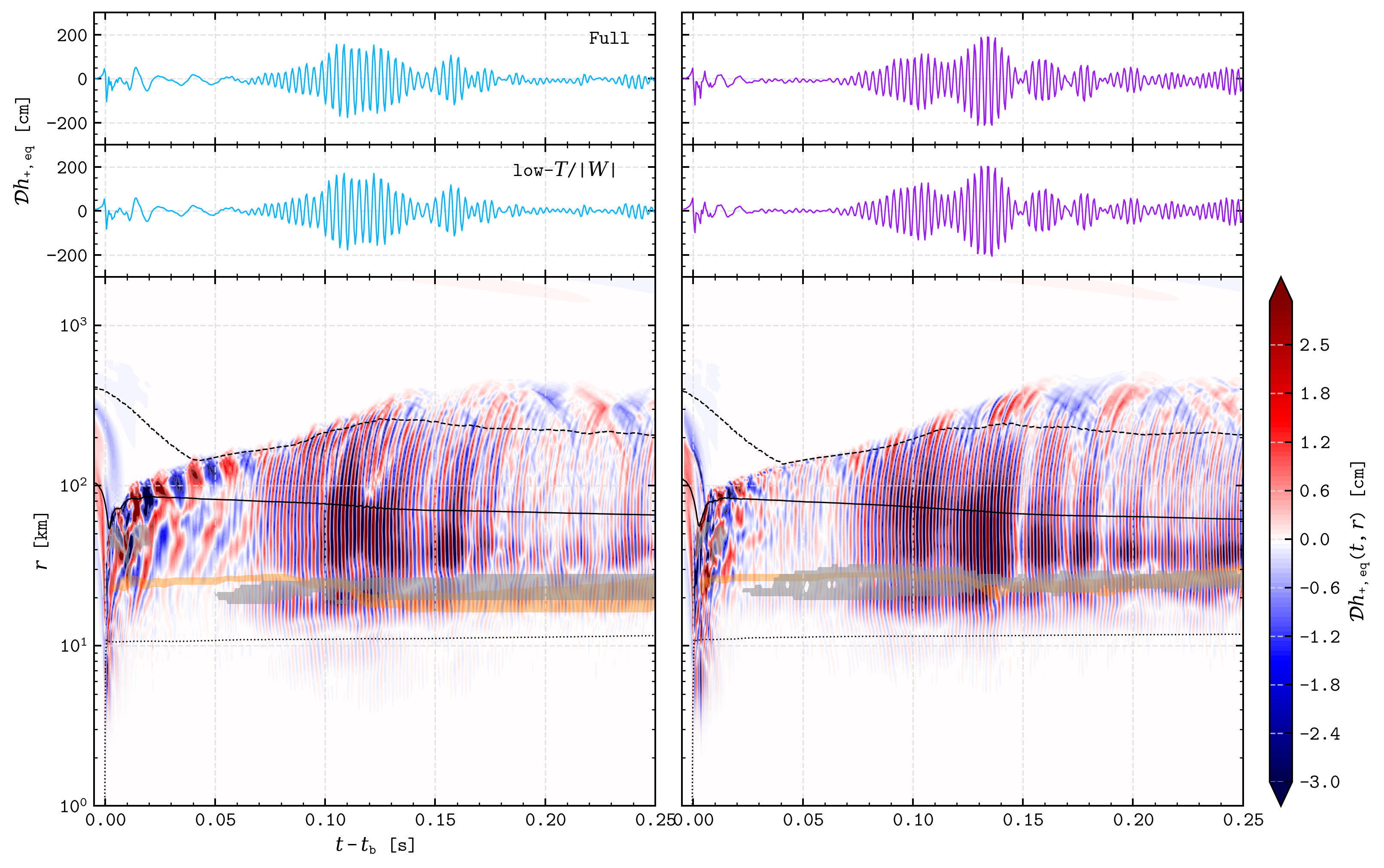}
    \caption{Evolution of the $+$ polarization of the \gls{GW} amplitude for an observer on the equatorial plane for models \texttt{TS} (left) and \texttt{SL} (right). The two upper panels show the time evolution of the \gls{GW} amplitude for the entire simulation domain (top panels), and for the region between the isosurfaces of $\unit[10^{14}]{g/cm^3}$ and $\unit[10^{9}]{g/cm^3}$ (middle panels).  
    The bottom panels show the space-time evolution of $\mathcal{D}h_{+,\textnormal{eq}}(r, t)$. Solid, dashed and dotted lines have the same meaning as in the left panel of Figure~\ref{fig:barcode}. Gray shades mark convective regions on the equatorial plane (i.e., where $N^2 < 0$) within the \gls{PNS}. Orange shades indicate the corotation region. The waveforms and space-time diagrams are computed from the simulation data with a temporal resolution of $\unit[0.5]{ms}$, which limits the resolution of frequencies to $\unit[1]{kHz}$. As a result, discrepancies are present, particularly during the bounce and early post-bounce phases, when higher-frequency components are dominant (see Figure~\ref{fig:GWs_strains}). However, since the \gls{GW} emission associated with the \gls{LTWI} is confined to frequencies well below $\unit[1]{kHz}$, the decomposition remains reliable for analysing the instability-driven signal.}
    \label{fig:GW_deco}
\end{figure}
To visualize the spatial origin and temporal development of the emission, we compute the \gls{GW} amplitude as a function of time and radius. We focus on the $+$ polarization for an equatorial observer in two representative cases: model \texttt{TS}, where the spiral modes terminate, and model \texttt{SL}, where they continue with reduced strength. For this analysis,  we refer to the last row of Figure~\ref{fig:GW_deco}, while the total amplitude and the contribution from the region enclosed by the $\unit[10^{14}]{g/cm^3}$ and $\unit[10^{9}]{g/cm^3}$ isosurfaces are displayed in the first and second rows, respectively.

The bounce signal, associated with the formation of the \gls{PNS}, marks the onset of \gls{GW} emission. During the few tens of milliseconds following the bounce signal, several modes emerge. These include prompt convection between $\unit[25]{km}$ and $\unit[60]{km}$, driven by negative entropy and $Y_e$ gradients, and oscillations of the innermost $\unit[10]{km}$ of the \gls{PNS}. These modes correspond to the ones described for the fast-rotating model in \citet{Cusinato2025a}.

After prompt convection ceases, model \texttt{TS} enters a $\sim\unit[50]{ms}$ phase in which the \gls{GW} signal is dominated by low-frequency modes propagating from the centre toward the \gls{PNS} surface (corresponding to mode~(4) in \cite{Cusinato2025a}). In contrast, in model \texttt{SL}, these modes are either disrupted or strongly suppressed by the onset of \gls{PNS} convection in the equatorial region (gray regions in the bottom row of Figure~\ref{fig:GW_deco}). These convective motions excite fast oscillating modes that propagate outward and pile up at the shock, interrupting the previously dominant low-frequency component.

In both models, the \gls{LTWI} begins to influence the \gls{GW} signal only after \gls{PNS} convection has developed. Once the corotation region (orange shading in the bottom row of Figure~\ref{fig:GW_deco}) enters the convectively unstable layer, the instability leads to a high amplification of the \gls{GW} signal. The modes of the resulting signal consist of fast oscillations at $\sim\unit[400]{Hz}$ originating at radii of $\sim\unit[20]{km}$, and spanning the entire \gls{PNS}. These modes penetrate below the  $\unit[10^{14}]{g/cm^3}$ isodensity surface, where they are quickly damped and therefore this part contributes negligibly to the observable signal. The modes due to the \gls{LTWI} also possess outward-propagating components that cross the \gls{PNS} boundary, slow down, but retain much of their amplitude before eventually being partially reflected and dissipated at the shock. In model \texttt{SL}, the corotation region appears to enter the convective zone earlier than the actual onset of the instability; this is likely due to our approximate identification of convective regions which is based on an angle-averaged $N^2$ over  $\theta\in[-30^\circ,30^\circ]$, which may overestimate the outer radial boundary of the convective layer (see Section~\ref{sec:corotation}).

Following the peak development of the instability, models \texttt{TS} and \texttt{SL} differ in their evolution. In model \texttt{TS}, the \gls{GW} enhancement associated with the \gls{LTWI} ceases, leaving only weak modes spanning the entire \gls{PNS} and propagating outwards. On the other hand, in model \texttt{SL}, the corotation region remains fully embedded in the convective zone and similar modes to the maximum activity continue for longer and with them the enhanced emission, albeit at a reduced amplitude. 

Finally, we find that most of the sub-kilohertz \glspl{GW} are emitted from the region bounded by the $\unit[10^{14}]{g/cm^3}$ and $\unit[10^{9}]{g/cm^3}$ isosurfaces.  The matching score defined in Equation~\eqref{eq:matching} between the \gls{GW} extracted from this region and the signal from the full domain yields $\mathcal{M}\geq 0.9$ for all polarisations and observer orientations. 

\subsubsection{GW EEMD decomposition}
\label{sec:EEMD_GWs}
\begin{figure}[t]
    \centering
    \includegraphics[width=1\linewidth]{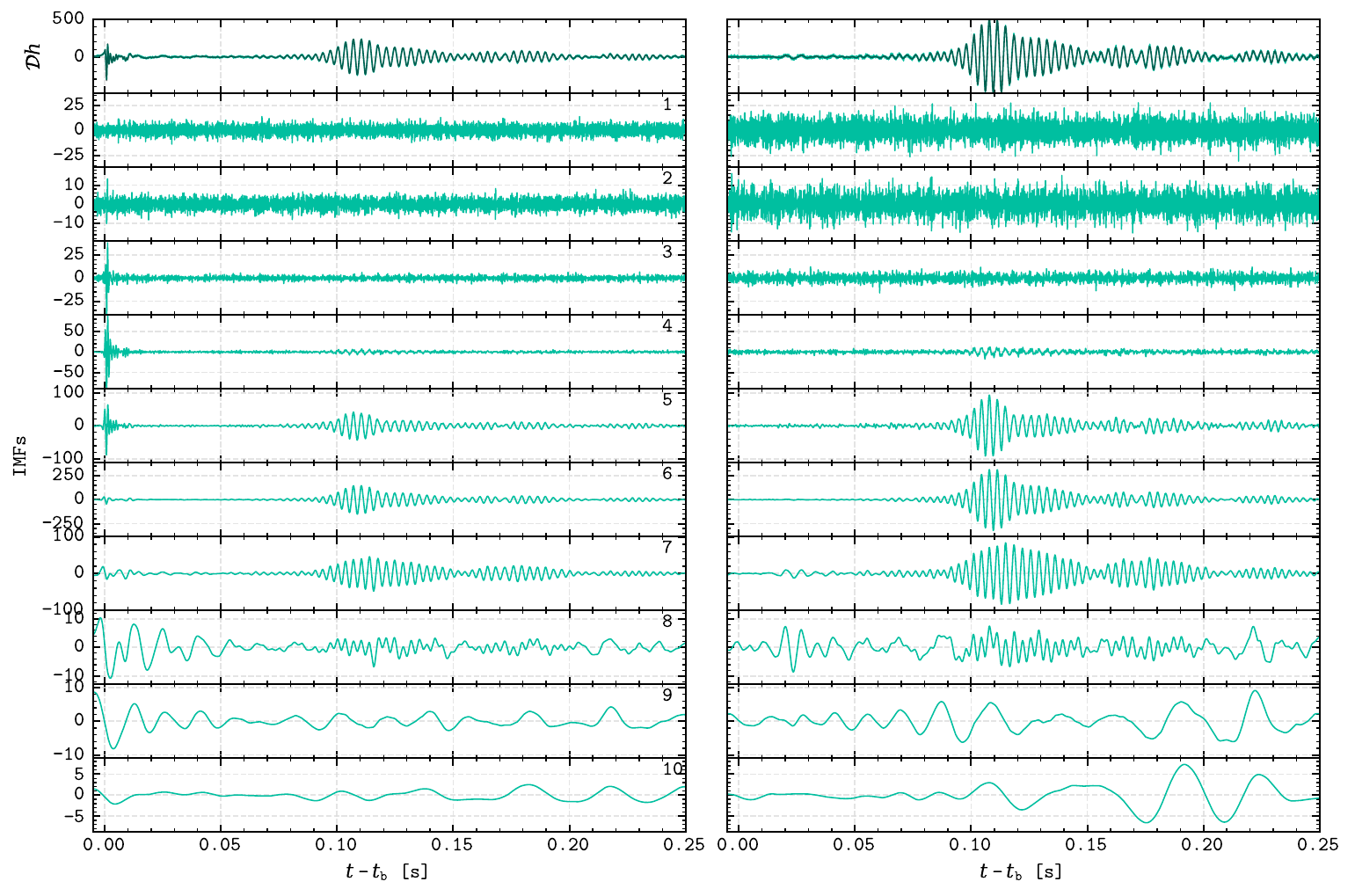}
    \caption{Time evolution of the total \gls{GW} $+$ polarization for an observer on the equatorial plane (left), and $\times$ polarization for an observer  on the polar axis and corresponding \gls{EEMD} in ten \glspl{IMF} for model \texttt{BB}. The black lines in the top row represent the original strain, the green lines the sum of the \glspl{IMF}.}
    \label{fig:GW_EEMD}
\end{figure}
We apply the procedure described in Section~\ref{sec:eemd} to decompose our signals into 10 \glspl{IMF}. Prior to the \gls{EEMD} we remove the residual, computed with a single  \gls{EMD}, to prevent a slow drift of the reconstructed \gls{GW} amplitude toward positive values that indicates jet formation.  
In practice, since none of our models have exploded or formed jets during the simulated time, the residual contribution to the waveform is negligible.

Figure~\ref{fig:GW_EEMD} shows the resulting \gls{IMF} decomposition: the $+$ polarization for an equatorial observer in the left panel and the $\times$ polarization for a polar observer in the right panel. The two polarisations display qualitatively similar \gls{IMF} behaviour. Specifically, \glspl{IMF} 1 and 2 reflect residual components of the added white noise that were not fully averaged out by the ensemble procedure. \glspl{IMF} 3 and 4 are very weak (peak $\mathcal{D}h\leq\unit[10]{cm}$) and, aside from a small contribution of the bounce signal in $h_{+,\mathrm{eq}}$, they contribute negligibly to the overall strain. Similarly, \glspl{IMF} 8-10 are weak ($\mathcal{D}h\leq\unit[10]{cm}$) and do not substantially affect the total signal.

\glspl{IMF} 5–7 are the dominant components, with maximum amplitudes above $\unit[50]{cm}$. By computing the matching score defined in Equation~\eqref{eq:matching} between the original waveform and the reconstruction from each IMF, we find that the component associated with the \gls{LTWI} is well captured by \gls{IMF} 6 alone. Across the full simulation sample, for all observer directions and polarisations, \gls{IMF} 6 matches the full signal generated by the \gls{LTWI} at $\mathcal{M}\geq0.99$, with only minor corrections when \glspl{IMF} 5 and 7 are included.

\subsection{Neutrino signal}
\label{sec:neutrinos}
\begin{figure}[t]
    \centering
    \includegraphics[width=\linewidth]{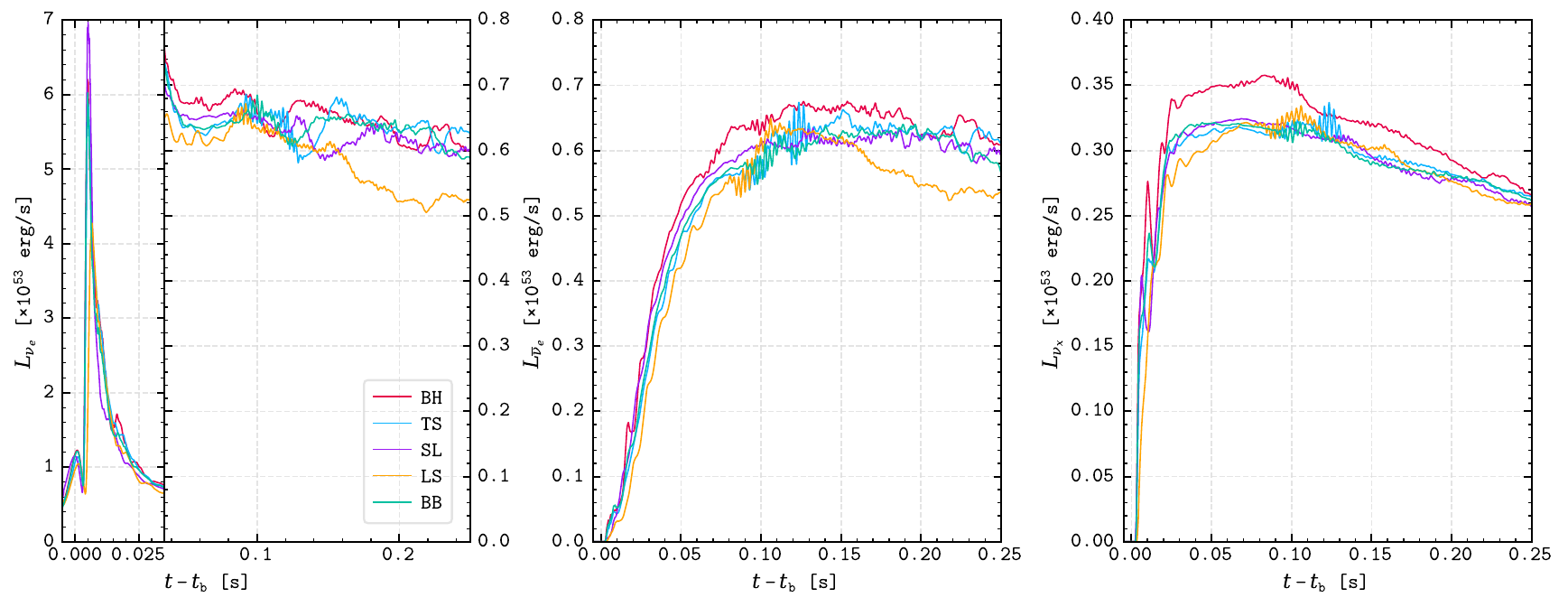}
    \caption{Time evolution of the neutrino luminosity for electron neutrinos, $\nu_e$, (left), electron antineutrinos, $\overline{\nu}_e$, (middle), and a single heavy lepton neutrino flavour, $\nu_x$, (right).
    }
    \label{fig:nu_curves}
\end{figure}

Figure~\ref{fig:nu_curves} shows the total neutrino luminosities extracted at $\unit[500]{km}$ for electron neutrinos ($L_{\nu_e}$, left panel), electron antineutrinos ($L_{\overline{\nu}e}$, middle panel), and heavy-lepton neutrinos ($L_{\nu_x}$, right panel).

Previous studies have found that the neutronization burst shows only a moderate sensitivity to the progenitor structure and nuclear \gls{EOS} \cite{Kachelriess2005, Hudepohl2010}. In contrast, our models exhibit a significant variation in the amplitude of the initial $L_{\nu_e}$ breakout peak, spanning $\unit[4\times10^{53}-7\times10^{53}]{erg/s}$. This suggests that \gls{EOS}-dependent effects may be more pronounced in our setup. These differences likely reflect variations in the \gls{PNS} structure and in the location of the neutrinosphere at shock breakout, and may additionally be enhanced by multidimensional effects absent in 1D simulations (e.g., \cite{Hudepohl2010}).

We find that neither the effective stiffness parameter nor the $B_1$ and $B_2$ proxies based on cold \gls{EOS} properties (see Table~\ref{tab:eoss_parameters}) provide a reliable predictor of the neutronization burst peak across our set of models. Nevertheless, these quantities do capture a robust limiting behaviour, as they consistently identify model \texttt{LS} as producing the weakest burst.

The two  \gls{EOS} with the smaller initial values of the Love number ($\kappa_{[1,15]}$ models \texttt{LS} and \texttt{BH}; see  Table~\ref{tab:quantitative_res}) are indeed associated with the lowest breakout luminosities, while intermediate-stiffness models yield stronger emission. Stiffer \gls{EOS} provide greater pressure support at a given density, leading to a less compact core at bounce. This may result in weaker shock formation and reduced compression at shock breakout, thereby producing a weaker prompt $\nu_e$ burst. 

The $B_1$ and $B_2$ proxies reproduce some aspects of the behaviour, correctly identifying model \texttt{LS} as the weakest case and placing \texttt{BH}, \texttt{BB}, and \texttt{SL} in the intermediate-to-high range. However, they also show clear discrepancies: in particular, they systematically overestimate the burst strength of model \texttt{TS} and incorrectly rank \texttt{BB} above \texttt{SL}.

Overall, this indicates that the neutronization burst peak is not primarily controlled by cold \gls{EOS} properties, but instead by dynamical quantities emerging during collapse and bounce. These likely include the evolution of the electron fraction, the thermal response of the \gls{EOS} at sub-saturation densities, and the detailed structure of the core at shock breakout. The fact that all indicators robustly identify model \texttt{LS} as producing the weakest burst likely reflects its comparatively low pressure support in these regimes, leading to a more compact core and reduced shock strength.

Most of the light curves are similar to each other, but there are a few outliers.
The electron neutrino and anti-neutrino luminosities of model \texttt{LS} show a noticeable drop around $\unit[150]{ms}$ to a level about $\sim20\%$ lower than in the other models. This phenomenon is caused by an interplay between the \gls{EOS} and the rotation. The stiffer \gls{EOS} of model \texttt{LS} has a smaller and less massive \gls{PNS} compared to the other models. Additionally, the accretion rate diminishes at $\sim\unit[120]{ms}$ lowering the neutrino luminosities. The drop in the neutrino luminosities also coincides with the dissipation of the spiral structure, which eliminates a source of \gls{PNS} destabilisation.
The heavy-lepton neutrino luminosity ($L_{\nu_x}$) for model \texttt{BH} dominates over the other models by $\sim 15\%$. This behaviour may be caused by the inclusion of the $\Lambda$-hyperons at high densities, which enlarge the heavy-lepton neutrinosphere. Although a larger decoupling radius would typically imply lower temperatures, in this case the temperature remains comparable to the other models, leading to an overall increase in the emitted neutrino luminosity. 

Models \texttt{BB}, \texttt{LS}, and \texttt{TS} develop fast luminosity oscillations at around $\unit[100]{ms}$ after bounce, with a characteristic frequency of $\sim\unit[350]{Hz}$. These oscillations correspond to the onset of large-scale non-axisymmetric modes within the \gls{PNS}, and they are visible in all three neutrino flavours.

\begin{figure}[t]
    \centering
    \includegraphics[width=\linewidth]{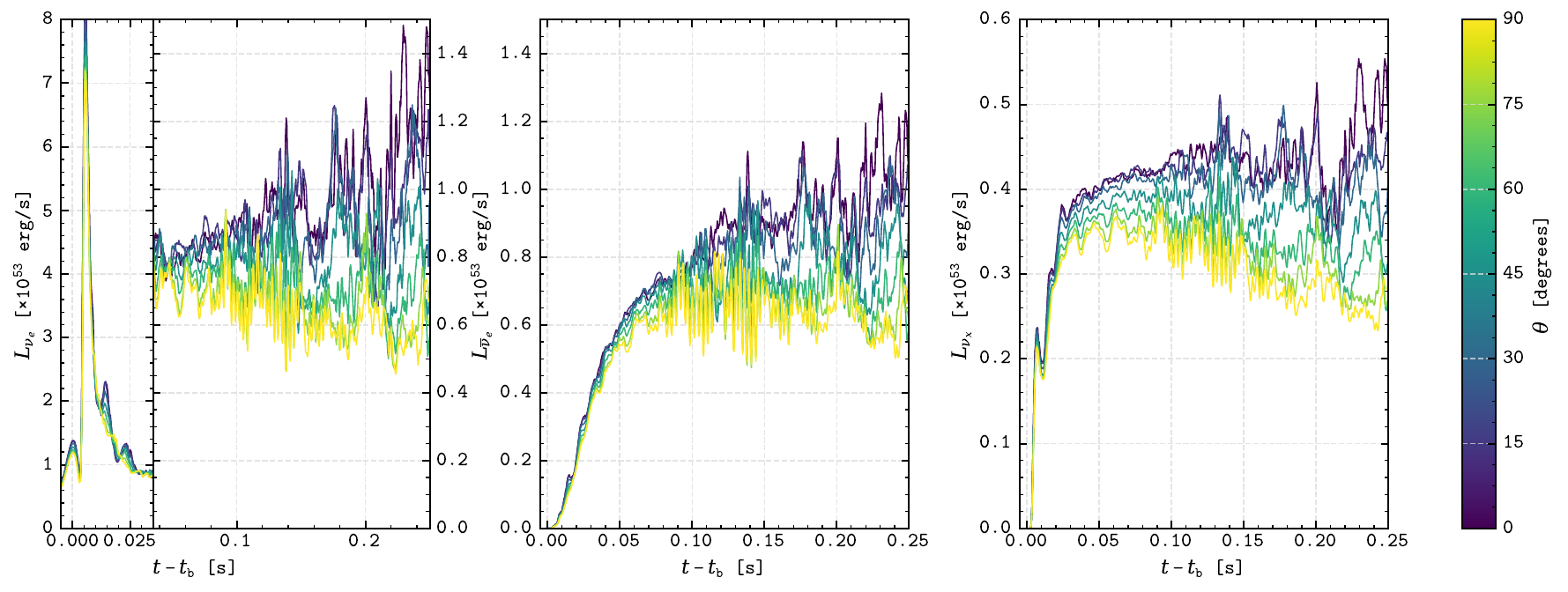}
    \caption{Time evolution of the isotropic equivalent neutrino luminosity emitted for different values of the colatitudinal angle $\theta$ for model \texttt{SL}. Panels from left to right show electron neutrinos, $\nu_e$, electron antineutrinos, $\overline{\nu}_e$, and heavy lepton neutrinos, $\nu_x$.}
    \label{fig:nu_angle}
\end{figure}

Since the neutrinospheres reside within the region affected by the instability, they are expected to be perturbed once the \gls{LTWI} mode sets in. However, the angle-integrated luminosities can mask the $\phi$-dependent variations associated with spiral modes. Therefore, 
to assess the angular dependence of the neutrino-signal modulation resulting from the \gls{LTWI}, we compute the isotropic equivalent neutrino luminosity,
\begin{equation}
    \label{eq:lum_nu_eq}
    L_{\nu_i}^{\rm iso}(\theta, \phi) = 4 \pi r^2 F_{\nu_i}^r(r, \theta, \phi),
\end{equation}
where $F_{\nu_i}^r(r,\theta,\phi)$ is the radial neutrino flux and $\nu_i$ denotes the neutrino flavour. We extract these luminosities at $r=\unit[500]{km}$ for a fixed azimuth $\phi = 0^\circ$, while varying the polar angle $\theta$ from $0^\circ$ to $90^\circ$. In Figure~\ref{fig:nu_angle} we compare the isotropic equivalent luminosities for model \texttt{SL}.

We find that although the integrated luminosity shows no clear signature of non-axisymmetric instability, the equivalent luminosities evaluated in specific directions do. 
In particular, viewing angles between $75^\circ$ and $90^\circ$ display more persistent and regular oscillations, consistent with the predominantly equatorial nature of the \gls{LTWI}, while toward the poles the signal appears more intermittent and lacks clear long-lived modulation.
This is a direct consequence of the \gls{LTWI}, which develops on the equatorial plane, and the spiral modes lead to larger temporal fluctuations in the position and the neutrinospheres.
Additionally, the neutrino luminosities are more intense at the poles than at the equator for all flavours. This reflects the rotationally induced flattening of the neutrinospheres: in the equatorial plane, the larger radius places the decoupling region at lower temperatures compared to the poles, resulting in a reduced equatorial emission. 

\begin{figure}[t]
    \centering
    \includegraphics[width=1\linewidth]{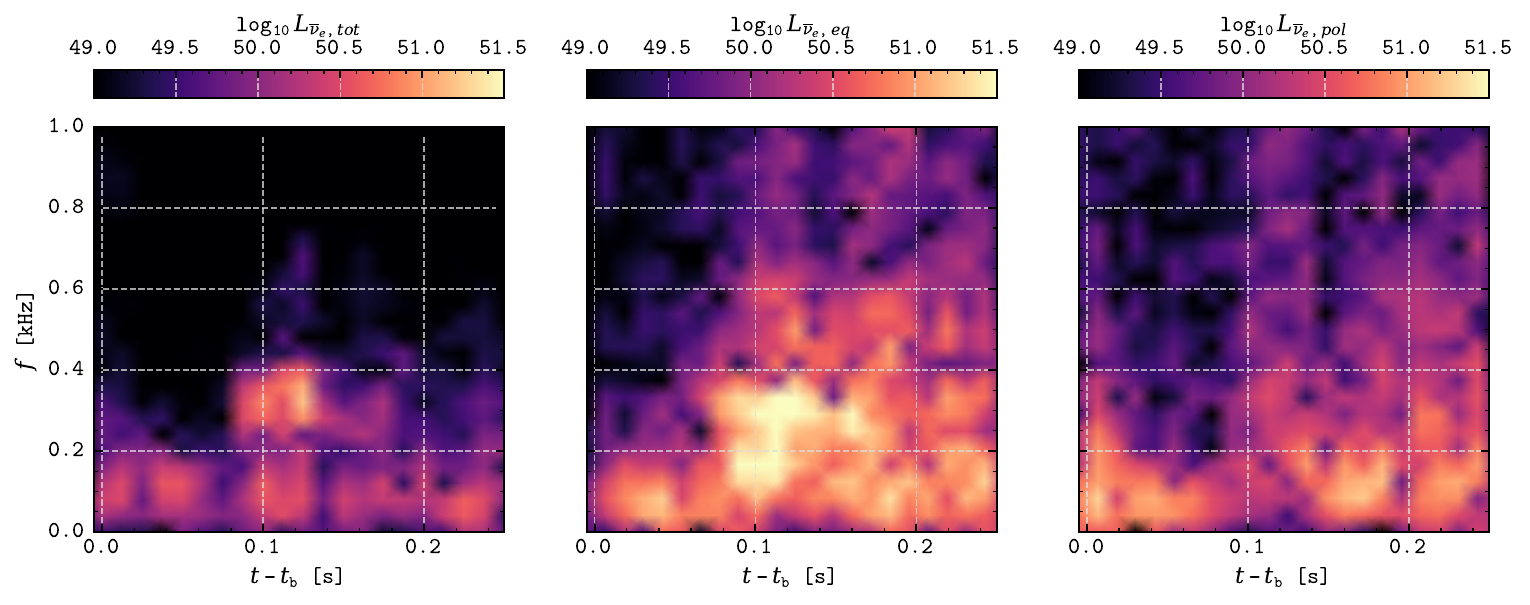}
    \caption{Spectrograms of the electron antineutrino luminosity for model \texttt{TS}: the total luminosity (left), the isotropic equivalent  luminosity from the equator (middle), and the isotropic equivalent luminosity from the pole (right). In all spectrograms a  band-pass filter in the range $f=[50,\,1000]\unit[]{Hz}$ has been applied.
    }
    \label{fig:nu_spectro}
\end{figure}

In Figure~\ref{fig:nu_spectro} we show the spectrograms of the angle-integrated neutrino luminosity (left), the equatorial isotropic equivalent luminosity (middle), and the polar isotropic equivalent luminosity (right) for $\overline{\nu}_e$ of model \texttt{TS}. 
The integrated luminosity exhibits a clear feature at a frequency just below $\unit[400]{Hz}$, lasting from $\sim\unit[80]{ms}$ to $\sim\unit[130]{ms}$, which matches the dominant $(\ell,m) = (2,2)$ mode identified in Section~\ref{sec:sph_harm_deco}. The equatorial luminosity reveals this same mode, which persists for a longer duration ($\unit[75-140]{ms}$) as well as an additional prominent feature below $\unit[200]{Hz}$ associated with the $(\ell,m) = (1,1)$ mode (see Table~\ref{tab:quantitative_res}). This lower-frequency contribution is likely suppressed in the integrated luminosity because its non-axisymmetric emission pattern largely cancels out under angle integration. In contrast, the polar emission shows no evidence of any \gls{LTWI}–driven modulation. This behaviour is fully consistent with the predominantly equatorial character of the \gls{LTWI} discussed in Section~\ref{sec:eq_symm}. Equivalent spectrograms for $\nu_e$ and $\nu_x$ and other models display the same qualitative trends (not shown).

\section{Discussion}
\label{sec:discussion}
In this section we discuss the prospects for detecting the \glspl{GW} and neutrino signals emitted by our models.
\subsection{GWs: Prospects of detection}
\label{sec:gws_detection}
\begin{figure}[t]
    \centering
    \includegraphics[width=\linewidth]{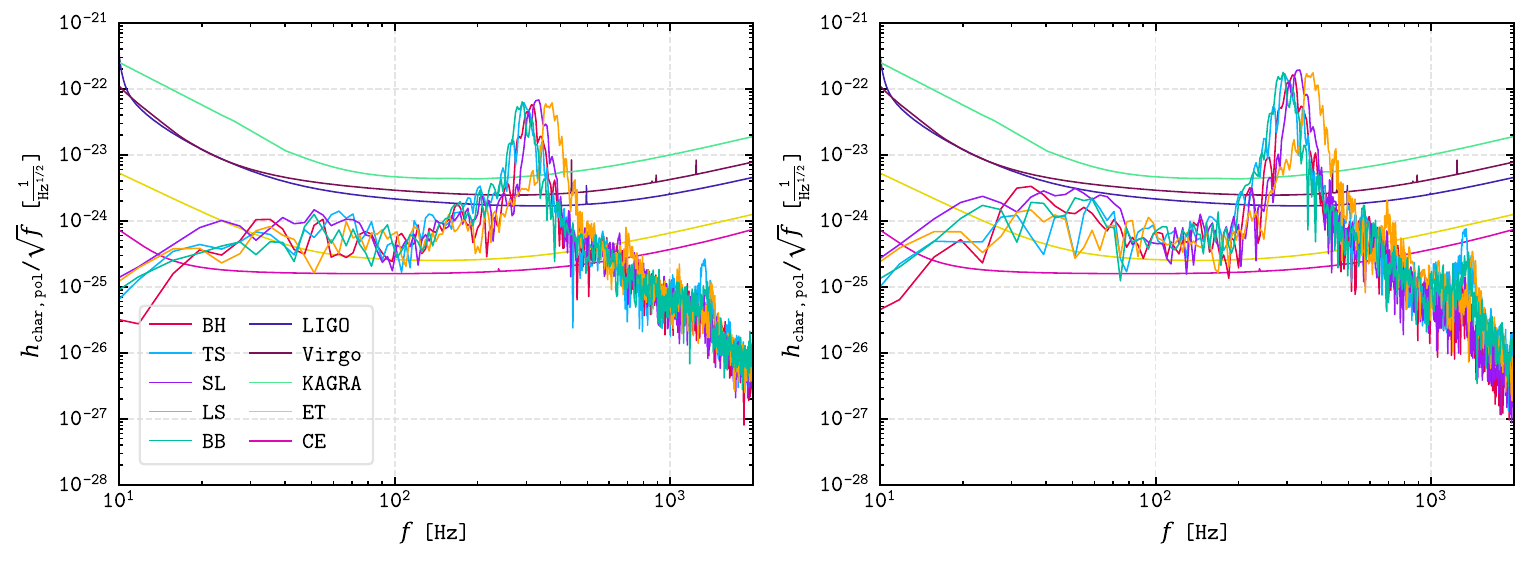}
    \caption{Characteristic \gls{GW} spectra for equatorial (left panel) and polar (right) observers, assuming a source distance of $\unit[1]{Mpc}$, compared with the design sensitivity curves of current- and next-generation interferometers.}
    \label{fig:GWs_det}
\end{figure}
We assess the detectability of the \gls{GW} emission by computing the characteristic strain \cite{Moore2015} for polar and equatorial observers,
\begin{equation}
    \label{eq:characteristic_strain}
    h_\textnormal{char}(f) = 2f\sqrt{|\tilde h_+(f)|^2 + |\tilde h_\times(f)|^2},
\end{equation}
where $\tilde{h}_{+/\times}(f)$ denote the Fourier transforms of the \gls{GW} at a distance $\mathcal{D}$.

Figure~\ref{fig:GWs_det} shows $h_\mathrm{char}/f^{1/2}$  (Equation~\eqref{eq:characteristic_strain}) for equatorial and polar 
observers, compared with the design \glspl{ASD} curves of the current-generation detectors, \gls{LIGO}~\cite{Aasi15,LVK20,LVKASDs}, Virgo~\cite{Acernese15,LVK20,LVKASDs}, and \gls{KAGRA}~\cite{Akutsu19,LVK20,LVKASDs}, as well as the projected \glspl{ASD} of third-generation detectors such as the \gls{ET} \cite{Hild11,ETASD} and \gls{CE} \cite{Srivastava22,CEASD}.

All models show a  maximum emission peak around $\unit[330]{Hz}$, which is associated with the $(\ell,m)=(2,2)$ mode generated by the \gls{LTWI} (Table~\ref{tab:quantitative_res}). This frequency lies close to the most sensitive band of both current- and next-generation detectors. The equatorial and polar spectra are broadly similar because the emission is dominated by the same non-axisymmetric instability. However, the equatorial emission is weaker since the symmetry about the equatorial plane suppresses the $h_\times$ polarization for equatorial observers.

Models \texttt{TS} and \texttt{LS} also show secondary peaks at approximately twice and three times the dominant frequency, corresponding to the excitation of higher-order spiral modes with $(\ell,m)=(3,3)$ and $(\ell,m)=(4,4)$.

To quantify detectability, we compute the maximum distance at which each model yields a \gls{SNR} of 8, adopting the optimal matched-filter \gls{SNR}. These distances assume an optimal matched-filter search with known signal templates and should therefore be regarded as optimistic estimates of the achievable detection horizon. We define the \gls{SNR} as in \citet{Moore2015},
\begin{equation}
    \label{eq:res_snr}
    \varrho = \sqrt{\int_0^{+\infty}\de\ln f \left(\frac{h_{\rm char}(f)}{h_n(f)}\right)^2},
\end{equation}
where $h_n(f)^2=f \textnormal{ASD}(f)^2$, in which we used  the \gls{ASD} relative to the detector noise.
The maximum distances at which each model achieves $\varrho=8$ for both polar and equatorial orientations are listed in Table~\ref{tab:det_GWs}.

The equatorial symmetry which suppresses the $h_\times$ polarization for equatorial observers reduces the detection distance by more than a factor of two relative to the polar direction. Among our models, \texttt{SL} reaches the largest detection distances:  $\unit[2.89]{Mpc}$ for polar observers with advanced \gls{LIGO}, and up to $\unit[26.1]{Mpc}$ with \gls{CE}. This is mainly due to its slightly stronger and broader dominant emission peak (Figure~\ref{fig:GWs_det}). In current detectors, all models are detectable well beyond the Milky Way for both orientations, with the most favourable (polar) line of sight reaching Andromeda \cite{Karachentsev2006}. In third-generation detectors, they become detectable throughout the Local Group, consistent with recent predictions for next-generation observatories \cite{Choi2024}.

The distances presented in this work exceed those reported by e.g., \cite{Powell2023} by factors of $\sim(3-5)$ (polar) and $\sim(8-10)$ (equatorial). This arises from both the much higher amplitudes of our models and the more favourable location of the dominant emission peak within the optimal sensitivity band of the detectors.

\begin{table}
    \centering
    \begin{tabular}{cc|ccccc}
    \hline
                                        & & \texttt{BH}        & \texttt{TS}        & \texttt{SL}        & \texttt{LS}        &  \texttt{BB} \\\hline\hline
    \multirow{ 2}{*}{\gls{LIGO}}  & {pol} & $\unit[2.44]{}$ & $\unit[2.37]{}$ & $\unit[2.89]{}$ & $\unit[2.62]{}$ & $\unit[2.50]{}$\\
                                  & {eqt} & $\unit[0.890]{}$  & $\unit[0.875]{}$  & $\unit[1.03]{}$ & $\unit[0.932]{}$  & $\unit[0.904]{}$\\\hline
    \multirow{ 2}{*}{Virgo}       & {pol} & $\unit[1.66]{}$ & $\unit[1.61]{}$ & $\unit[1.96]{}$ & $\unit[1.74]{}$ & $\unit[1.71]{}$\\
                                  & {eqt} & $\unit[0.605]{}$  & $\unit[0.598]{}$  & $\unit[0.700]{}$  & $\unit[0.623]{}$  & $\unit[0.619]{}$\\\hline
    \multirow{ 2}{*}{\gls{KAGRA}} & {pol} & $\unit[0.862]{}$  & $\unit[0.848]{}$  & $\unit[1.00]{}$ & $\unit[0.863]{}$  & $\unit[0.905]{}$\\
                                  & {eqt} & $\unit[0.313]{}$  & $\unit[0.314]{}$  & $\unit[0.358]{}$  & $\unit[0.310]{}$  & $\unit[0.325]{}$\\\hline
    \multirow{ 2}{*}{\gls{ET}}    & {pol} & $\unit[14.1]{}$ & $\unit[13.9]{}$ & $\unit[16.3]{}$ & $\unit[14.0]{}$ & $\unit[14.88]{}$\\
                                  & {eqt} & $\unit[5.13]{}$ & $\unit[5.17]{}$ & $\unit[5.85]{}$ & $\unit[5.03]{}$ & $\unit[5.35]{}$\\\hline
    \multirow{ 2}{*}{\gls{CE}}    & {pol} & $\unit[22.1]{}$ & $\unit[21.5]{}$ & $\unit[26.1]{}$ & $\unit[23.3]{}$ & $\unit[22.79]{}$\\
                                  & {eqt} & $\unit[8.10]{}$ & $\unit[8.04]{}$ & $\unit[9.38]{}$ & $\unit[8.41]{}$ & $\unit[8.28]{}$\\\hline
    \end{tabular}
    \caption{\label{tab:det_GWs} Maximum source distance (in Mpc) at which our models  achieve an optimal \gls{SNR} of 8 in current- and next-generation ground based detectors.}
\end{table}

\subsection{Neutrinos: Prospects of detection}
\label{sec:nu_detection}
Galactic \gls{CCSN} explosions are expected to emit neutrinos which are detectable by current- and next-generation neutrino experiments. In this study, we focus on neutrino observatories that, at the time of writing, are operational, are of different types, or provide complementary detection channels. For this reason, we consider \gls{SK} \cite{Abe2016}, IceCube \cite{Abbasi2011, Kopke2011}, \gls{JUNO} \cite{An2016}, and \gls{SNO+} \cite{Kraus2006}. 

\gls{SK} and IceCube are $\unit[32.5]{kt}$ and $\unit[3.5]{Mt}$ water (or ice) Cherenkov detectors, respectively, and are primarily sensitive to electron antineutrinos via the \gls{IBD} reaction,
\begin{equation}
    \overline{\nu}_e+p\to e^++n.
\end{equation}

\Gls{JUNO} and \gls{SNO+} are liquid scintillator neutrino observatories consisting of spherical volumes containing $\unit[20]{kt}$ and $\unit[780]{t}$ of linear alkylbenzene, respectively. The former is most sensitive to $\overline{\nu}_e$ via \gls{IBD}, while the latter is primarily sensitive to $\nu_e$ via charged-current interactions on carbon,
\begin{equation}
    \nu_e + {}^{12}\textnormal{C} \to {}^{12}\textnormal{N} + e^- .
\end{equation}

Since neutrino flavours may oscillate during their propagation from the collapsing star to the detector, we consider three scenarios: no flavour conversion, and adiabatic flavour conversion due to \gls{MSW} effects assuming either the normal or inverted neutrino mass hierarchy. The neutrino and antineutrino fluxes at Earth are given by \cite{Dighe2000}:
\begin{equation}
    \begin{split}
        F_e &= p_mF^0_e + (1-p_m)F^0_x,\\
        \overline{F}_e &= \overline{p}_m\overline{F}^0_e + (1-\overline{p}_m)F^0_x,\\
        F_\mu+F_\tau &=  (1-p_m)F^0_e + (1+p_m)F^0_x,\\
        \overline{F}_\mu+\overline{F}_\tau &=   (1-\overline{p}_m)\overline{F}^0_e + (1+\overline{p}_m)F^0_x,
    \end{split}
\end{equation}
where $F^0$ and $\overline{F}^0$ denote the neutrino and antineutrino fluxes at the source in the absence of flavour conversion, the subscript $x$ represents a heavy-lepton neutrino species, and $p_m$ and $\overline{p}_m$ are the neutrino and antineutrino survival probabilities, respectively, with $m=(n,i)$ corresponding to the normal or inverted mass hierarchy. The neutrino and antineutrino survival probabilities are given by
\begin{equation}
    \begin{split}
        p_n&=\sin^2\theta_{13},\\
        \overline{p}_n&=\cos^2\theta_{12}\cos^2\theta_{13},
    \end{split}\qquad
     \begin{split}
        p_i&=\sin^2\theta_{12}\cos^2\theta_{13},\\
        \overline{p}_i&=\sin^2\theta_{13},
    \end{split}
\end{equation}
where $\theta_{12}$ and $\theta_{13}$ are the neutrino mixing angles, set to $\sin^2\theta_{12}=0.297$ and $\sin^2\theta_{13}=0.025$ \cite{Capozzi2017}.

Neutrino flavour conversion is simulated using \texttt{SNEWPY} \cite{Baxter2022}, while detector effects and event rates are simulated with \texttt{SNOwGLoBES} \cite{Scholberg2021}.

\begin{figure}[t]
    \centering
    \includegraphics[width=\textwidth]{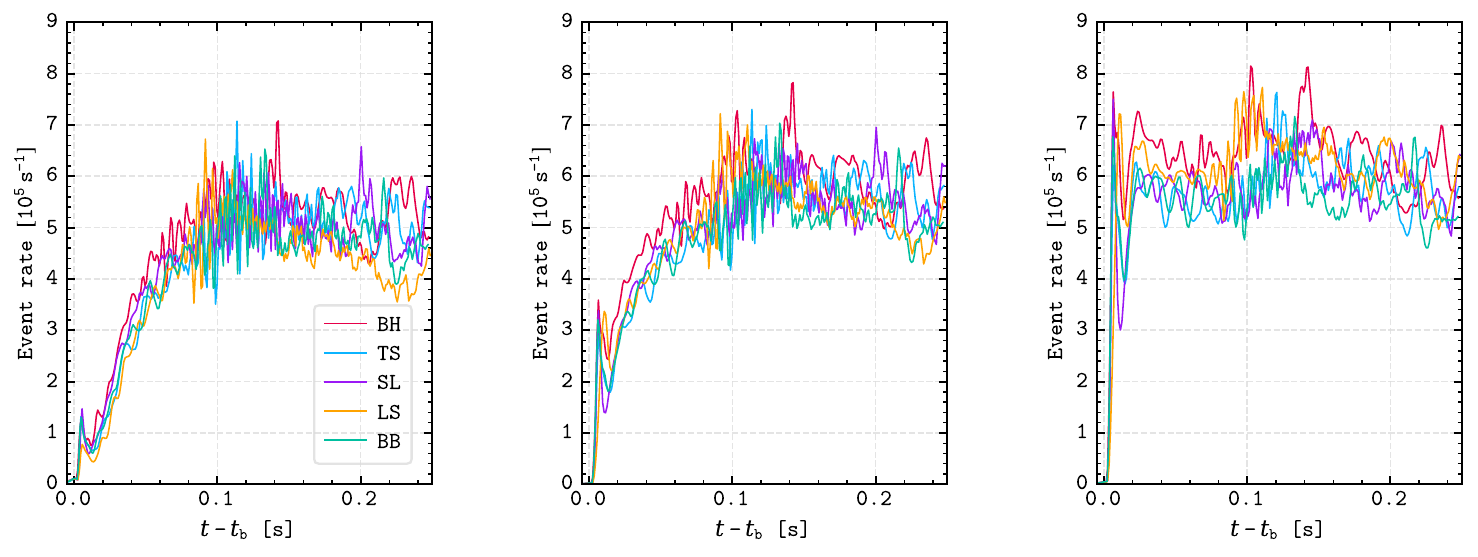}
    \caption{Neutrino event rates based on the equatorial, $(\theta,\,\phi) = (\frac{\pi}{2},\,0)$, isotropic equivalent luminosities for the main detection channel in IceCube, assuming no neutrino oscillations (left panel), adiabatic \gls{MSW} with normal  (middle panel) and inverted (right panel) mass order at a distance of $\unit[10]{kpc}$. 
    }
    \label{fig:nu_count}
\end{figure}

Figure~\ref{fig:nu_count} shows the neutrino event rates expected in IceCube at a reference distance of $\unit[10]{kpc}$ for our set of models, assuming no flavour conversion and adiabatic \gls{MSW} conversion with normal and inverted mass ordering (left, middle, and right panels, respectively). For all oscillation scenarios, the predicted event rates are of order $\sim\unit[5\times10^5]{events/s}$, consistent with the findings of \cite{Choi2025}. The overall event rate depends on the assumed flavour-conversion model, with the inverted-hierarchy \gls{MSW} case showing a more sustained, constant event rate immediately following the bounce compared to the other models. 

Importantly, the rapid oscillations observed in the equatorial neutrino luminosity (Figure~\ref{fig:nu_angle}) persist across all flavour conversion models in the detected event rates. The modulation associated with the spiral mode survives the flavour-conversion and detector-response pipeline in the ideal event-rate signal. At Galactic distances, the predicted event rates are sufficiently high that such a modulation could be observable, depending on its fractional amplitude and on the adopted time binning. However, the maximum distance at which the modulation itself can be identified is expected to be smaller than the distance at which the burst can be detected, and requires a dedicated statistical analysis including Poisson fluctuations and detector backgrounds. Assuming that of order $10^3$ detected events are required to identify the high-frequency modulation, the corresponding distance scale is $\sim 10^2\,{\rm kpc}$ for the event rates considered here, rather than the larger distance at which the overall burst may still be detectable.

\begin{figure}[t]
    \centering
    \includegraphics[width=\textwidth]{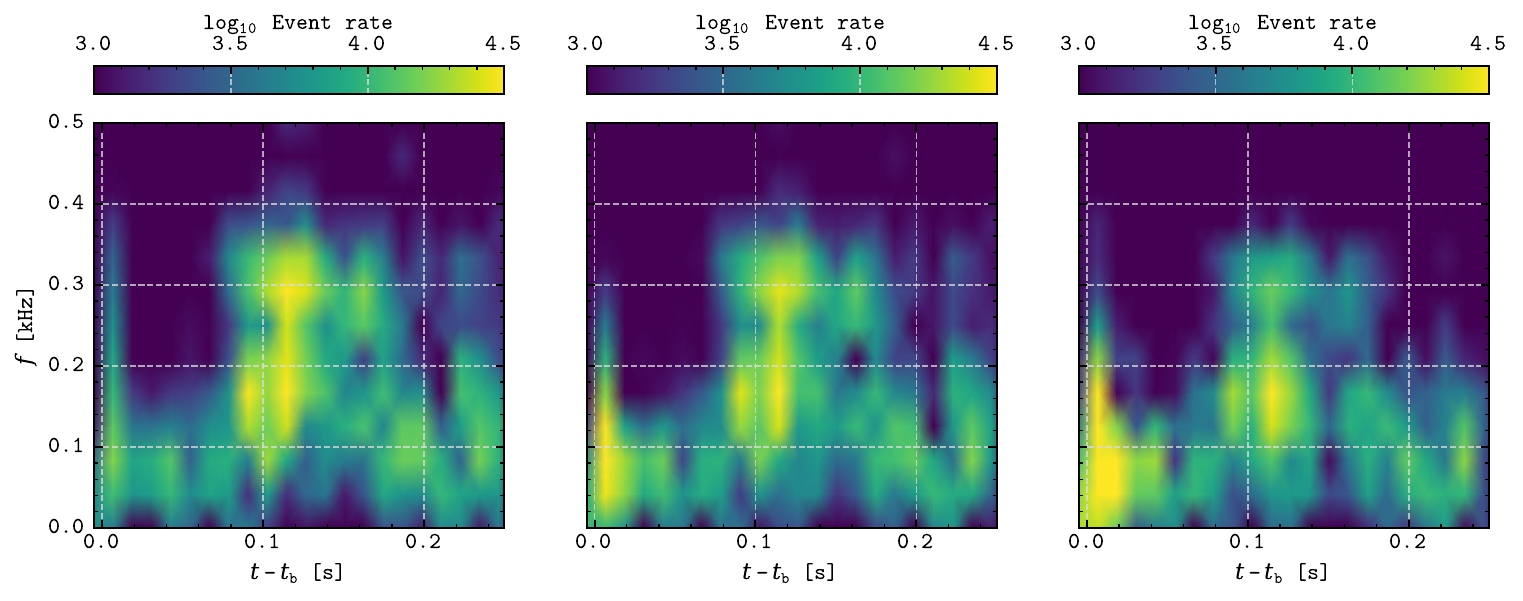}
    \caption{Spectrograms of the neutrino event rate of the equatorial neutrino luminosities for model \texttt{TS} simulated in the main detection channel in IceCube, assuming no neutrino oscillations (left panel), adiabatic \gls{MSW} with normal  (middle panel) and inverted (right panel) mass order at a distance of $\unit[10]{kpc}$. The neutrino event rate has a non-zero mean causing an excess of power in the low frequencies, therefore in all spectrograms a high-pass filter at $f=\unit[50]{Hz}$ has been applied. 
    }
    \label{fig:nu_count_spectro}
\end{figure}

The time-frequency maps of the equatorial neutrino event rates for model \texttt{TS}, in Figure~\ref{fig:nu_count_spectro} further confirm the imprints of the \gls{LTWI}. All the assumed flavour-conversion scenarios show clear imprints of the spiral modes at frequencies just below $\unit[200]{Hz}$ and $\unit[400]{Hz}$ (see Table~\ref{tab:quantitative_res}) persisting from approximately $\unit[75]{ms}$ to $\unit[140]{ms}$. The dominant features  in the event-rate spectrograms are consistent with the \gls{LTWI} modulation shown in Figure~\ref{fig:nu_spectro}, demonstrating a potentially robust multimessenger correlation for the \gls{LTWI}. 
While flavour conversion affects the overall event-rate normalisation, the characteristic frequencies and temporal evolution of the instability-induced modulations are largely insensitive to the oscillation model.

\begin{table}[t]
    \centering
    \begin{tabular}{ccc|ccccc}
    \hline
        & & & \texttt{BH} &	\texttt{TS} & \texttt{SL} &	\texttt{LS} & \texttt{BB} \\\hline\hline
        \multirow{9}{*}{IceCube} & \multirow{3}{*}{avg} & No & $\unit[724]{}$ & $\unit[692]{}$ & $\unit[704]{}$ & $\unit[670]{}$  & $\unit[687]{}$ \\
        &  & \gls{MSW} NMO & $\unit[784]{}$ & $\unit[740]{}$ & $\unit[751]{}$ & $\unit[739]{}$  & $\unit[735]{}$ \\
        &  & \gls{MSW} IMO & $\unit[844]{}$ & $\unit[790]{}$ & $\unit[797]{}$ & $\unit[812]{}$  & $\unit[786]{}$ \\
        & \multirow{3}{*}{eqt} & No & $\unit[677]{}$ & $\unit[653]{}$ & $\unit[658]{}$ & $\unit[625]{}$  & $\unit[642]{}$ \\
        &  & \gls{MSW} NMO & $\unit[734]{}$ & $\unit[701]{}$ & $\unit[705]{}$ & $\unit[670]{}$  & $\unit[690]{}$ \\
        &  & \gls{MSW} IMO & $\unit[793]{}$ & $\unit[751]{}$ & $\unit[751]{}$ & $\unit[777]{}$  & $\unit[741]{}$ \\
        & \multirow{3}{*}{pol} & No & $\unit[821]{}$ & $\unit[768]{}$ & $\unit[775]{}$ & $\unit[742]{}$  & $\unit[771]{}$ \\
        &  & \gls{MSW} NMO & $\unit[883]{}$ & $\unit[817]{}$ & $\unit[823]{}$ & $\unit[812]{}$  & $\unit[820]{}$ \\
        &  & \gls{MSW} IMO & $\unit[944]{}$ & $\unit[866]{}$ & $\unit[868]{}$ & $\unit[885]{}$  & $\unit[869]{}$ \\\hline
        \multirow{9}{*}{\gls{JUNO}} & \multirow{3}{*}{avg} & No & $\unit[83]{}$ & $\unit[80]{}$ & $\unit[80]{}$ & $\unit[77]{}$  & $\unit[79]{}$ \\
        &  & \gls{MSW} NMO & $\unit[82]{}$ & $\unit[78]{}$ & $\unit[79]{}$ & $\unit[77]{}$  & $\unit[78]{}$ \\
        &  & \gls{MSW} IMO & $\unit[77]{}$ & $\unit[73]{}$ & $\unit[74]{}$ & $\unit[74]{}$  & $\unit[73]{}$ \\
        & \multirow{3}{*}{eqt} & No & $\unit[78]{}$ & $\unit[76]{}$ & $\unit[75]{}$ & $\unit[73]{}$  & $\unit[74]{}$ \\
        &  & \gls{MSW} NMO & $\unit[77]{}$ & $\unit[74]{}$ & $\unit[74]{}$ & $\unit[73]{}$  & $\unit[73]{}$ \\
        &  & \gls{MSW} IMO & $\unit[73]{}$ & $\unit[70]{}$ & $\unit[69]{}$ & $\unit[70]{}$  & $\unit[69]{}$ \\
        & \multirow{3}{*}{pol} & No & $\unit[93]{}$ & $\unit[88]{}$ & $\unit[88]{}$ & $\unit[84]{}$  & $\unit[88]{}$ \\
        &  & \gls{MSW} NMO & $\unit[92]{}$ & $\unit[86]{}$ & $\unit[86]{}$ & $\unit[81]{}$  & $\unit[86]{}$ \\
        &  & \gls{MSW} IMO & $\unit[86]{}$ & $\unit[80]{}$ & $\unit[80]{}$ & $\unit[80]{}$  & $\unit[81]{}$ \\\hline
        \multirow{9}{*}{\gls{SK}} & \multirow{3}{*}{avg} & No & $\unit[86]{}$ & $\unit[83]{}$ & $\unit[84]{}$ & $\unit[80]{}$  & $\unit[82]{}$ \\
        &  & \gls{MSW} NMO & $\unit[85]{}$ & $\unit[72]{}$ & $\unit[82]{}$ & $\unit[80]{}$  & $\unit[81]{}$ \\
        &  & \gls{MSW} IMO & $\unit[81]{}$ & $\unit[77]{}$ & $\unit[77]{}$ & $\unit[77]{}$  & $\unit[76]{}$ \\
        & \multirow{3}{*}{eqt} & No & $\unit[81]{}$ & $\unit[79]{}$ & $\unit[78]{}$ & $\unit[75]{}$  & $\unit[77]{}$ \\
        &  & \gls{MSW} NMO & $\unit[80]{}$ & $\unit[77]{}$ & $\unit[77]{}$ & $\unit[76]{}$  & $\unit[76]{}$ \\
        &  & \gls{MSW} IMO & $\unit[76]{}$ & $\unit[73]{}$ & $\unit[72]{}$ & $\unit[74]{}$  & $\unit[72]{}$ \\
        & \multirow{3}{*}{pol} & No & $\unit[97]{}$ & $\unit[92]{}$ & $\unit[92]{}$ & $\unit[88]{}$  & $\unit[92]{}$ \\
        &  & \gls{MSW} NMO & $\unit[96]{}$ & $\unit[90]{}$ & $\unit[90]{}$ & $\unit[88]{}$  & $\unit[90]{}$ \\
        &  & \gls{MSW} IMO & $\unit[90]{}$ & $\unit[84]{}$ & $\unit[84]{}$ & $\unit[84]{}$  & $\unit[84]{}$ \\\hline
        \multirow{9}{*}{\gls{SNO+}} & \multirow{3}{*}{avg} & No & $\unit[16]{}$ & $\unit[16]{}$ & $\unit[16]{}$ & $\unit[15]{}$  & $\unit[15]{}$ \\
        &  & \gls{MSW} NMO & $\unit[16]{}$ & $\unit[15]{}$ & $\unit[15]{}$ & $\unit[15]{}$  & $\unit[14]{}$ \\
        &  & \gls{MSW} IMO & $\unit[15]{}$ & $\unit[14]{}$ & $\unit[15]{}$ & $\unit[15]{}$  & $\unit[15]{}$ \\
        & \multirow{3}{*}{eqt} & No & $\unit[15]{}$ & $\unit[15]{}$ & $\unit[15]{}$ & $\unit[14]{}$  & $\unit[14]{}$ \\
        &  & \gls{MSW} NMO & $\unit[15]{}$ & $\unit[15]{}$ & $\unit[15]{}$ & $\unit[14]{}$  & $\unit[14]{}$ \\
        &  & \gls{MSW} IMO & $\unit[14]{}$ & $\unit[14]{}$ & $\unit[14]{}$ & $\unit[14]{}$  & $\unit[14]{}$ \\
        & \multirow{3}{*}{pol} & No & $\unit[18]{}$ & $\unit[17]{}$ & $\unit[17]{}$ & $\unit[17]{}$  & $\unit[17]{}$ \\
        &  & \gls{MSW} NMO & $\unit[18]{}$ & $\unit[17]{}$ & $\unit[17]{}$ & $\unit[17]{}$  & $\unit[17]{}$ \\
        &  & \gls{MSW} IMO & $\unit[17]{}$ & $\unit[16]{}$ & $\unit[16]{}$ & $\unit[16]{}$  & $\unit[16]{}$ \\
        \hline
    \end{tabular}
    \caption{\label{tab:det_nu} Maximum distance (in kpc) at which  our models yield 25 neutrino detections in the first $\unit[250]{ms}$ for current neutrino experiments and no neutrino flavour oscillations, adiabatic \gls{MSW} with normal (NMO) and inverted (IMO) mass order.}
\end{table}

Finally, we quantify the detectability distances for polar, equatorial, and angle-averaged neutrino emission assuming no flavour conversion and adiabatic \gls{MSW} conversion with normal and inverted mass ordering. We define the detectability distance as the maximum distance at which at least 25 neutrino events are expected over the $\unit[250]{ms}$ duration of the simulations. This choice is historically motivated, as it corresponds to the order of magnitude of the total number of neutrinos detected from SN~1987A, and is adopted here as a common benchmark to facilitate a direct comparison between detectors, emission geometries, and flavour-conversion scenarios, rather than as a formal detection-significance criterion. Additionally, we note that more sophisticated detectability criteria which take into account time-dependent backgrounds, will lower the detectability estimates. Therefore, the resulting distances for IceCube, \gls{SK}, \gls{JUNO}, and \gls{SNO+} summarised in Table~\ref{tab:det_nu} should be regarded as optimistic upper limits for burst detection. 
IceCube achieves the largest detection distances, reaching $\sim\unit[700-800]{kpc}$, due to its large instrumented volume. \gls{SK} and \gls{JUNO}, which have comparable target masses and are both dominated by the \gls{IBD} channel, yield similar detection distances of approximately $\unit[80]{kpc}$. \gls{SNO+}, with its smaller mass and primary sensitivity to $\nu_e$, reaches maximum distances of order $\sim\unit[15]{kpc}$. Among our models, \texttt{BH} and \texttt{LS} show the largest and smallest detectability distances, respectively, consistent with their higher and lower neutrino luminosities and mean energies. As already noted in Section~\ref{sec:neutrinos}, polar emission  yields larger detection distances  than equatorial and average emission for all detectors. The detection horizon is also affected by the assumed flavour-conversion scenario: \gls{SK} and \gls{JUNO} are most favourable in the absence of oscillations, whereas IceCube and \gls{SNO+} achieve larger distances for adiabatic \gls{MSW} conversion with inverted mass ordering, reflecting the different flavour sensitivities of the detectors.

We compare the detection distances for \glspl{GW} and neutrino emission (Tables~\ref{tab:det_GWs} and \ref{tab:det_nu}). For neutrinos, we adopt the distances obtained for \gls{SK}, as its detection is dominated by resolved inverse beta decay events, providing a relatively direct link between the detected counts and the underlying neutrino flux. For \glspl{GW}, we consider the sensitivity of a single Advanced \gls{LIGO} detector representative of the current generation.
We find that the \gls{GW} signal is detectable at distances $\sim 8$ to $\sim 30$ times larger than those for neutrinos, depending on the viewing angle. In contrast, the neutrino signal remains effectively limited to distances comparable to that of SN~1987A. This highlights the broader reach of \glspl{GW} for probing rapidly rotating \gls{CCSN} events, despite their stronger dependence on source orientation.

\subsection{Influence of the EOS on the detection distance}
The nuclear \gls{EOS} introduces only modest variations (of order $\sim 5\%$) in the overall neutrino detectability horizon, even though it affects several quantities that determine the detailed properties of the signal.

Among the models, the total emitted neutrino energy varies only moderately, $E_\nu \simeq (5.9-6.6)\times10^{52}\,{\rm erg}$, which corresponds to a spread of about $\Delta E_\nu/E_\nu\sim 10\%$. 
Since the number of detected events scales approximately as $N_{\rm det}\propto E_\nu/(4\pi D^2)$, the maximum detection distance behaves as $D_{\rm max}\propto \sqrt{E_\nu}$. 
\begin{equation}
\frac{\Delta D_{\rm max}}{D_{\rm max}}
\simeq \frac{1}{2}\frac{\Delta E_\nu}{E_\nu}
\sim 5\%.
\end{equation}
Hence, differences among the considered \glspl{EOS} affect the detection distance only at the few–percent level.

The different \glspl{EOS} influence the \gls{PNS} structure, in particular its radius at early times, which in turn modifies the location and temperature of the neutrinosphere. In our models the \gls{PNS} radius varies in the range $R_{\rm PNS}\sim 59-66\,{\rm km}$.
Because the detection cross section of the dominant \gls{IBD} channel scales approximately as $\sigma_{\rm IBD}\propto E_\nu^2$ \citep{Vogel-Becom_1999PhRvD..60e3003,Strumia_2003PhLB..564...42}, even modest changes in the mean neutrino energy can produce noticeable variations in the event rate. Thus \gls{EOS}–dependent differences in the \gls{PNS} compactness may translate into small but measurable differences in the detected neutrino signal.

The \gls{LTWI} produces time-dependent modulations in the neutrino emission. The characteristic frequency of this variability is set by the spiral mode frequency, $f_{\rm LTWI}\sim 150-180\,{\rm Hz}$, which corresponds to millisecond-scale luminosity variations. The amplitude of these modulations is typically only a few percent of the total luminosity, so their detectability depends critically on the event statistics of the detector. Large-volume detectors such as IceCube are particularly sensitive to such
high-frequency variations because their enormous effective detection volume
provides very high counting statistics and millisecond time resolution
\citep{Abbasi2011,Lund_2012PhRvD..86j5031}.

Finally, the \gls{EOS} affects the duration of the \gls{LTWI} phase. In the models analysed here the instability lasts between $\sim40$ and $\sim170\,{\rm ms}$ depending on the \gls{EOS}. For a characteristic frequency $f_{\rm LTWI}\sim170\,{\rm Hz}$, this corresponds to a number of oscillation cycles
$N_{\rm cycles}\simeq f_{\rm LTWI}\,\Delta t \sim 7-30$.
A longer-lasting instability therefore increases the likelihood that periodic modulations could be extracted from the neutrino signal. In summary, while the \gls{EOS} has only a modest impact on the overall neutrino detection horizon, it influences several aspects of the signal, including the neutrinosphere temperature, the average neutrino energy, and the amplitude and duration of the \gls{LTWI}-driven variability,  which may provide additional diagnostics of the dense-matter physics in a nearby \gls{CCSN}.

\section{Conclusion}
\label{sec:conclusion}

We presented the first $\unit[250]{ms}$ of evolution from five rapidly rotating \gls{CCSN} simulations based on the same $\unit[35]{M_\odot}$ Wolf-Rayet progenitor, adopting identical initial magnetic-field and rotational configurations while varying the high-density \gls{EOS}. The selected \glspl{EOS}, namely BBSk3, BHB$\Lambda\phi$, LS220, SLy4, and TSO, span a broad range of stiffness, correlated in this paper to the value of the tidal Love number, symmetry-energy slopes, and maximum neutron-star masses, and include both \gls{RMF} and Skyrme-based descriptions of dense matter. All \glspl{EOS}, except LS220, are consistent with current astrophysical and experimental constraints, while LS220 enables direct comparison with a large body of previous \gls{CCSN} studies. The goal of this work was to investigate the impact of the nuclear \gls{EOS} on the development of non-axisymmetric instabilities during the early post-bounce phase, and on their associated multimessenger emission.

Our analysis shows that the high-density \gls{EOS} already affects the dynamics prior to core bounce, significantly modifying the stratification and rotational structure of the inner core. In particular, the \gls{EOS} alters both the rotational profile and the amount of differential rotational energy available at bounce, thereby influencing the initial conditions for the post-bounce evolution and the \gls{LTWI} growth.

Despite these \gls{EOS}-dependent differences, all models retain sufficient differential rotation and develop overlapping corotation and convectively unstable regions inside the \gls{PNS}, conditions that favour the onset of the \gls{LTWI}. Unlike previous studies of the instability \cite{Shibagaki2020,Takiwaki2021,Bugli23}, our simulations also include weak magnetic fields, whose strength is not sufficient to suppress the development of \gls{LTWI}. In all models, the \gls{LTWI} develops through the growth of large-scale non-axisymmetric spiral structures, dominated by low-order azimuthal modes ($m=1,\,2$) in the equatorial plane, and generates quasi-periodic modulations in both \glspl{GW} and neutrino emission lasting for approximately $\unit[100]{ms}$. The characteristic \gls{GW} and neutrino frequencies remain quasi-constant during the active phase of the instability, with typical values of $\sim\unit[330]{Hz}$ and $\sim\unit[170]{Hz}$, respectively. The characteristic \gls{GW} frequency is approximately twice the neutrino modulation frequency, consistent with the frequency-doubling relation reported by \citet{Takiwaki2018}. This behaviour can be explained by the neutrino signal tracing the rotation of the dominant non-axisymmetric spiral structure, while the \gls{GW} emission, being quadrupolar, oscillates at twice the pattern frequency. The \glspl{GW} and neutrino frequencies lie above the typical \gls{SASI} range ($\lesssim\unit[100]{Hz}$) and below the dominant \gls{PNS} oscillation frequencies ($\gtrsim\unit[500]{Hz}$), placing the \gls{LTWI} in a comparatively distinct region of the multimessenger spectrum. In contrast to previous studies \cite{Shibagaki2020,Takiwaki2021,Bugli23}, we do not observe a clear phase of increasing mode frequency during the activity of the instability.

We find that the peak frequency of the \gls{GW} emission associated with the \gls{LTWI} correlates approximately linearly with the stiffness of the nuclear \gls{EOS} as measured by $\kappa_2$, with stiffer \glspl{EOS} producing higher frequencies. This trend likely arises because the \gls{EOS} modifies the compactness and rotational structure of the \gls{PNS}, thus affecting the characteristic pattern frequency of the spiral modes. These results suggest that future \gls{GW} observations of rapidly rotating \glspl{CCSN} may provide direct constraints on the properties of dense matter. At the same time, a broader exploration of progenitor rotation rates will be necessary to disentangle \gls{EOS} effects from possible degeneracies with the rotational configuration of the collapsing core.

In addition, we find that the \gls{GW} component associated with the instability can be effectively isolated through the \gls{EEMD} decomposition of the strain signal. Together with the results of \citet{Cusinato2025a}, this further supports the usefulness of \gls{EEMD} techniques for associating specific frequency components of the \gls{GW} signal with their underlying emission regions.

Finally, we find that the nuclear \gls{EOS} also affects the amplitude of the neutrino breakout burst, while the onset of the \gls{LTWI} induces characteristic quasi-periodic modulations in the neutrino luminosities. These signatures primarily originate from neutrinos emitted near the equatorial region and exhibit distinct frequency components associated with the dominant non-axisymmetric modes. Corresponding features are also visible in the simulated event rates of next-generation neutrino detectors, highlighting the intrinsically multimessenger nature of the instability and the possibility of jointly probing its dynamics through \gls{GW} and neutrino observations.

Overall, our results indicate that the \gls{LTWI} is a robust feature of rapidly rotating \glspl{CCSN} across a broad range of nuclear \glspl{EOS}, while its characteristic multimessenger signatures retain measurable sensitivity to the properties of dense matter. Future work extending this analysis to a broader range of progenitors, rotation rates, magnetic-field strengths, and longer post-bounce evolution times will be necessary to assess the generality of these trends and their detectability with current- and next-generation multimessenger observatories.
In particular, this study considered only a single rapidly rotating Wolf-Rayet progenitor, limiting our ability to explore how progenitor properties such as initial mass, metallicity, compactness, and angular-momentum distribution affect the development of the \gls{LTWI} and its multimessenger signatures. Extending the analysis to a wider set of progenitors and to a larger sample of nuclear \glspl{EOS} will therefore be crucial to probe the robustness of the trends identified here.

Additionally, although our simulations include magnetic fields, the adopted pre-collapse configuration remains dynamically weak. Stronger magnetic-field configurations may substantially modify the rotational structure of the \gls{PNS} and the growth of non-axisymmetric instabilities. Exploring the interplay between magnetic stresses, differential rotation, and the \gls{LTWI} therefore represents an important direction for future investigations.

Finally, numerical resolution is known to affect the post-bounce dynamics of \glspl{CCSN}, including turbulence, shock evolution, and explosion properties \cite{Nagakura2019,Melson2020,Varma2026}. Higher-resolution simulations will then be required to assess the quantitative robustness of the instability growth, amplitude, and associated multimessenger emission reported in this work.

\ack{We thank the organisers and participants of \emph{SN 2025gw: First IGWN Symposium on Core Collapse Supernova Gravitational Wave Theory and Detection} held in Warsaw in July 2025 for inspiring discussions on the subject. MAA thanks the Yukawa Institute for Theoretical Physics at Kyoto University. Discussions during the YITP long-term workshop YITP-T-26-02 on "Multi-Messenger Astrophysics in the Dynamic Universe" were useful to complete this work. }

\funding{We acknowledge support from grants PID2021-127495NB-I00 and PID2025-171322NB-C22 funded by MCIN/AEI/10.13039/501100011033 and the European Union, as well as from the Astrophysics and High Energy Physics programme of the Generalitat Valenciana (ASFAE/2022/026), funded by MCIN and the European Union NextGenerationEU (PRTR-C17.I1), and from the Prometeo excellence programme grant CIPROM/2022/13 funded by the Generalitat Valenciana. MC acknowledges the support through the Generalitat Valenciana via the grant CIDEGENT/2019/031. MO was supported by the Ram\'on y Cajal programme of the Agencia Estatal de Investigación (RYC2018-024938-I).
The computations have been performed on servers Lluisvives and Tirant-4 (grant AECT-2025-2-0002) of the Servei d'Informàtica de la Universitat de València and on the Red Española de Supercomputación (RES) on MareNostrum (grants AECT-2025-1-0012 and AECT-2025-2-0006). MAA acknowledges partial funding of the YITP for attending the long-term workshop YITP-T-26-02.}

\data{The data are available upon reasonable request to the authors.}

\printbibliography

\end{document}